\documentclass[12pt,a4paper]{article}  
 \usepackage[skins,theorems]{tcolorbox}

\newtcolorbox{whitebox}{colback=white,colframe=black,boxrule=0.5mm,arc=4mm,auto outer arc}

\tcbset{highlight math style={enhanced,
  colframe=red,colback=white,arc=0pt,boxrule=1pt}}
  \usepackage[bookmarksopen, bookmarksnumbered, bookmarksopenlevel=2]{hyperref}
  \usepackage{tikz}
  \usepackage{tikz-3dplot}
  \usepackage{multirow}
 \usetikzlibrary{calc}
 \usetikzlibrary{decorations} %
 \usepackage[UKenglish]{babel}
 \usepackage[toc,page]{appendix}
 \usepackage{amsmath}
 \usepackage{amssymb}
 \usepackage{graphicx}
 \usepackage{hhline}
 \usepackage[bf]{caption}
\usepackage{cite}
\usepackage[vcentermath]{youngtab}
\usepackage{geometry}
\usepackage{slashed}
\usepackage{color}
\usepackage{stackrel}
\usepackage{tikz-cd} 
\usepackage{tkz-euclide}
\usepackage{cancel} 
\usepackage{subcaption}
\usepackage[normalem]{ulem}
\usepackage{mdframed}
\usepackage{adjustbox}
\usepackage{tcolorbox}
\usepackage{enumitem}
\usepackage{empheq}
\usepackage{arydshln}

\newenvironment{eqn*}{\begin{equation*}\begin{aligned}}{\end{aligned}\end{equation*}\noindent}

\newtheorem{claim}{Claim}

\newcommand{\bqa}{\begin{eqnarray}}
\newcommand{\eqa}{\end{eqnarray}}

\hypersetup{
    pdftitle={},
    pdfauthor={},
    pdfsubject={}
}
\numberwithin{equation}{section}
\numberwithin{table}{section}

\makeatletter

\definecolor{BF}{HTML}{f903d7}

\newtheorem{definition}{Definition}

\newcommand{\be}{\begin{equation}}
\newcommand{\ee}{\end{equation}}
\newcommand{\beq}{\begin{equation}}
\newcommand{\eeq}{\end{equation}}
\newcommand{\ba}{\begin{aligned}}
\newcommand{\ea}{\end{aligned}}

\newcommand{\bea}{\begin{eqnarray}}
\newcommand{\eea}{\end{eqnarray}}

\newcommand{\cO}{\mathcal{O}}
\newcommand{\cT}{\mathcal{T}}
\newcommand{\cE}{\mathcal{E}}

\newcommand{\cC}{\mathcal{C}}
\newcommand{\cD}{\mathcal{D}}

\newcommand{\cN}{\mathcal{N}}
\newcommand{\cW}{\mathcal{W}}

\newcommand{\cA}{\mathcal{A}}

\newcommand{\cB}{\mathcal{B}}

\newcommand{\cI}{\mathcal{I}}

\newcommand{\cR}{\mathcal{R}}
\newcommand{\cS}{\mathcal{S}}

\newcommand{\cV}{\mathcal{V}}

\newcommand{\cM}{\mathcal M}

\newcommand\bi{\begin{itemize}}
\newcommand\ei{\end{itemize}}

\renewcommand{\a}{{\alpha}}

\renewcommand{\l}{{\lambda}}

\def\Im{\mathop{\mathrm{Im}}\nolimits}

\def\Re{\mathop{\mathrm{Re}}\nolimits}

\def\Tr{\mathop{\mathrm{Tr}}\nolimits}

\def\unit{{1\kern-.65ex {\rm l}}}
\def\1{{1\kern-.65ex {\rm l}}}

\def\CE{{\cal E}}

\def\ii{{\rm i}}

\tcbset {
	base/.style={
		arc=0mm, 
		bottomtitle=0.5mm,
		boxrule=0mm,
		colbacktitle=black!10!white, 
		coltitle=black, 
		fonttitle=\bfseries, 
		left=2.5mm,
		leftrule=1mm,
		right=3.5mm,
		title={#1},
		toptitle=0.75mm,
		breakable
	}
}

\newtcolorbox{subbox}[1]{
	colframe=black!30!white,
	base={#1}
}

\newcount\hour \newcount\minute
\hour=\time \divide \hour by 60
\minute=\time
\def\now{%
\ifnum \hour<13
  \ifnum \hour=0 \advance \hour by 12 \number\hour:\else \number\hour:\fi%
     \ifnum \minute<10 0\fi%
     \number\minute%
\ A.M.%
\else \advance \hour by -12 \number\hour:%
  \ifnum \minute<10 0\fi%
  \number\minute%
  \ P.M.%
\fi%
}

\makeatother

\begin{document}

\begin{flushright}
 \small{ZMP-HH/26-5}
\end{flushright}

\vskip 40 pt
\begin{center}
{\large \bf
Quantum obstructions for $\cN=1$ infinite distance limits -- \vspace{2mm} \\
Part II: Kähler obstructions 
} 

\vskip 11 mm

Lukas Kaufmann,  Jeroen Monnee, Timo Weigand, and Max Wiesner

\vskip 11 mm
\small \textit{II. Institut f\"ur Theoretische Physik, Universit\"at Hamburg, Notkestrasse 9,\\ 22607 Hamburg, Germany} \\[3 mm]
\small \textit{Zentrum f\"ur Mathematische Physik, Universit\"at Hamburg, Bundesstrasse 55, \\ 20146 Hamburg, Germany  }   \\[3 mm]

\end{center}

\vskip 7mm

\begin{abstract}
We continue our analysis of quantum corrections in the complex structure moduli space of four-dimensional Type IIB/F-theory compactifications with ${\cal N}=1$ supersymmetry. We find that limits in the complex structure moduli space of F-theory generically induce a strong backreaction on other sectors of the theory, reflecting the non-factorisation of the field space in genuine $\mathcal{N}=1$ theories at the quantum level. Our focus is on quantum corrections to the K\"ahler moduli in F-theory on Calabi--Yau fourfolds and proceeds in two independent ways: A detailed analysis of the worldsheet theory of candidate EFT strings for pure complex structure infinite distance limits reveals a mismatch with expectations based on the classical effective action and points to a quantum obstruction of the limit. Complementary to this, we confirm, in large classes of theories, the existence of significant complex structure dependent quantum corrections to the action of BPS instantons which at tree-level are governed by the K\"ahler moduli. As the quantum corrections become uncontrolled at large complex structure, they require a co-scaling of the K\"ahler moduli to maintain perturbative control. As a result, the naive, classical effective action does not provide an accurate description of pure large complex structure regimes. We comment on  possible implications for string phenomenology, specifically with regard to model building and moduli stabilisation. 
\end{abstract}

\vfill

\thispagestyle{empty}
\setcounter{page}{0}
 \newpage
\tableofcontents
\vspace{25pt} 
\setcounter{page}{1}

\section{Introduction}

Quantum effects are known to play a key role for the structure of string vacua.
 For example, in the framework of Type IIB/F-theory compactifications to four dimensions, quantum effects are essential ingredients in the stabilisation of the volume moduli~\cite{Kachru:2003aw,Balasubramanian:2005zx,Conlon:2005ki} and hence are central to any attempt  
 of identifying string vacua with a realistic phenomenology or cosmology. 
  At the same time, if quantum effects become uncontrolled, the very framework of a perturbative expansion of the effective action loses its meaning unless a dual description encodes all quantum effects in its classical formulation.

 Quantum corrections to the effective action are often severely constrained thanks to non-renormalisation theorems.
  For instance, in theories with ${\cal N}=2$ supersymmetry, 
 the vector multiplet sector of the theory
 is protected against spacetime quantum corrections.
   Furthermore, regions at infinite distance in the vector multiplet moduli space feature parametrically small gauge couplings and are hence natural starting points to analyse the effective action.
  However, non-renormalisation theorems become substantially less constraining once supersymmetry is broken to at most four supercharges, as required in phenomenologically more realistic contexts. 
 Strikingly, even the 
 ${\cal N}=1$ descendants of the asymptotic vector multiplet regimes can now be subject to substantial quantum corrections.
  In \cite{Paper1}, we have explicitly analysed this more general phenomenon in the concrete context of Type IIB orientifolds and their complex structure moduli space. 
Our analysis has shown how quantum effects in the Type IIB string coupling $g_s$ substantially change the asymptotics of this moduli space. 
 As a drastic example, entire regions at infinite distance in the vector multiplet moduli space of a 4d ${\cal N}=2$ compactification of Type IIB string theory can be removed from the quantum corrected moduli space once supersymmetry is broken to ${\cal N}=1$: The relevant modulus now parametrises only a finite-distance rather than an infinite distance direction.
 This effect, while appearing radical from the point of view of the effective action alone, is automatically incorporated in the geometry of F-theory, which encodes the $g_s$-exact open-closed moduli space of Type IIB compactifications with 7-branes. In particular, as shown in~\cite{Alim:2009rf,Alim:2009bx,Grimm:2009ef,Alim:2010za}, the F-theory complex structure moduli space incorporates the 7-brane position moduli and the associated superpotential~\cite{Lerche:1998nx,Lerche:2001cw,Lerche:2002ck,Jockers:2008pe,Jockers:2009mn}.

  In the present work, we continue the analysis of \cite{Paper1} with a focus on 
  quantum corrections to K\"ahler moduli
 in asymptotic regions of the complex structure moduli space of four-dimensional F-theory compactifications with minimal supersymmetry. Earlier works on the complex structure moduli spaces of Calabi--Yau fourfolds and concrete examples can be found, e.g., in~\cite{Greene:1993vm,Mayr:1996sh,Klemm:1996ts,Intriligator:2012ue,Gerhardus:2016iot,Cota:2017aal,Grimm:2019ixq,Grimm:2023lrf,vandeHeisteeg:2024lsa,Grimm:2024fip}.
  The quantum corrections which we will identify not only affect the study of the asymptotic physics in far-off regions of the moduli space, for example from the point of view of the Distance Conjecture~\cite{Ooguri:2006in} or the Emergent String Conjecture~\cite{Lee:2019oct}, but they also appear to become essential in the context of moduli stabilisation and for analysing cosmological solutions in extreme regimes of the moduli space. 

The general principle underlying the importance of quantum effects in the regime of large complex structure is simple: The action of BPS instantons which are classically controlled by $g_s$ or the K\"ahler moduli can receive 
complex structure dependent threshold corrections. At large complex structure, these may exceed the tree-level contribution. This signals a departure from a perturbatively controlled regime
and might, in particular, lead to unsuppressed instanton effects. To restore computational control, the modulus governing the tree-level action -- $g_s$ in the case of the effects studied in \cite{Paper1} or the K\"ahler moduli for the present paper -- must be co-scaled appropriately as the complex structure is taken to infinity. 
 This picture also explains why, by contrast, large volume limits in 4d ${\cal N}=1$ Type II orientifolds/F-theory \cite{Lee:2019jan,Klaewer:2020lfg,Lanza:2021udy,Cota:2022yjw} are not quantum obstructed.
 For large complex structure limits, however, the effects may be significant and, as already pointed out above, can  remove  classical infinite distance directions in the full quantum corrected moduli space \cite{Paper1}. 

In the sequel, we will oftentimes refer to the effects studied in \cite{Paper1} and the present paper as $g_s$ or K\"ahler corrections because we are primarily interested in quantum obstructions to limits in the pure complex structure moduli space which can be avoided, at best, by a co-scaling of $g_s$ and the K\"ahler moduli. From the point of view of the effective action, however, these obstructions have their origin in complex structure dependent corrections to BPS quantities governed classically by $g_s$ or the K\"ahler moduli, as explained above.

Computing quantum corrections to the moduli space geometry for a general 4d ${\cal N} = 1$ theory is a difficult problem that has not been solved in full generality; see, however, \cite{Becker:2002nn,vonGersdorff:2005bf,Berg:2005ja,Berg:2007wt,Berg:2011ij,Grimm:2013gma,Grimm:2013bha,Berg:2014ama,Minasian:2015bxa,Weissenbacher:2019mef,Klaewer:2020lfg,Cicoli:2021rub,Wiesner:2022qys,Cvetic:2024wsj,Casas:2025qxz} for a selection of results in this direction. Due to this difficulty, we are lead to follow alternative strategies to analyse the effect of the quantum corrections. In \cite{Paper1}, the geometry of F-theory came to the rescue as it resums all $g_s$ effects exactly. 
For corrections to the K\"ahler moduli, no such \emph{dea ex machina} is in sight.

As outlined in more detail in Section \ref{ssec:Kahleroverview}, we therefore pursue two complementary approaches to identify corrections to the K\"ahler moduli which become relevant in the large complex structure regime. In one approach, we study the worldsheet theory of so-called EFT strings \cite{Lanza:2021udy}, which govern the possible infinite distance limits of the supergravity effective action. EFT strings in various 4d compactifications of string theory have recently played a central role in the investigations of~\cite{Marchesano:2022avb,Martucci:2022krl,Marchesano:2022axe,Martucci:2024trp,Grieco:2025bjy,Hassfeld:2025uoy}. Here, we focus on the less explored EFT strings associated with complex structure limits in Type IIB/F-theory and in particular on their worldsheet theory. These strings can be analysed by understanding the limits as semi-stable degenerations of the compactification space \cite{Hassfeld:2025uoy, Monnee:2025ynn}.
In the ${\cal N}=1$ context studied in this paper, we observe a systematic mismatch between the microscopics of the candidate EFT strings and the expectations from the classical effective field theory: The classical effective action fixes the scaling of the EFT string tension and the quantum gravity cutoff in the infinite distance regimes. Via an application of the Emergent String Conjecture~\cite{Lee:2019oct} these scaling then determine the features of the dual frame that would have to emerge if the infinite distance limits were unobstructed. Although the Emergent String Conjecture plays an important role in our analysis, the way we employ it differs significantly from previous tests in concrete string theory setups, see\cite{Lee:2018urn,Lee:2019oct,Lee:2019apr,Marchesano:2019ifh,Baume:2019sry,Lee:2019jan,Klaewer:2020lfg,Xu:2020nlh,Alvarez-Garcia:2021pxo,Lee:2021qkx,Lee:2021usk,Basile:2022zee,Rudelius:2023odg,Alvarez-Garcia:2023gdd,Alvarez-Garcia:2023qqj,Chen:2024cvc,Blumenhagen:2023yws,Aoufia:2024awo,Hassfeld:2025uoy,Grieco:2025bjy,Monnee:2025ynn,Monnee:2025msf}: These previous works successfully identified the tower of states predicted by the Emergent String Conjecture, thereby confirming the latter in the considered infinite distance limit. Here, we instead use the constraints implied by the Emergent String Conjecture for the microscopics of the candidate EFT strings. A violation of these constraints then signals a quantum obstruction of the classical infinite distance limit under consideration. This mismatch between the classical effective action and the properties of the candidate EFT strings can only be resolved by superimposing the complex structure limit with an additional limit for the K\"ahler moduli, very much in the spirit of the discussion above. This is an indirect sign of what we call a quantum K\"ahler obstruction of the pure complex structure limit.

  This indirect but powerful approach to detect quantum K\"ahler obstructions can be further reinforced via a complementary strategy, namely by explicitly identifying the complex structure dependent quantum corrections to the action of BPS instantons. As stressed already, the computation of these corrections is currently not possible in full generality. Nonetheless, we can show that in large classes of complex structure limits the leading complex-structure-dependent corrections to the BPS instanton action imply a K\"ahler obstruction of these limits. This lends further support to the more generally applicable approach to detect K\"ahler obstructions via the microscopics of the candidate EFT strings. The analysis hence illustrates that arguments developed in the context of the Swampland program~\cite{Vafa:2005ui}, see~\cite{Palti:2019pca,vanBeest:2021lhn,Agmon:2022thq} for reviews, can be used to infer technical features of quantum gravitational theories for which a direct computation is challenging given the current techniques. 
    
  The upshot is that infinite distance regions in the complex structure moduli space at finite values of the Einstein-frame K\"ahler moduli are beyond the regime of perturbative control. That such a scenario is possible has been anticipated in~\cite{Seo:2025ejt}. The results of this work show that it is indeed realized. In particular, this loss of perturbative control must be taken into account in attempts to stabilise the moduli in large complex structure regimes, where the structure of the flux induced superpotential is particularly amenable to explicit computations. 

Let us now summarise the content of our analysis in more detail.
In Section \ref{sec:overview}, we give a general overview of the notions of $g_s$  and K\"ahler obstructions to complex structure limits (see in particular Definitions \ref{def:2fusionheunconnected} and \ref{def:2fusionheunconnectedribbon}) and explain in greater depth our two complementary strategies to detect these. The main results are captured in Claims \ref{claim:explicitobstructions}  and 
\ref{claim:EFTstringobstruction}.

In Section \ref{ssec:CY4degenerations}, we first review the classification of complex structure infinite distance limits for F-theory on elliptic Calabi-Yau fourfolds as limits of type II, III, IV and V~\cite{Grimm:2019ixq}. We then divide these into two classes depending on whether or not the generic elliptic fiber degenerates: Regular-fiber limits can be thought of as limits where only the base degenerates, while in I$_n$-type limits the elliptic fiber degenerates in codimension-zero at least over some components of the base. In Section \ref{ssec:countingFM}, we explain how to count the worldsheet modes of EFT strings associated with these limits via duality with M-theory. An outline of the counting strategy for general complex structure EFT strings in Type IIB compactifications on Calabi-Yau threefolds, based on their semi-stable degeneration \cite{Hassfeld:2025uoy, Monnee:2025ynn}, is provided in Appendix \ref{app:WSmodesIIB}.
 The counting procedure developed in Section  \ref{ssec:countingFM} is the basis for our first strategy to detect K\"ahler obstructions to complex structure limits by observing that the candidate EFT strings do not behave as they should if they were to induce a well-defined infinite distance limit in the complex structure moduli space.

 In Section \ref{sec:Sentype}, we give a general argument along these lines for all I$_n$-type limits: In these cases, the existence of vector multiplets on the EFT string candidates points to an inconsistency of the naive complex structure limits. 
 For the standard Sen-limit~\cite{Sen:1996vd,Clingher:2012rg}, this is in agreement with expectations from the effective field theory, as observed already in \cite{Lanza:2020qmt}, and can be resolved by a co-scaling of the Einstein-frame K\"ahler moduli.
Furthermore,  this argument excludes all pure limits of type V because they necessarily involve a degeneration of the elliptic fiber.

 In Section \ref{sec:regularfiber}, we analyse regular-fiber limits. As we explain, for limits of type II and III, the candidate EFT strings fail to satisfy the constraints imposed by the effective action of being critical or six-dimensional supercritical strings, respectively. These results are also in agreement with our second, complementary strategy based on the explicit computation of quantum corrections, which is possible in certain cases. For a special class of type II limits
(in which the F-theory elliptic fourfold is also K3-fibered with base ${\cal B}_2$), duality with the heterotic string allows us to explicitly confirm the existence of quantum corrections to the volume of ${\cal B}_2$ whose magnitude depends on the complex structure moduli. 
 Similarly, for certain type III limits for which the F-theory fourfold is fibered by a Calabi--Yau threefold $X_3$, 
 a chain of dualities implies strong quantum corrections to the volume of the base of $X_3$ viewed as an elliptic fibration. 
The only limits for which we cannot directly point to an inconsistency are regular fiber limits of type IV, but we view this more as a technical obstacle rather than as pointing to special behaviour of this class of limits.

In Section~\ref{ssec:mirror}, we briefly discuss Type IIA orientifolds which are mirror dual to the perturbative Type IIB orientifolds/ F-theory setups studied in the rest of our work. In particular, we comment on how the obstructions to complex structure limits in perturbative Type IIB orientifolds translate to the mirror dual Type IIA side.

Finally, in Section \ref{sec:pheno} we discuss the consequences of our findings for moduli stabilisation and the computation of the effective action. The key message is that even in regions of large complex structure, additional quantum effects in the K\"ahler moduli can arise and imply a significant deviation from the naive low-energy supergravity approximation which must be addressed.

\section{General overview and a simple example}\label{sec:overview}
This work continues the analysis in the companion paper~\cite{Paper1} investigating the interplay between complex structure degenerations and weak coupling limits for orientifolds of Type IIB Calabi--Yau threefold compactifications. The conclusion of~\cite{Paper1} is that the \emph{perturbative} Type IIB effective action alone can miss crucial quantum effects that obstruct a large class of complex structure deformations in the sense of
\begin{definition}\label{def:2fusionheunconnected}
    Consider Type IIB string theory compactified on an orientifold of a Calabi--Yau threefold $V$ with orientifold action $\Omega$ leading to O7/O3-planes and consider an infinite distance limit in the complex structure moduli space $\cM_{\rm c.s.}(V/\Omega)$. Such a limit is called $g_s{\rm\text{-}obstructed}$ if in the $g_s$-corrected moduli space $\cM_Q(V/\Omega)$ the following two conditions cannot be satisfied simultaneously when taking the limit:
    \begin{enumerate}
        \item The Type IIB string coupling remains finite, $g_s>0$, along the limit. 
        \item The 4d $\cN=1$ effective action derived from Type IIB string theory on $V/\Omega$ remains under perturbative control in $g_s$. 
    \end{enumerate}
\end{definition}
These obstructions become apparent when studying the actual geometry of the orientifolded Calabi--Yau threefold and in particular its uplift to F-theory. 

To study the infinite distance regimes in the $g_s$-corrected 4d $\cN=1$ moduli space, we directly consider the lift to F-theory on an elliptically fibered Calabi--Yau fourfold
\begin{equation}
\cE\hookrightarrow W\to\cB_3 \,.
\end{equation}
Its complex structure moduli space, $\cM_{\rm c.s.}(W)$,  encodes all pure $g_s$ corrections, perturbative and non-perturbative, to the perturbative Type IIB effective action. We therefore focus on $\cM_{\rm c.s.}(W)$ in F-theory
and investigate whether there are additional obstructions to infinite distance limits in this moduli space due to $\alpha'$ and mixed $g_s$, $\alpha'$ corrections.

\subsection{Complex structure and K\"ahler sector mixing}\label{ssec:Kahleroverview}
 At the classical level, the $\cN=1$ moduli space $\cM_{\rm cl}^{\rm F}(W)$ is given by the product
\begin{equation}\label{eq:Mcl-Ftheory}
    \cM_{\rm cl}^{\rm F}(W)=\cM_{\rm c.s.}(W)\times\cM_{\rm K}(W)\,,
\end{equation}
where the complex structure moduli space $\cM_{\rm c.s.}(W)$ is the $g_s$-corrected Type IIB moduli space encoding the Type IIB axio-dilaton, the complex structure deformations of the orientifold and the 7-brane moduli.\footnote{In the presence of spacetime-filling D3-branes there is an additional factor corresponding to the position moduli of these D3-branes along $\cB_3$. We do not discuss this component of the moduli space.} The second factor is the classical K\"ahler moduli space of $W$ that is spanned by the Einstein-frame volumes of divisors of the base $\cB_3$. The factorisation as in \eqref{eq:Mcl-Ftheory} is realized in the large volume limit for the base $\cB_3$ of $W$.

While F-theory geometrises all pure $g_s$ corrections through the elliptic fibration of $W$, any finite volume effects (perturbative as well as non-perturbative) constitute quantum effects also from the F-theory perspective. Thus, moving away from infinite Einstein-frame volumes, $\alpha'$ as well as mixed $g_s$, $\alpha'$ corrections can spoil the classical factorisation in~\eqref{eq:Mcl-Ftheory}. In particular, this means that not all infinite distance limits in $\cM_{\rm c.s.}(W)$ can be necessarily realized in the fully quantum-corrected moduli space $\cM_{\rm Q}^{\rm F}(W)$ at constant values for the K\"ahler moduli. Such limits are then obstructed in the sense of

\begin{definition}\label{def:2fusionheunconnectedribbon}
    Consider F-theory compactified on an elliptically fibered Calabi--Yau fourfold $\cE \hookrightarrow W \to \cB_3$. An infinite distance limit in the factor $\cM_{\rm c.s.}(W)$ of the classical $\cN=1$ moduli space $\cM_{\rm cl}^F(W)$ is called {\rm K\"ahler\text{-}obstructed} if in the quantum-corrected moduli space the following two conditions cannot be satisfied simultaneously when taking the limit:
    \begin{enumerate}
        \item The classical Einstein-frame volume $\cV_D^{(0)}$ of all effective divisors $D\in {\rm Eff}^1(\cB_3)$ remains finite, $\cV_{D}^{(0)}<\infty$. 
        \item The 4d $\cN=1$ effective action as derived from the classical geometry of $W$ remains under perturbative control in $\alpha'$. 
    \end{enumerate}
\end{definition}
This definition can be viewed in analogy to Definition~\ref{def:2fusionheunconnected} of $g_s$ obstructions reviewed above. While the question whether or not a classical infinite distance limit in $\cM_{\rm c.s.}(V/\Omega)$ is $g_s$-obstructed can be decided by lifting the limit to the F-theory moduli space, a similar strategy is not available for classical infinite distance limits in $\cM_{\rm c.s.}(W)$: This would require a lift to a theory that is both $g_s$- and $\alpha'$-exact. At present, no such theory is known, meaning that no general geometric argument can act as a diagnostic for K\"ahler obstructions to a limit in $\cM_{\rm c.s.}(W)$.

Instead, we propose two different methods for detecting K\"ahler obstructions. The first approach is based on explicit knowledge of quantum corrections to divisor volumes:

\begin{claim}\label{claim:explicitobstructions}
    Consider F-theory compactified on an elliptically fibered Calabi--Yau fourfold $\cE \hookrightarrow W \to \cB_3$ and an infinite distance divisor $\Delta \subset \cM_{\rm c.s.}(W)$ corresponding to $z\to \ii \infty$ for a suitable coordinate $z$ on $\cM_{\rm c.s.}(W)$. Suppose there exists a divisor $D\in {\rm Eff}^1(\cB_3)$ for which the leading corrections to the quantum volume are of the form 
    \begin{equation}\label{eq:quantumvolume}
        \cV_D = \cV_D^{(0)} + a \left(\Im z\right)^\alpha + \dots \,,\qquad 0\neq a\in\mathbb{R}\,,\;0<\a \in\mathbb{R}\,. 
    \end{equation}
    Then, maintaining perturbative control over the 4d $\cN=1$ effective action derived from the classical geometry of $W$ in the $\Im z \to \infty$ limit requires a co-scaling of the classical divisor volume $\cV_D^{(0)}\succsim \left(\Im z\right)^\alpha$. In other words, if there exists a divisor $D\in {\rm Eff}^1(\cB_3)$ whose quantum corrected volume behaves as in \eqref{eq:quantumvolume} near $\Delta$, the corresponding limit in $\cM_{\rm c.s.}(W)$ is K\"ahler-obstructed. 
\end{claim}

In the effective supergravity, the quantum correction \eqref{eq:quantumvolume} can be interpreted as a field redefinition of the modulus. This, however, does not mean that the quantum corrections of modulus ${\cal V}_D$ are innocuous. First, if $a < 0$, as e.g. in the example of Section \ref{sssec:typeIIIexplicit}, the corrected modulus tends to zero in the limit ${\rm Im}(z) \to \infty$, leading to an unsuppressed BPS instanton whose action is controlled by the redefined modulus. Second, even for $a > 0$, the classical supergravity action derived on the basis of the factorisation of the moduli space is not applicable. 
Indeed, the corrections of the form~\eqref{eq:quantumvolume} explicitly break the factorisation of the classical F-theory moduli space~\eqref{eq:Mcl-Ftheory} and are therefore genuine $\cN=1$ effects. In Section~\ref{sec:regularfiber} we discuss specific cases of infinite distance limits in $\cM_{\rm c.s.}(W)$ for which corrections to divisor volumes of the form \eqref{eq:quantumvolume} can be explicitly computed; these are (a) threshold corrections to the gauge coupling of a dual heterotic string~\cite{Dixon:1990pc}, and (b) D-brane instanton corrections to the contact potential in a dual Type IIA setup~\cite{Alexandrov:2008gh}. These two examples already highlight one important restriction of this method: A duality frame must be known in which  the relevant quantum corrections can be computed explicitly. This is notoriously hard to achieve in theories with minimal supersymmetry. In fact, the corrections in cases (a) and (b) mentioned above make explicit use of {\it extended} (worldsheet or spacetime) supersymmetry in the respective duality frames. 

To avoid the technical difficulties of computing the corrections to general divisor volumes, we propose another method for diagnosing K\"ahler obstructions to a given infinite distance limit in $\cM_{\rm c.s.}(W)$. Assuming the classical factorisation~\eqref{eq:Mcl-Ftheory}, for a limit purely in the complex structure moduli space $\cM_{\rm c.s.}(W)$, the F-theory K\"ahler potential can be approximated as
\begin{equation}
    K_F\approx K_{\rm c.s.}(z)=-\log\left[\int_W\Omega_4\wedge\bar{\Omega}_4\right]\,.
\end{equation}
Close to infinite distance singularities in $\cM_{\rm c.s.}(W)$, $K_F$ enjoys approximate shift symmetries as asymptotically $K_{\rm c.s.}$ only depends on the saxions $s^i=\Im z^i$, but is invariant under shifts of the axions $a^i=\Re z^i$. This can be seen more explicitly using the growth theorem for the limiting mixed Hodge structure associated to the infinite distance singularity, see the discussion around~\eqref{eq:grwoththeorem}.

As discussed in~\cite{Lanza:2021udy}, given this approximate shift-symmetry, the axions can be dualised into two-form potentials $B_{2,i}$. The objects charged electrically under these two-forms are strings. In four dimensions, strings are (real) codimension-two objects implying that their backreaction induces a logarithmic profile that changes the asymptotics of the theory. The strings carrying charge $\mathbf{e}= (e^1,\dots, e^{h^{3,1}})$ under the two-forms $B_{2,i}$ can then be described as cosmic string solutions to the 4d $\cN=1$ effective action of the form~\cite{Greene:1989ya}
\begin{equation}\label{EFTstringsolution}
    {\rm d}s^2 = -{\rm d}t^2 +{\rm d} x^2 + e^{-K_{\rm c.s.}} {\rm d}u{\rm d}\bar{u} \,, \qquad z^k(u) = \frac{e^k}{2\pi \ii} \log\frac{u}{u_0}\,,
\end{equation}
where $(t,x)$ are the coordinates along the string and $u$ parametrises the complex plane transverse to the string such that the string is localised at $u=0$. The warp factor is given by the K\"ahler potential, which depends on $u$ through $z^k$. A string with charge $\mathbf{e}\in \mathbb{N}_{0}^{h^{3,1}}$ induces a backreaction that realizes an asymptotic singularity in $\cM_{\rm c.s.}(W)$ as the string core is approached. Such strings have been dubbed \emph{EFT strings} in~\cite{Lanza:2021udy}. These EFT strings are the four-dimensional analogue of supergravity strings in higher dimensions that are either critical strings or form black strings for high charges. In~\cite{Lanza:2021udy} it was further conjectured that every infinite distance limit in the moduli space of 4d $\cN=1$ low-energy effective theories of quantum gravity (along which all saxionic moduli are either constant or scale to infinity at a fixed rate) can be realized through the backreaction of such an EFT string. 

Here, we use the EFT string perspective to study the fate of classical infinite distance limits in $\cM_{\rm c.s.}(W)$ at the quantum level. We therefore make use of a connection between the EFT strings and the dual frame that emerges asymptotically as predicted by the Emergent String Conjecture~\cite{Lee:2019oct}. According to this conjecture, the dual theory arising in an infinite distance limit is either a weakly coupled, critical string theory or a higher-dimensional theory, and the tower of states predicted by the Distance Conjecture~\cite{Ooguri:2006in} corresponds to the excitation states of the critical string or a KK-tower in the respective cases. Key for the connection between EFT strings and the Emergent String Conjecture is the relation between the tension of the EFT string and the quantum gravitational scales associated with the limit. The latter are given by the mass scale, $M_{\rm tower}$, of the tower of light states predicted by the Distance Conjecture~\cite{Ooguri:2006in} and the quantum gravity cutoff $\Lambda_{\rm QG}$, or species scale~\cite{Dvali:2007hz,Dvali:2009ks,Dvali:2010vm}. Instead, the tension of the EFT string is simply given by\footnote{As discussed in~\cite{Hassfeld:2025uoy}, we are actually interested in loop-configurations of this EFT string in order to regulate its backreaction at infinity. In the following we will (implicitly) always work with such a configuration.} 
\begin{equation} \label{EFT-tension}
    \frac{T_{\rm EFT}}{M_{\rm Pl}^2} = -e^k \frac{\partial K_F}{\partial s^k}\,. 
\end{equation}
There are four possibilities for the hierarchy of the three scales:
\begin{enumerate}
    \item \textbf{Critical string:}\\
    The EFT string is a critical Type II or heterotic string that becomes weakly coupled and tensionless such that asymptotically 
    \begin{equation}\label{eq:hierarchy1}
        \sqrt{T_{\rm EFT}} \sim \Lambda_{\rm QG} \sim M_{\rm tower}\,.
    \end{equation} 
    \item \textbf{Higher-dimensional supergravity string:}\\
    The EFT string is a supergravity string in the higher-dimensional theory to which the 4d theory decompactifies asymptotically. In this case, the mass scales satisfy 
    \begin{equation}\label{eq:hierarchy2}
            \sqrt{T_{\rm EFT}} \sim \Lambda_{\rm QG} \gg M_{\rm tower}\,.
    \end{equation}
    \item \textbf{Higher-dimensional defect:}\\
    If the limit is a decompactification limit to a higher-dimensional theory, a hierarchy of the form 
    \begin{equation}\label{eq:hierarchy3}
        \sqrt{T_{\rm EFT}} \gg \Lambda_{\rm QG} \gg M_{\rm tower}\,,
    \end{equation}
    signals that the EFT string is a higher-dimensional defect in the asymptotic higher-dimensional theory.
    \item \textbf{Non-perturbative brane of weakly coupled string theory:}\\ 
     If the limit is an emergent string limit, a hierarchy of the form 
    \begin{equation}\label{eq:hierarchy4}
        \sqrt{T_{\rm EFT}} \gg \Lambda_{\rm QG} \sim M_{\rm tower}\,,
    \end{equation}
    signals that the critical, emergent string is not the EFT string but that the latter is a non-perturbative brane configuration of the dual weakly coupled string theory. An example are wrapped D-branes in perturbative Type II string theory. 
\end{enumerate}
As we discuss in Sections~\ref{sec:Sentype} and~\ref{sec:regularfiber}, the hierarchy between the tension of the EFT strings and the quantum gravity cutoff in the asymptotic regimes in $\cM_{\rm c.s.}(W)$ can be determined at the level of the effective action. Assuming that the classical effective action is a good approximation to the low-energy theory in infinite distance limits of $\cM_{\rm c.s.}(W)$, the tension of the corresponding candidate EFT strings cannot be parametrically larger than the quantum gravity cutoff. Therefore, in infinite distance complex structure limits only the first two possibilities can be realised. To see this, recall that a good estimate for the quantum gravity cutoff $\Lambda_{\rm QG}$ is given by the coefficient of the Gauss--Bonnet term in the higher-derivative effective action~\cite{vandeHeisteeg:2022btw,vandeHeisteeg:2023ubh,Martucci:2024trp}, which schematically reads
\begin{equation}
    S_{\rm h.d.} \supset \int d^4 x  \,F_{\rm GB}(\phi) R^2 \,.
\end{equation}
Here $F_{\rm GB}(\phi)$ encodes the dependence of this coupling on the scalar fields $\phi$ in the theory. The quantum gravity scale is then approximated by~\cite{vandeHeisteeg:2022btw,vandeHeisteeg:2023ubh} 
\begin{equation}\
    \frac{\Lambda_{\rm QG}}{M_{\rm Pl}} \sim \frac{1}{\sqrt{F_{\rm GB}(\phi)}}\,. \label{eq:FGBscaling}
\end{equation}
In Type IIB compactifications on Calabi--Yau threefolds, the dependence of $F_{\rm GB}$ on the complex structure moduli is encoded in the genus-one free energy, which can be computed from the topological string~\cite{Bershadsky:1993ta,Bershadsky:1993cx}. Crucially, at infinite distance in the complex structure moduli space, $F_{\rm GB}$ grows linearly in the complex structure moduli such that the quantum gravity cutoff scales as the square root of the EFT string tension. More generally, in 4d $\cN=1$ theories the Gauss--Bonnet term is generically (and at most) linear in the saxions in the vicinity of infinite distance limits as argued in~\cite{Martucci:2024trp}. Therefore, also in F-theory $F_{\rm GB}$ must be linear in the complex structure moduli of the fourfold. This implies that the tension of the candidate EFT string is of the order of the quantum gravity scale. 
By comparison with~\eqref{eq:hierarchy1} and~\eqref{eq:hierarchy2}, consistency of the classical effective action demands that an EFT string probing an infinite distance in the classical moduli space $\cM_{\rm c.s.}(W)$ has to be a critical string or a supergravity string in a higher-dimensional theory of quantum gravity, in agreement with the Emergent String Conjecture \cite{Lee:2019oct}. As we will discuss in detail in Sections~\ref{sec:Sentype} and~\ref{sec:regularfiber}, this poses certain constraints on the worldsheet theory on the string, see e.g.~\cite{Kim:2019vuc,Katz:2020ewz}. For the classical infinite distance limits to be consistent, these constraints must be satisfied by the worldsheet theory on the candidate EFT strings.

In the infinite distance limit, the Calabi--Yau fourfold $W$ splits into a normal crossing variety $W_0$ made up of multiple components intersecting normally.  The worldsheet theory on the candidate EFT strings that probe infinite distance limits in the classical moduli space $\cM_{\rm c.s.}(W)$ can be inferred from the geometry of the normal crossing variety $W_0$ arising in this limit. This is possible using a similar strategy as employed in~\cite{Hassfeld:2025uoy} for Type IIB compactified on Calabi--Yau threefolds.
The worldsheet theory can then be directly compared to the above expectations from the effective field theory; in particular, we argued above that the EFT string worldsheet theory has to be (super-)critical and must satisfy the constraints put on critical strings or supergravity strings in higher-dimensional theories. If instead, the worldsheet theory of the candidate EFT string is subcritical or violates the supergravity string constraints, this indicates a K\"ahler obstruction to the classical infinite distance limit in $\cM_{\rm c.s.}(W)$: 

\begin{claim}\label{claim:EFTstringobstruction}
    Consider F-theory compactified on an elliptically fibered Calabi--Yau fourfold $\cE \hookrightarrow W \to \cB_3$ and an infinite distance divisor $\Delta \subset \cM_{\rm c.s.}(W)$ along which the Calabi--Yau fourfold undergoes a semi-stable degeneration $W\to W_0$. If the worldsheet theory of the candidate EFT string derived from the geometry of $W_0$ is incompatible with the classical 4d $\cN=1$ effective action in the vicinity of $\Delta$, there are two options for the uplift $\hat{\Delta}$ of $\Delta$ in the quantum corrected 4d $\cN=1$ moduli space $\cM^{\rm F}_{\rm Q}(W)$: 
    \begin{enumerate}
        \item[$i.$] To reach $\hat{\Delta}\subset \cM^{\rm F}_{\rm Q}(W)$, an additional limit in the K\"ahler moduli space of $\cB_3$ must be taken. In this case, the EFT string responsible for the combined limit is a bound state of the putative EFT string associated with $\Delta$ and D3-branes wrapping suitable movable curves inside $\cB_3$. This results in a modified limit different from the original one, for which the inconsistency between the EFT string and the asymptotic physics can be avoided. 
        \item[$ii.$] The divisor $\hat{\Delta}\subset \cM^{\rm F}_{\rm Q}(W)$ can be reached without taking a limit in the K\"ahler moduli space, but the classical geometry of the degeneration $W\to W_0$ cannot be used to infer the worldsheet theory on the EFT string. This, in turn, means that the geometric description of F-theory as a compactification on the geometric Calabi--Yau fourfold $W$ is invalid. In other words, the limit exists at finite $\cV_{\cB_3}$, but the supergravity approximation of the 4d $\cN=1$ effective theory is not under perturbative control in $\alpha'$. 
    \end{enumerate}
    Case $i.$ violates condition~1 of Definition~\ref{def:2fusionheunconnectedribbon}, whereas case $ii.$ violates condition~2. We conclude that infinite distance limits in $\cM_{\rm c.s.}(W)$ for which the candidate EFT string worldsheet theory is incompatible with the classical 4d $\cN=1$ effective action are K\"ahler-obstructed. 
\end{claim}
In Sections~\ref{sec:FandEFT}-\ref{sec:regularfiber} we discuss in more detail how to infer the worldsheet theory on candidate EFT strings associated with complex structure degenerations of elliptic Calabi--Yau fourfolds. We also analyse explicitly in which cases this worldsheet theory does not match the expectations of the Emergent String Conjecture, thus signaling a K\"ahler obstruction of the limit.

\subsection{A simple example: K\"ahler obstructions in O-Type B orientifolds}\label{ssec:simple-ex}
As a simple example of a K\"ahler-obstructed limit, we consider an orientifold of $V={\rm K3}\times T^2$. In~\cite{Paper1}, we discussed two kinds of orientifolds based on this model that were labelled O-type A and O-type B. Here, we focus on the O-type B orientifold:  The orientifold action is given by $\Omega_{\rm B}=\Omega_p(-1)^{F_L}\rho$, where $\Omega_p$ is worldsheet parity, $F_L$ the left-moving spacetime fermion number and $\rho$ is an anti-symplectic involution on the ${\rm K3}$, i.e., $\rho^\ast(\omega_{\rm K3})=-\omega_{\rm K3}$ for $\omega_{\rm K3}$ the holomorphic two-form on K3. The fixed point locus of $\rho$ consists of a union of curves $C_{\rm I}$ inside the ${\rm K3}$, corresponding to O7-planes wrapping $C_{\rm I}\times T^2$. 

We further make the simplifying assumption that the K3 surface is itself elliptically fibered with generic fiber $\cE$. The model considered here is then dual to the O-type A model discussed in~\cite{Paper1} by T-dualizing both the $T^2$ factor and the elliptic fiber $\cE$ of the K3. The classical vector multiplet moduli space $\cM_{\rm VM,cl}(V/\Omega_{\rm B})$ contains as a subspace
\begin{equation}\label{eq:MVMcl}
    \cM_{\rm VM,cl}(V/\Omega_{\rm B}) \supset \cM_{S}\times \cM_T \times \cM_U\,. 
\end{equation}
Here, $U$ is the complex structure parameter of the torus, whereas $S$ and $T$ denote the (Einstein-frame) volumes of the divisors $D_\cE=\cE\times T^2$ and $D_b=\mathbb{P}^1_b\times T^2$ with $\mathbb{P}^1_b$ the base of the elliptically fibered K3. As the model is T-dual to the model analysed in~\cite{Paper1}, the quantum corrections to the K\"ahler potential can be extracted from Appendix~A of~\cite{Paper1}. Before discussing the explicit corrections, we argue -- similar to what is done in Sections~\ref{sec:Sentype} and~\ref{sec:regularfiber} -- that the K\"ahler obstruction to the limit can in fact be inferred from the zero mode structure on the EFT string realizing this limit at its core via an application of Claim~\ref{claim:EFTstringobstruction}.

To this end, we recall that before orientifolding, the limit $U\to\ii\infty$ is a (semi-stable) Tyurin degeneration\footnote{After base change, the $T^2$ develops a singularity of Kodaira type I$_2$, which is resolved into the union of two rational curves intersecting at two points $p_{1,2}$. The double surface of the semi-stable degeneration consists of two copies of the K3, $V_1\cap V_2=({\rm K3}\times\{p_1\})\cup({\rm K3}\times\{p_2\})$. 
As discussed in~\cite{Hassfeld:2025uoy}, considering both components in the zero mode counting on the EFT string results in a double counting. In the following we focus on one K3-component of the double surface.} of the threefold $V$: $V$ splits into the union of two threefolds $V_1 \cup V_2$ which intersect over a K3 surface, called the double surface of the degeneration.
 From~\cite{Hassfeld:2025uoy} it is known that the worldsheet theory of the EFT string realizing the limit $U\to\ii\infty$ at its core is that of a critical heterotic string. The zero modes on this string arise from the reduction of the Type IIB supergravity fields along the double surface as well as geometric modes corresponding to the position of the double surface inside $V_1\cup V_2$ and the location of the string in the 4d spacetime. The scalars on the string worldsheet are given by
\begin{equation}\label{eq:K3T2-WS}
    (|\mathbf{z}_0|,b_0,\Tilde{b}_0,b_1,b_2,b_3,\Phi,{\rm arg}(\mathbf{z}_0)).
\end{equation}
Here $\mathbf{z}_0$ is the complex scalar parametrising the normal directions of the EFT string in the 4d spacetime and $\Phi$ is the real position of the string along the degenerate $T^2$. The modes $b_0$ and $\Tilde{b}_0$ are obtained via reductions of $B_2$ and $C_2$, respectively. Instead, $b_1$ and $b_2$ arise from $C_4$ reduced over a $(2,2)$-sublattice of the transcendental lattice $\Lambda_{\rm trans}({\rm K3})$ and $b_3$ from $C_4$ reduced over a $(1,1)$-sublattice of the polarization lattice $\Lambda_{\rm pol}({\rm K3})$. These eight real scalars make up the bosonic part of a hypermultiplet in the $\cN=(0,8)$ worldsheet theory of the ($\frac{1}{2}$BPS) EFT string.\footnote{In contrast to the $\cN=(0,4)$ worldsheet theory inspected in~\cite{Hassfeld:2025uoy}, there are no interactions between the fields in the hypermultiplet due to the enhanced supersymmetry.} 

Performing the orientifold projection $\Omega_{\rm B}$ projects out some of the Type IIB supergravity fields and hence also some of the zero modes that localise on the EFT string worldsheet. To make this more concrete, we start with the worldsheet part of the orientifold action which acts on the Type IIB fields as
\begin{equation}
    \Omega_p(-1)^{F_L}: \qquad \begin{array}{lll}
        g\mapsto g\,,&\phi\mapsto\phi\,,& B_2\mapsto-B_2\,,\\
         C_0\mapsto C_0\,,& C_2\mapsto-C_2\,,& C_4\mapsto C_4\,.
    \end{array}
\end{equation}
As $B_2$ and $C_2$ are projected out from the 10d spectrum, so are the associated zero modes $b_0$ and $\Tilde{b}_0$ on the string worldsheet. This already shows that the conformal anomaly of the string is below criticality. 

The remaining six scalars must assemble into 2d $\cN=(0,4)$ (twisted) hypermultiplets,\footnote{The universal zero modes listed in~\eqref{eq:K3T2-WS} are non-chiral scalars in 2d. The only 2d $\cN=(0,4)$ multiplets containing full scalars are (twisted) hypermultiplets containing four scalars each, see e.g.~\cite{Lawrie:2016axq}.} meaning that two more scalars must be projected out. To determine which, we consider the geometric action of the orientifold projection: Holomorphicity of $\rho$ implies that $\Lambda_{\rm pol}({\rm K3})$ is invariant such that $b_3$ is not projected out. On the other hand, since the orientifold involution $\rho$ is anti-symplectic,  $\Lambda_{\rm trans}({\rm K3})$ is anti-invariant, i.e.,~the zero modes $b_{1,2}$ are projected out.\footnote{Assume that $\omega\in\Lambda_{\rm trans}({\rm K3})$ is invariant under $\rho$. However, since $\rho$ is anti-symplectic, $\omega$ needs to be a $(1,1)$-form, i.e., an element in the Picard lattice of ${\rm K3}$. This is a contradiction to $\omega\in\Lambda_{\rm trans}({\rm K3})$ unless $\omega=0$.} 
Therefore, the remaining scalars $(|\mathbf{z}_0|,b_3,\Phi,{\rm arg}(\mathbf{z}_0))$ form the bosonic part of an $\cN=(0,4)$ (twisted) hypermultiplet on the EFT string realizing the orientifolded limit.

We conclude that the candidate EFT string corresponding to the limit $U\to\ii\infty$ in the classical Type IIB $\Omega_{\rm B}$-orientifold complex structure moduli space is {\it subcritical}. By the arguments underlying Claim~\ref{claim:EFTstringobstruction}, this means that the O-type B limit is K\"ahler-obstructed. This means that, if we insist on keeping the K\"ahler moduli at constant values, the effective action must be significantly corrected in the $U\to \ii \infty$ limit. This expectation is confirmed in the following. Let us first notice that the limit lifts trivially to the F-theory moduli space in the sense that there are no pure $g_s$ corrections becoming unsuppressed in the limit. This is seen most easily by noticing that with respect to the limit $U\to\ii\infty$ the orientifold projection $\Omega_B$ is of O-type B, as introduced in~\cite{Paper1}. Instead, there are D$3$-brane instantons correcting the K\"ahler potential of the $\cN=2$ vector multiplet moduli space.

As discussed above, our model is related through four T-dualities along the generic elliptic fiber $\cE$ as well as the $T^2$-factor of $V$ to the $\Omega_{\rm A}$-orientifold analysed in~\cite{Paper1}. The D$(-1)$-instantons becoming unsuppressed in the O-type A limit map to D$3$-branes wrapping $\cE\times T^2$, whereas the D$3$-brane instantons wrapping the ${\rm K3}$ in the O-type A orientifold map to D$3$-branes wrapping the $T^2$-factor of $V$ as well as the base $\mathbb{P}^1_b$ of the ${\rm K3}$.\footnote{Notice that two T-dualities along the fiber $\cE$ of ${\rm K3}$ relate the setup considered here to the Type I compactification considered in~\cite{Camara:2008zk,Antoniadis:1996vw,Berg:2005ja}.} Since the complex structure of the $T^2$ changes in a modular covariant way under the four T-dualities, the analysis in~\cite{Paper1} implies a correction due to D$3$-brane instantons on the divisor $D_\cE=\cE\times T^2$ of $V$ with a large $\Im(U)$-expansion given by
\begin{equation}
    \cV_{D_\cE} = \cV_{D_\CE}^{(0)} - \text{Im}\,U +\dots \,.
\end{equation}
This is of the general form~\eqref{eq:quantumvolume} with $a=-1$ and $\alpha=1$. In fact, the same holds for D$3$-brane instantons on the divisor $D_b=\mathbb{P}^1_b\times T^2$ as follows from the discussion in Appendix~A of~\cite{Paper1}. %
By Claim~\ref{claim:explicitobstructions}, the limit is thus {\it K\"ahler-obstructed} in the sense of Definition~\ref{def:2fusionheunconnectedribbon}. This simple example shows in particular that the criterion for K\"ahler obstructions following from Claim~\ref{claim:EFTstringobstruction} based on the properties of the EFT strings goes hand-in-hand with Claim~\ref{claim:explicitobstructions}: If the EFT string associated with an infinite distance behaviour does not have the properties as required by the Emergent String Conjecture, there are corrections of the form~\eqref{eq:quantumvolume} to the classical divisor volumes that render the classical effective action invalid. 

Let us stress that this example does not correspond to a setup with 4d $\cN=1$ supersymmetry but in fact has enhanced 4d $\cN=2$ supersymmetry even after orientifolding. Still, we observe K\"ahler obstruction to the $U\to \ii \infty$ limit. This reflects that the fundamental origin for such obstructions is not minimal supersymmetry in 4d, but the orientifold action; the latter breaks the classical factorisation of the moduli space as in~\eqref{eq:MVMcl}, which only holds in the 4d $\cN=4$ parent theory. This example thus illustrates the importance of taking into account supersymmetry breaking effects when studying the effective action of string theory compactifications.

\section{F-theory complex structure limits and EFT strings}\label{sec:FandEFT}
To see how  the findings of the previous section generalise beyond this simple example, we now turn to
infinite distance limits in the F-theory complex structure moduli space and study whether additional quantum effects -- (non-)perturbative in both $\alpha'$ and $g_s$ -- render the original infinite distance limits K\"ahler-obstructed in the sense of Definition~\ref{def:2fusionheunconnectedribbon}.
 As preparation we first review in Section~\ref{ssec:CY4degenerations} the structure of the 4d $\cN=1$ effective action obtained from F-theory compactifications and introduce the classical infinite distance limits in the complex structure moduli space of a Calabi--Yau fourfold. The geometric classification of these limits parallels the discussion of semi-stable degenerations of Calabi--Yau threefolds, see~\cite{Monnee:2025ynn,Paper1}. Since for F-theory compactifications the elliptic fibration of the Calabi--Yau fourfold is crucial, we will distinguish two different types of degenerations for elliptically fibered Calabi--Yau fourfolds which we dub I$_n$\emph{-type} and \emph{regular-fiber} degenerations, respectively. Their fate in the quantum-corrected F-theory moduli space is discussed separately in Sections~\ref{sec:Sentype} and~\ref{sec:regularfiber}. The analysis  will be based on the two complementary lines of arguments as outlined in the two Claims~\ref{claim:explicitobstructions} and~\ref{claim:EFTstringobstruction} to determine a K\"ahler obstruction.

The worldsheet theory on the candidate EFT strings realizing classical infinite distance limits in the complex structure moduli space is  essential for Claim~\ref{claim:EFTstringobstruction}. For candidate EFT strings arising in general asymptotic limits in the complex structure moduli space of Calabi--Yau $n$-folds, the worldsheet theory has previously not been discussed in the literature. A first step in this direction was achieved in~\cite{Hassfeld:2025uoy}, where the worldsheet theory was derived for EFT strings arising in the vector multiplet moduli space of Type IIB compactified on Calabi--Yau threefolds in type II limits. Applying a similar logic to complex structure degenerations of Calabi--Yau fourfolds, we derive the worldsheet theories of F-theory candidate EFT strings in Section~\ref{ssec:countingFM}. For completeness, we also derive the worldsheet theory on EFT strings inducing infinite distance limits of types III and IV in the vector multiplet moduli space of Type IIB Calabi--Yau threefold compactifications. Since this discussion is not essential to the analysis of F-theory complex structure limits, we have relegated it to Appendix~\ref{app:WSmodesIIB}.

\subsection{Complex structure degenerations of Calabi--Yau fourfolds}\label{ssec:CY4degenerations}
The 4d $\cN=1$ low-energy effective action obtained from F-theory has been derived in~\cite{Grimm:2010ks}. Here and in the following, we will denote by $W$ the underlying elliptically fibered Calabi--Yau fourfold, with $\pi:\cE \hookrightarrow W \to \cB_3$ the projection to the base threefold $\cB_3$. The classically massless scalar fields in the  effective action are associated with the complex structure deformations $z^i$, $i=1,\dots, h^{3,1}(W)$, of $W$ and the K\"ahler moduli of $\cB_3$ and form part of 4d $\cN=1$ chiral multiplets.\footnote{For simplicity, we ignore the axionic fields associated with $H^{2,1}(W)$.} The K\"ahler moduli 
\begin{equation}\label{def:TaF}
    T_a = \frac12 \int_{D_a }\left(C_4 + \ii J_{\cB_3}^2\right)\,
\end{equation}
of $\cB_3$ classically correspond to the complexified Einstein-frame volumes of the generators $D_a$, $a=1,\dots, h^{1,1}(\cB_3)$, of the cone of effective divisors of $\cB_3$. Here, $J_{\cB_3}$ is the K\"ahler form on $\cB_3$ in the 10d Einstein-frame.

The factorisation \eqref{eq:Mcl-Ftheory} of the classical chiral multiplet moduli space is reflected in the classical F-theory K\"ahler potential given by
\begin{equation}\label{eq:KFclass}
    K_F = K_{\rm c.s.} (z) + K_{\rm K}(T) = -\log\left[\int_W \Omega_4\wedge \bar{\Omega}_4\right] -2 \log\left[\int_{\cB_3} J_{\cB_3}^3\right]\,.
\end{equation}
Here, $\Omega_4$ is the unique $(4,0)$-form on $W$ and the second term should be viewed as a function of $\Im\,T_a$ defined in~\eqref{def:TaF}. The complex structure parameters $z^i$ parametrise the split of the middle cohomology of $W$ as 
\begin{equation}
    H^4_{\rm hor}(W,\mathbb{C}) = H^{4,0}(W,\mathbb{C})\oplus H^{3,1}(W,\mathbb{C})\oplus H^{2,2}_{\rm hor}(W,\mathbb{C}) \oplus H^{1,3}(W,\mathbb{C}) \oplus H^{0,4}(W,\mathbb{C})\,,
\end{equation}
where horizontal refers to the part of the middle cohomology that is generated by complex structure variations of $W$. The classical metric on $\cM_{\rm c.s.}(W)\subset \cM^{\rm F}_{\rm cl}(W)$ is then derived as 
\begin{equation}
    G_{i\bar{\jmath}} = \partial_{z^i} \partial_{\bar{z}^j} K_F =  -\partial_{z^i} \partial_{\bar{z}^j}\left(\log \left[\int_W \Omega_4\wedge \bar{\Omega}_4\right]\right)\,,
\end{equation}
where we used that $K_K(T)$ is classically independent of the complex structure moduli $z^i$. 

As stressed several times by now, $\alpha'$ and mixed $g_s$, $\alpha'$ effects can become unsuppressed in the vicinity of singular divisors in the complex structure moduli space of the F-theory fourfold and spoil the factorisation \eqref{eq:Mcl-Ftheory}. In this way, infinite distance limits in the fourfold complex structure moduli space can be further corrected in the full F-theory field space. This in particular applies to Type IIB O-type B orientifolds, which, as explained in~\cite{Paper1}, survive the (classical) F-theory lift.

For now we stick to this classical factorisation and focus on infinite distance limits in the complex structure sector of $W$. We are interested in asymptotic regimes in $\cM_{\rm c.s.}(W)$ which correspond to normal-crossing singularities $\Delta_{k_1\dots k_n}\subset \cM_{\rm c.s.}(W)$. In local complex coordinates $u_k$ on $\cM_{\rm c.s.}(W)$ these are given by 
\begin{equation}
    \Delta_{k_1\dots k_n} = \{u_{k_1}=\dots=u_{k_n}=0\}\,. 
\end{equation}
In the vicinity of $\Delta_{k_1\dots k_n}$, we can then define the covering coordinates 
\begin{equation}
    z^{k_i} \equiv a^{k_i} + \ii s^{k_i}= \frac{1}{2\pi \ii }\log u_{k_i}\,,
\end{equation}
in which the singularity corresponds to $z^{k_1},\dots, z^{k_n} \to \ii \infty$. To describe the effective action derived from the K\"ahler potential $K_{\rm c.s.}(z)$ in the vicinity of a singularity $\Delta_{k_1\dots k_n}$, we use the limiting mixed Hodge structure associated to $H^4_{\rm hor}(W,\mathbb{C})$, see \cite{Grimm:2019ixq,vandeHeisteeg:2024lsa}. In short, in the vicinity of $\Delta_{k_1\dots k_n}$ we have the splitting 
\begin{equation}
    H^4_{\rm hor}(W,\mathbb{C}) = \bigoplus_{0\leq p,q\leq 4} I^{p,q} (\Delta_{k_1 \dots k_n})\,.
\end{equation}
For the classification of infinite distance limits, the dimensions
\begin{equation}
    i^{p,q} = \text{dim}\left(I^{p,q}(\Delta_{k_1\dots k_n})\right)
\end{equation}
are the key input. For Calabi--Yau fourfolds, the $i^{p,q}$ are severely constrained. In particular, only one of the $i^{4,q}$ for $q\in\{0,\dots,4\}$ is non-zero. The primary classification of complex structure degenerations is determined by the value of $q$ for which $i^{4,q}\neq 0$. In this way, we distinguish between type I ($q=0$), type II ($q=1$), type III ($q=2$), type IV ($q=3$), and type V ($q=4$) degenerations for Calabi--Yau fourfolds.\footnote{In addition to $i^{4,q}$, there are two further independent non-zero $i^{p,q}$ which determine the secondary singularity type. These will not play a major role in our analysis and we refer to~\cite{Grimm:2019ixq} for the definition of the secondary singularity type for Calabi--Yau fourfolds.} From the limiting mixed Hodge structure associated to the limit one can infer the asymptotic growth of the K\"ahler potential via the growth theorem \cite{schmid,CKS}. To this end, consider a growth sector approaching the singularity $\Delta_{k_1\dots k_n}$ defined as 
\begin{equation}
    \cR_{k_1 \dots k_n} = \left\{z^i=a^i+\ii s^i\,\Bigg|\,\frac{s^{k_1}}{s^{k_2}},\dots, \frac{s^{k_{n-1}}}{s^{k_n}}, s^{k_n}>\gamma \right\}\,,\quad \gamma>1\,. 
\end{equation}
In this growth sector, the K\"ahler potential of $\cM_{\rm c.s.}(W)$ scales as 
\begin{equation}\label{eq:grwoththeorem} 
    e^{-K_{\rm c.s.}} = \int_W \Omega_4\wedge\bar{\Omega}_4 \sim \prod_{i=1}^n \left(\frac{s^{k_{i}}}{s^{k_{i+1}}}\right)^{d_{k_1\dots k_i}}\,,
\end{equation}
where $d_{k_1\dots k_i}\in\{1,\dots,4\}$ encodes the type of the ${\rm codim}_\mathbb{C}=i$ singularity $\Delta_{k_1\dots k_i}\supset\Delta_{k_1\dots k_n}$ with $d_{k_1\dots k_i}=1,2,3,4$ corresponding to a type II/III/IV/V singularity, respectively.

For our analysis of the fate of asymptotic limits in $\cM_{\rm c.s.}(W)$, this algebraic treatment of the singularities $\Delta\subset\cM_{\rm c.s.}(W)$ in terms of the limiting mixed Hodge structure is not sufficient. Instead, we also require geometric input about the geometry of $W$ as we approach the singularities. For the purposes of this paper, it is enough to focus on $\text{codim}_\mathbb{C}=1$ singularities in $\cM_{\rm c.s.}(W)$. We can hence focus on a single local coordinate $u$ such that $u=0$ corresponds to the singularity under consideration. As in the threefold case, we consider a family of Calabi--Yau fourfolds 
\begin{equation}\begin{aligned} \label{eq:4foldW}
W_u \ \hookrightarrow & \  \ \mathcal{W} \cr 
&\ \ \downarrow\cr 
& \ \  \mathbf{D}\,
\end{aligned}\end{equation}
varying over the unit disk $\mathbf{D}=\{u\in \mathbb{C}||u|\leq 1\}$. All fibers $W_{u\neq0}$ are assumed to be smooth, whereas the central fiber $W_0$ is degenerate. In the remainder of this work, we make the assumption that the central fiber $W_0$ has been brought into a semi-stable form (as guaranteed by Mumford's semi-stable reduction theorem) such that it decomposes into a union of $N$ smooth components 
\begin{equation}
    W_0 = \bigcup_{i=1}^N W_i\,,
\end{equation}
intersecting transversally. We can define 
\begin{equation} \label{eq:Wi0ik}
    W_{i_0\cdots i_k} = W_{i_0}\cap\cdots \cap W_{i_k}\,,
\end{equation}
and 
 \begin{equation}\label{def:Wk+1-1}
    W^{(k+1)} = \bigsqcup_{i_0,\ldots,i_k} W_{i_0\cdots i_k}\,,\qquad 0\leq k\leq 4\,.
\end{equation}
The various types of limits now differ in the highest $k$ such that $W^{(k+1)}\neq \emptyset$. If the type of an infinite distance singularity $\Delta\subset \cM_{\rm c.s.}(W)$ is characterised by the integer $d$ as defined below~\eqref{eq:grwoththeorem}, we have $W^{(d+1)}\neq \emptyset$ and $W^{(k+1)}=\emptyset$ for $k>d$. As for threefolds, the manifolds $W_{i_0 \dots i_d}$ contributing to $W^{(d+1)}$ are then Calabi--Yau $(4-d)$-folds (here a point counts as Calabi--Yau 0-fold). Moreover, the dimension of the dual graph $\Pi(W_0)$ of the degeneration is again given by $d$, see~\cite{Paper1,Monnee:2025ynn} for a more detailed discussion of the dual graph in the analogue threefold case.

We can make an important refinement of the classification of degenerations of elliptically fibered Calabi--Yau fourfolds that is relevant for F-theory compactifications. Given the special role of the elliptic fiber $\cE$ in the F-theory context, we can distinguish between 
\begin{itemize}
    \item \emph{regular-fiber} limits, in which each double threefold $W_{i_0i_1}$ is itself elliptically fibered with generic fiber $\cE$,
    \item and I$_n$\emph{-type} limits, in which the generic fiber over at least one component of the base develops a non-split singularity of Kodaira type I$_n$.\footnote{In a Weierstrass model the only 
    singular fibers in codimension-zero which are of
    normal crossing type (and hence consistent with semi-stability of the degeneration) are Kodaira-type I$_n$. 
   }
\end{itemize}
For regular-fiber limits it follows that also all higher codimension loci $W^{(k+1)}$, $k\leq d$, are elliptically fibered with generic fiber $\cE$. Notice further that an I$_n$-type limit in which the base does not degenerate and the generic fiber is of non-split Kodaira type I$_2$ corresponds to the standard Sen-limit~\cite{Sen:1996vd, Clingher:2012rg}. By contrast, type V limits are always of I$_n$-type.

Finally, semi-stable degenerations have been analysed in detail in  \cite{Lee:2021qkx,Lee:2021usk} for F-theory Weierstrass models of elliptic K3 surfaces and in \cite{Alvarez-Garcia:2023gdd,Alvarez-Garcia:2023qqj} for elliptic threefolds.

\subsection{Worldsheet spectrum of EFT strings}
\label{ssec:countingFM}
In Section~\ref{ssec:Kahleroverview}, the relation between asymptotic limits in the complex structure moduli space of a Calabi--Yau fourfold $W$ and EFT strings was reviewed. In particular, we argued that with the help of the Emergent String Conjecture it is possible to detect pathologies in classical infinite distance limits in $\cM_{\rm c.s.}(W)$ by looking at the worldsheet theory of the candidate EFT string. In this section, we provide details on how to infer this worldsheet theory from the geometry of the normal crossing variety $W_0$ arising at an infinite distance singularity in $\cM_{\rm c.s.}(W)$.

Since we are interested in $\text{codim}_\mathbb{C}=1$ singularities in $\cM_{\rm c.s.}(W)$, we focus on elementary string charges for which only one charge $e^{k_0}\neq 0$ such that at the string core we realize the limit $z^{k_0} \to \ii \infty$. The worldsheet spectrum on the candidate EFT string can be determined via F-/M-theory duality. To see how this works, consider M-theory compactified on $W$. The analogue of the EFT string in F-theory is now a particle in the resulting 3d $\cN=2$ theory which we dub an \emph{EFT particle}. In the remainder of this section we drop the qualifier ``candidate'' for the classical string/particle solutions realizing infinite distance limit in $\cM_{\rm c.s.}(W)$. The backreaction of the EFT particle can be obtained from the solution~\eqref{EFTstringsolution} upon circle compactification of the $x$-coordinate. In other words, the 3d EFT particle can be obtained from the 4d EFT string by wrapping the string around the circle. The worldsheet theory on the EFT string compactified on the circle gives rise to a one-dimensional quantum mechanics. The solution of the 3d $\cN=2$ effective theory is locally of the form $ \cW \times \mathbb{R} $ with $\cW$ the total space of the family in~\eqref{eq:4foldW}. As before, we interpret the disk ${\bf D}$ as the plane transverse to the worldline of the EFT particle in 3d. Under the assumption that $\cW$ is itself a Calabi--Yau fivefold, the quantum mechanical theory localised at $\{u=0\}\subset \cW$ is in fact an $\cN=2$ super-quantum mechanics (SQM).\footnote{Similar to the case of Calabi--Yau threefold degenerations discussed in~\cite{Hassfeld:2025uoy,Monnee:2025ynn}, $\cW$ being a Calabi--Yau fivefold is not necessarily compatible with the central fiber having only simple normal crossing singularities. As in the cited references we expect that also here additional singularities in the central fiber do not alter the mode counting on the EFT strings.\label{fn:5fold-CY}} In the following, we show how the modes of the 1d SQM can be obtained by studying the modes of M-theory compactified on $\cW$ localised to $u=0\in \mathbf{D}$. The spectrum of the EFT string in F-theory is then obtained by performing the F-theory uplift for these localised quantum mechanics modes. 

The counting of degrees of freedom of the SQM is very similar to the counting of degrees of freedom on the EFT string in Type IIB complex structure degenerations that was pioneered in~\cite{Hassfeld:2025uoy} and for which we work out the generalisation in Appendix~\ref{app:WSmodesIIB}. The modes of the SQM can be split into geometric degrees of freedom and the degrees of freedom arising from the reduction of the M-theory $p$-forms along localised $(p-2)$-forms. 

\paragraph{Universal geometric mode.} Irrespective of the type of the degeneration, there is a universal geometric zero mode of the SQM associated with the position of the EFT particle in the spatial part of the 3d spacetime. We denote this mode by $\mathbf{z}_0=|\mathbf{z}_0| e^{\ii \arg{\mathbf{z}_0}}$. It provides two real scalars for the SQM, a non-compact scalar $|\mathbf{z}_0|$ and a compact scalar $\arg(\mathbf{z}_0)$. 

\paragraph{Internal geometric modes.} In addition, there are modes associated with the location of the double threefolds $V_{ij}=W_i\cap W_j$ inside the normal crossing variety $W_0$. The number of geometric modes is determined by the type of degeneration as it corresponds to the dimension of its dual graph $\Pi(W_0)$, which is given by the integer $d$ characterising the primary singularity type. Accordingly, there are $d$ real scalars arising from the internal geometric modes of the degeneration which we denote by $\Phi_i$, $i=1,\dots,d$.

\paragraph{Modes from $p$-forms.} 
There are also localised modes arising from a suitable reduction of the M-theory 
 three-form $\cC_3$ and six-form $\cC_6$.
 To extract these, one performs the following reduction along 
 the components $W_i$ of $W_0$:
\begin{equation}\begin{aligned}\label{Mtheoryexpansion}
    \cC_6 &= b^a \omega_a \wedge {\rm d}u \wedge {\rm d}{\bar u} \,,\quad \omega_a \in H^4(V_{ij})\oplus H^4(S_{ijk})\,, \\ 
    \cC_3 &= c^\alpha \gamma_\alpha \wedge  {\rm d}u \wedge {\rm d}{\bar u} \,,\quad \gamma_\alpha \in H^1(V_{ij}) \oplus H^1(S_{ijk}) \oplus H^1(C_{ijkl})\,.
\end{aligned}\end{equation}
Here, $V_{ij}$ denote the double threefolds $V_{ij}=W_i\cap W_j$, $S_{ijk}$ the triple surfaces and $C_{ijkl}$ the quadruple curves of the normal crossing variety $W_0$. Since the quadruple curves are contained in all triple surfaces, and each triple surface is contained in multiple double threefolds, the expansions as written in~\eqref{Mtheoryexpansion} in fact lead to an overcounting of modes. To take care of this and to identify the independent modes of the quantum mechanics theory, one has to take into account the linear relations in (co)homology reflecting the above inclusions. As this depends on the details of the degeneration, we will not discuss this here in generality. In Appendix~\ref{app:WSmodesIIB}, the analogous counting problem is analysed in some more detail for Calabi--Yau threefold degenerations.

Having established the counting of modes of the quantum mechanics theory associated with the degeneration of $W$ in M-theory, we can now discuss their uplift to F-theory. From the F-theory perspective, the M-theory dual arises by compactifying the 4d effective theory on an additional circle. In F-theory, the fivefold $\cW$ describes the total space of a BPS string solution. If $\cW$ is Calabi--Yau, the string worldsheet theory preserves 2d $\cN=(0,2)$ supersymmetry. The M-theoretic quantum mechanics is then obtained by wrapping this string on the additional circle. Since the elliptic fiber $\cE$ plays a central role for the uplift to F-theory, the worldsheet theory of the F-theory EFT string differs depending on whether the limit in $\cM_{\rm c.s.}(W)$ is a regular-fiber limit or an I$_n$-type limit as defined at the end of the previous section. We now discuss these two cases in turn. 

\subsubsection{Regular-fiber limits} \label{sec:regfibcounting}
\begin{figure}[t]
    \centering
    \begin{subfigure}{0.45\textwidth}
        \centering
        \includegraphics[width=\textwidth]{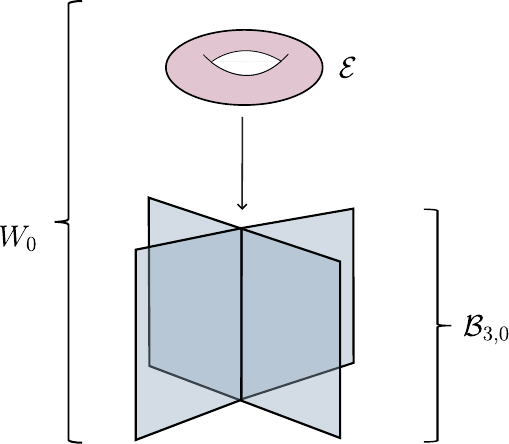}
        \caption{Regular-fiber limit}
        \label{fig:regular-fiber-limit}        
    \end{subfigure}
    \hfill
    \begin{subfigure}{0.45\textwidth}
        \centering
        \includegraphics[width=\textwidth]{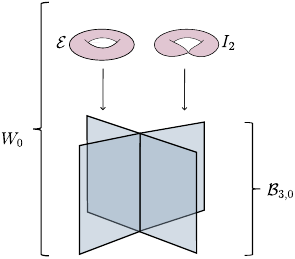}
        \caption{\emph{I}$_n$-type limit}
        \label{fig:generalized-Sen-limit}        
    \end{subfigure}
    \caption{The two classes of semi-stable degeneration limits $W\to W_0$ for the elliptically fibered Calabi--Yau fourfold $W$. In (a) the double threefold is itself elliptically fibered with generic smooth fiber $\cE$ corresponding to a regular-fiber limit, whereas in (b) the generic fiber over one base component degenerates to $\mathcal{E}_0$, corresponding to an I$_n$-type limit.}
    \label{fig:regular/generalized-Sen-limit}
\end{figure}

If all double threefolds $W_{i_0i_1}$ of $W_0$ are themselves elliptically fibered with generic fiber $\cE$, the degeneration of $W$ can be viewed as a degeneration of its base, $\cB_3\to\cB_{3,0}$. The situation is depicted in Figure~\ref{fig:regular-fiber-limit}. The zero modes of the M-theory SQM lift to F-theory as follows:
\begin{itemize}
    \item \textbf{Universal geometric mode:}\\
    The two real modes of the M-theory SQM associated with $\mathbf{z}_0$ lift to two real scalars of the F-theory string worldsheet.
    \item \textbf{Internal geometric modes:}\\
     Since the degeneration of $W$ is induced by a degeneration of $\cB_3$, the real modes $\Phi_i$ describing the internal geometric deformations of the intersections are the coordinates of the degeneration within $\cB_{3,0}$. These modes therefore survive the F-theory lift and give rise to real scalars on the worldsheet of the string in 4d.
    \item \textbf{Modes from $p$-forms:}\\
    We are left with the modes $c^\alpha$ and $b^a$ arising from localised $p$-forms in the expansion of $\cC_3$ and $\cC_6$ in~\eqref{Mtheoryexpansion}.
    \begin{itemize}    
        \item The modes $c^\alpha$ obtained from the reduction of $\cC_3$ lift to real scalar fields on the F-theory string worldsheet if the $\gamma_\alpha$ are one-forms on the elliptic fiber $\cE$. This is possible if either $V_{ij}$ or $S_{ijk}$ are trivially fibered by $\cE$ or $C_{ijkl}$ itself is $\cE$. In this case, the $c^\alpha$ lift to the real scalar modes obtained from $C_6$ and $B_6$ in the Type IIB language.
        \item  For the modes $b^a$ obtained from the reduction of $\cC_6$ we have to distinguish whether a 4-form $\omega_a$ appearing in the expansion~\eqref{Mtheoryexpansion} is vertical or not with respect to the elliptic fibration $\pi: \cE \hookrightarrow W^{(d+1)}\to\pi(W^{(d+1)})$. If $\omega_a$ is \emph{not} vertical, the associated mode $b^a$ does not lift to a scalar mode on the F-theory string worldsheet. Correspondingly, the entire multiplet associated with $b^a$ in the M-theory SQM does not lift to a multiplet of 2d $\cN=(0,2)$ theory on the string in F-theory.\footnote{This can be compared to the lift of $\cC_6$ axions in the 3d $\cN=2$ effective action of M-theory on an elliptically fibered Calabi--Yau fourfold $W$. The axion in the 3d effective action obtained by reducing $\cC_6$ over the section of $W$ is the axionic partner of the volume modulus of the elliptic fiber and hence does not lift to a modulus in the 4d $\cN=1$ effective action obtained from F-theory on $W$.} The remaining $b$-modes associated with vertical 4-forms lift to modes on the F-theory string worldsheet. The modes obtained from $H^4(V_{ij})$ have definite chirality in the 2d worldsheet theory of the string in F-theory. To see this, we notice that in the Type IIB language, the vertical modes of $\cC_6$ become the modes of the self-dual $C_4$-form of Type IIB. Since the F-theory lift effectively projects the $\pi$-vertical forms in $H^4(V_{ij})$ to two-forms on the base $\cB_{2,ij}$ of $V_{ij}$, the chirality of the $b$-modes can be read off from the intersection form on $H^2(\cB_{2,ij})$. Instead, the $\pi$-vertical forms in $H^4(S_{ijk})$ give rise to real scalar fields upon reducing $\cC_6$ over them. 
    \end{itemize}
\end{itemize}
Together with their fermionic partners, the above bosonic modes fill out complete supermultiplets of the 2d $\cN=(0,2)$ supersymmetry on the worldsheet. In the case of regular-fiber limits we obtain chiral and Fermi multiplets in this way. 

As a simple illustration, we now apply this general procedure to a trivial fibration $\cE \times V$ where the Calabi--Yau threefold $V$ undergoes an infinite distance degeneration and show how the above counting of modes in F-theory reproduces the Type IIB counting. 

\paragraph{Example: Type II Limit in trivial fibration.} Consider M-theory compactified on $\cE\times V$, where $\cE$ is an elliptic curve and $V$ is a Calabi--Yau threefold viewed as the base of the trivial elliptic fibration. Suppose there exists a limit in $\cM_{\rm c.s.}(V)$ in which $V$ undergoes a Tyurin degeneration, i.e., a semi-stable degeneration $V_z\hookrightarrow\cV\rightarrow{\bf D}$ for which the central fiber $V_0$ splits as
\begin{equation}
    V_z\rightarrow V_0=V_1\cup_Z V_2
\end{equation}
with $V_{1,2}$ Fano threefolds and $Z=V_1\cap V_2$ a K3 surface. In this case all triple surfaces and quadruple curves vanish. From~\cite{Hassfeld:2025uoy} it is known that the EFT string corresponding to this infinite distance limit is a critical heterotic string. To see this from the F-theory perspective, we follow the general discussion above and write $\omega_a=\omega_\cE\wedge\hat{\omega}_a$ for the vertical 4-forms $\omega_a\in H^4(\cE\times Z)$ appearing in~\eqref{Mtheoryexpansion}. Here $\omega_\cE$ is the 2-form on the elliptic fiber $\cE$ and $\hat{\omega}_a$ a basis of 2-forms on the K3 surface $Z$. Accordingly, reducing $\cC_6$ over this basis yields zero-modes in the 1d SQM that lift to $3$ right-moving and $19$ left-moving scalars on the 2d F-theory worldsheet. 
Since $H^1(Z)=0$, there are only two 1d scalars coming from the reduction of $\cC_3$ which correspond to the two 1-forms $\gamma_a\in H^1(\cE)$, $a=1,2$, on the elliptic fiber $\cE$. Both of these scalars lift to real scalars on the F-theory worldsheet as argued above.
 From the perspective of Type IIB string theory on $V$, these correspond to the two scalars obtained from the reduction of the Type IIB 6-forms $(B_6,C_6)$ along the unique element of $H^4(Z)$~\cite{Hassfeld:2025uoy}.
Together with the scalars ${\bf z}_0$ and $\Phi_1$ describing, respectively, the normal modes of the string in the 4d spacetime and the location of the degeneration inside $V_0$ we have thus confirmed the counting of~\cite{Hassfeld:2025uoy} from the F-theory perspective. 

\subsubsection{\texorpdfstring{I$_n$}{In}-type limits}\label{sssection:EFT-string-Sen}
If instead we consider an I$_n$-type limit, the F-theory lift is more subtle, since now $\cE$ itself degenerates over some base component(s) as illustrated in Figure~\ref{fig:generalized-Sen-limit}. As we now explain, EFT strings in fact provide a physical interpretation for the symmetry algebra associated with the codimension-zero degenerate fibers since this will play an important role for our arguments in Section \ref{sec:Sentype}.

Note first that while the lift of $\mathbf{z}_0$ and the modes arising from $\cC_6$ and $\cC_3$ is as in the previous case, a difference occurs for the lift of the modes $\Phi_i$. These do not all lift to real scalar fields on the F-theory string worldsheet. One or several combinations of the coordinates $\Phi_i$ parametrise the deformations of the M-theory EFT particle inside the elliptic fiber $\cE$ normal to the degeneration.
Without loss of generality, we label these coordinates as $\Phi_I$ so that the remaining $\Phi_{i\neq I}$ parametrise the location of the degeneration in the base $\cB_{3,0}$ of $W_0$. 
In the F-theory lift, the volume of all fibral components goes to zero. As a result, the coordinates $\Phi_I$ do not lift to real scalars on the F-theory string. Instead, in the M-theoretic SQM the coordinates $\Phi_I$  correspond to the Coulomb branch parameters of vector multiplets. The scalars $\Phi_I$ thus become part of  non-dynamical vectors on the F-theory string. By supersymmetry, these vectors reside in vector multiplets of the 2d $\cN=(0,2)$ theory on the F-theory string which contain as dynamical degrees of freedom a left-moving Weyl fermion. 
Furthermore, there are additional massless modes on the F-theory string worldsheet that are generically massive in the M-theory SQM. These are charged under the gauge theory corresponding to the vector multiplets.
 Their origin in M-theory is from M2-branes wrapping additional fibral curves $\cC^f_{\rm loc}$ that are localised on the degenerate fiber $W_0$ of $\cW$, including the extra fibral curves in codimension-zero over ${\cal B}_{3,0}$. Since all fibral curves shrink in the F-theory limit, the M2-branes wrapping $\cC^f_{\rm loc}$ give rise to massless, charged modes on the F-theory string that provide the W-bosons of the non-Abelian gauge theory associated with the degeneration of the elliptic fiber in codimension-zero. As an illustration, we now apply the above general discussion of the mode counting to the standard Sen-limit. 

\paragraph{Example: Standard Sen-limit.} In the standard Sen-limit of an elliptically fibered Calabi--Yau fourfold $\cE\hookrightarrow W\to \cB_3$, the elliptic fiber $\cE$ degenerates in codimension-zero on $\cB_3$ in a non-split I$_2$ singularity. After resolving the singularity~\cite{Clingher:2012rg}, the generic fiber factorises into the union of two rational curves, $\cE \to C_{\rm I}+ C_{\rm II}$. Let $C_{\rm I}$ be the proper transform of the elliptic fiber after resolution and $C_{\rm II}$ the exceptional curve. The curves $C_{\rm I}$ and $C_{\rm II}$ intersect over two points. The two intersection points are exchanged by a monodromy around codimension-one loci in $\cB_3$ and define a bi-section that is a double cover of the base $\cB_3$. Due to the branching of the fiber in codimension-one, this double cover of $\cB_3$ is a Calabi--Yau threefold~\cite{Clingher:2012rg}. The standard Sen-limit is hence an example of a type II singularity for the fourfold for which the two components $W_1$ and $W_2$ are given by the fibration of the rational curves $C_{\rm I}$ and $C_{\rm II}$ over $\cB_3$. Since the intersection points of $C_{\rm I}$ and $C_{\rm II}$ are exchanged under monodromy, the moduli space of $C_{\rm II}$ is the bi-section, i.e., the Calabi--Yau threefold. Wrapping M2-branes on $C_{\rm II}$ therefore gives rise to vector multiplets that are localised on the bi-section and can be interpreted as the W-bosons of a non-Abelian $Sp(1)$ gauge group on the 2d string worldsheet. The Coulomb branch parameter for this gauge group in M-theory is identified with the single internal geometric scalar mode $\Phi_1$ in the SQM of the EFT particle. In the geometry, $\Phi_1$ can be identified with the blow-up K\"ahler modulus resolving the I$_2$ singularity in codimension-zero. The existence of this $Sp(1)$ gauge group on the 2d string worldsheet is in accordance with the expectation from the Type IIB orientifold description of the Sen-limit: Here the EFT string comes from a D7-brane wrapping the Calabi-Yau threefold and indeed carries gauge group $Sp(1)$~\cite{Lanza:2021udy}.

In addition, the elliptic fiber $\cE$ degenerates in codimension-one and higher on the base $\cB_3$. From the Type IIB perspective, these are the loci of the D7-branes and O7-planes. 
This degeneration in codimension-one is already present away from the Sen-limit, which we take to appear over a divisor $u=0$ in the complex structure moduli space. At $u=0$, the codimension-one degeneration worsens, reflecting an enhancement of the Kodaira type due to the collision with the codimension-zero degeneration.
 For example, let us assume that in the Type IIB orientifold, the tadpole is cancelled locally by an $SO(8)$ stack of D7-branes on top of the O7-plane. For $u\neq 0$, this leads to an  I$_0^*$ singularity over a divisor $D_{\rm O7}$ on the base that corresponds to the lift of the O7-plane.
Resolving the associated I$_0^*$ singularity in F-theory yields four exceptional curves $C^i_{SO(8)}$, $i=1,\dots,4$. 
Since these additional fibral curves are not localised to $u=0$, M2-branes wrapping $C^{i}_{SO(8)}$ do not give rise to localised modes on the F-theory string worldsheet, but are simply the W-bosons of the \emph{spacetime} $SO(8)$ gauge group. 
 At $u=0$ and over $D_{\rm O7}$, the I$_0^*$ fibers collide with the I$_2$ fibers which appear over generic points of the base at $u=0$. There are therefore 
additional modes coming from M2-branes wrapping linear combinations of 
the resulting fibral curves.
Since these combinations involve curves that are localised to $u=0$, they yield localised multiplets on the F-theory string worldsheet that are charged under both the worldsheet $Sp(1)$ and spacetime $SO(8)$ gauge groups. Depending on the details of the geometry, these modes yield either chiral or Fermi-multiplets on the string worldsheet. In fact, the net chiral index of these zero modes can be computed as~\cite{Schafer-Nameki:2016cfr}
\begin{equation}\label{chiS}
    \chi(S)=\sum_{i=0}^2(-1)^ih^i(S,L\otimes \sqrt{K_S})\,,
\end{equation}
where $S=D_{\rm O7}$ is the divisor in $\cB_3$ wrapped by the D7-branes and $L$ is the vector bundle associated to worldvolume flux. Charged chiral multiplets contribute positively to $\chi(S)$, while charged Fermi multiplets give a negative contribution.\footnote{For $S$ a spin manifold, the worldvolume flux can be chosen to be trivial, $L=\cO_S$. By Hirzebruch's signature theorem, the index~\eqref{chiS} is then given by
\begin{equation}
    \chi(S)=-\frac{1}{8}\sigma(S)\,,
\end{equation}
with $\sigma(S)=b_2^+-b_2^-$ the signature of the intersection pairing on $H^2(S)$. Thus, for D7-branes wrapping a spin four-manifold $S$ with negative signature, we can conclude that there are charged chiral multiplets, e.g. $S={\rm K3}$ has $\sigma({\rm K3})=-16$. Even for surfaces with vanishing or positive signature, chiral multiplets generically exist. Indeed, an example of vanishing signature is given by $T^4$ for which $h^0(T^4,\sqrt{K}_{T^4})=1$.}

Notice that since the Calabi--Yau threefold $V_{12}= W_1\cap W_2$ is a double-cover of the base $\cB_3$, there are no vertical four-forms that give rise to uncharged chiral fields arising from $\cC_6$ on the string worldsheet. Similarly, $H^1(V_{12})=\emptyset$ such that also $\cC_3$ does not contribute any uncharged modes on the string.

\section{K\"ahler obstructions for \texorpdfstring{I$_n$}{In}-type limits}\label{sec:Sentype}
We are now in a position to present our arguments why infinite distance limits in $\cM_{\rm c.s.}(W)$ are K\"ahler-obstructed in the sense of Definition~\ref{def:2fusionheunconnectedribbon}. The I$_n$-type limits and the regular-fiber limits will be treated separately, in this section and Section \ref{sec:regularfiber}, respectively.  
 
 In Section~\ref{ssec:globalSen}, we illustrate that the two complementary lines of reasoning outlined at the beginning of Section~\ref{sec:FandEFT} both lead to the same conclusions. Specifically, we will argue that the standard Sen-limit is K\"ahler-obstructed, first based on the properties of the worldsheet theory on the candidate EFT string and then by identifying the relevant corrections to the effective action. We then follow, in Section \ref{ssec:localsen}, the approach based on the worldsheet theory on the candidate EFT strings in F-theory to argue that also I$_n$-type limits in $\cM_{\rm c.s.}(W)$ are K\"ahler-obstructed. 

\subsection{Standard Sen-limit}\label{ssec:globalSen}
Consider an elliptically fibered Calabi--Yau fourfold $W$ for which we can take a standard Sen-limit in $\cM_{\rm c.s.}(W)$ as the limit $z^1\to \ii \infty$. In this case the K\"ahler potential $K^F_{\rm c.s.}$ takes the asymptotic form\footnote{Here we assume that we take the standard Sen-limit without superimposing it with an additional base degeneration. Otherwise, as explained in~\cite{Paper1}, extra care has to be taken when computing the K\"ahler potential.}
\begin{equation}
    K^F_{\rm c.s.} = -\log(\Im z^1) - \log P(z^{j}) + \dots\,,
\end{equation}
where $P(z^j)$ is a degree-3 polynomial that can depend on all complex structure moduli other than $z^1$. In the following, we drop the superscript and write $z\equiv z^1$. From a Type IIB orientifold perspective, we can identify the modulus $z$ with the axio-dilaton 
\begin{equation}
    z = C_0 + \frac{\ii }{g_s}\,. 
\end{equation} 
The tension of the candidate EFT string realizing this limit, see \eqref{EFT-tension}, has classical tension
\begin{equation}
    \frac{T_{\rm Sen}}{M_{\rm Pl}^2} = (\Im z)^{-1} = g_s\to 0\,. 
\end{equation}
The small $g_s$ limit of Type IIB string theory must correspond to an emergent string limit with the role of the critical string played by the fundamental Type IIB string. After orientifolding, this string is non-BPS, but it should nonetheless be the lightest string in the spectrum in the $g_s\to 0$ limit. To compare the tension of the fundamental string with $T_{\rm Sen}$, we recall that the limit $z\to \ii \infty$ is taken at constant \emph{Einstein}-frame volume $\cV_E(\cB_3)$ of $\cB_3$ in order not to impose an additional limit in the F-theory K\"ahler moduli space. The tension of the fundamental Type IIB string is given by 
\begin{equation}
    \frac{T_{\rm F1}}{M_{\rm Pl}^2} \sim \frac{g_s^{1/2}}{\cV_{E}(\cB_3)}\,,
\end{equation}
such that, at constant $\cV_{E}(\cB_3)$, we realize the hierarchy $T_{\rm Sen} \ll T_{\rm F1}$. Thus, even though the $g_s\to 0$ limit is expected to be an emergent string limit for the F1-string, the tension of the candidate EFT string associated with the Sen-limit is parametrically below this scale. For this to be possible, the candidate EFT string would itself have to be a critical string and thus feature at least four neutral chiral multiplets on its worldsheet to account for the 10-dimensional spacetime of a critical string. However, as discussed in Section~\ref{sssection:EFT-string-Sen}, this is not the case: The candidate EFT string for the standard Sen-limit only has a single neutral chiral multiplet whose scalars correspond to the coordinates normal to the string in 4d. This excludes the candidate EFT string asscociated with the standard Sen-limit from being a critical string.\footnote{The same conclusion can be reached directly from the Type IIB orientifold perspective. As already discussed in~\cite{Lanza:2021udy}, the EFT string realizing the $g_s\to 0$ limit corresponds to a D7-brane wrapping the Calabi--Yau orientifold. Due to the absence of one-cycles, the only scalars on the resulting string worldsheet arise from the scalar in the 8d vector multiplet corresponding to the transverse coordinates of the string.} 
Thus, according to Claim~\ref{claim:EFTstringobstruction}, the classical limit $z\to \ii \infty$ has to be K\"ahler-obstructed. 
As anticipated in Claim~\ref{claim:EFTstringobstruction}, there are two options for how this can be cured to lead to an infinite distance limit in the quantum moduli space:
\begin{enumerate}
    \item An additional infinite distance limit in the K\"ahler moduli space of $\cB_3$ is taken. In this case, the EFT string responsible for the combined limit in the K\"ahler moduli space and $z\to \ii\infty$ would be a bound state of the string described above and D3-branes wrapping  movable curves inside $\cB_3$. The EFT strings arising in F-theory from wrapped D3-branes have been discussed in detail in~\cite{Lanza:2021udy,Cota:2022yjw}.
    For an EFT string obtained in F-theory by wrapping a D3-brane on a generator $C^a$ of the cone of movable curves on $\cB_3$, the tension in Planck units is given by 
    \begin{equation}
        \frac{T_{D3|_{C_a}}}{M_{\rm Pl}^2} \sim \frac{1}{\text{Re} \,T_a}\,,\qquad \text{Re}\,T_a = \frac{1}{2 g_s} \int_{D_a} J_s\wedge J_s\,.
    \end{equation}
    Here, $D_a$ is the generator of $\text{Eff}^1(\cB_3)$ dual to $C^a$ and $J_s$ the K\"ahler form on $\cB_3$ in string-frame.  An example of such a combined limit is to take $g_s\to 0$ without co-scaling the \emph{string-frame} K\"ahler form on $\cB_3$. The tension of the bound states of strings realizing such a combined limit scales as
    \begin{equation}
        \frac{T_{\rm bound}}{M_{\rm Pl}^2} \sim g_s \gg g_s^2 = \frac{M_{\rm IIB}^2}{M_{\rm Pl}^2}\,.
    \end{equation}
    Since this limit is the straightforward weak coupling limit of Type IIB string theory, the quantum gravity cutoff is set by the fundamental Type IIB string scale $M_{\rm IIB}$. As a result, in this combined limit we realize the hierarchy $T_{\rm EFT}\gg \Lambda_{\rm QG}$, more precisely the relation \eqref{eq:hierarchy4}; according to our general discussion, the EFT string need therefore not be a higher-dimensional supergravity or critical string. 
    In any event, the resulting limit in the quantum moduli space does not satisfy condition 1 in the Definition~\ref{def:2fusionheunconnectedribbon}. 
    \item A limit akin to $z\to \ii \infty$ at finite $\cV_{\cB_3}$ could exist, but the geometry of the degeneration of $W$ does not correctly capture the worldsheet degrees of freedom on the EFT string. This is in principle a viable option, but it would mean that the geometric description of F-theory as a compactification on the geometric background given by $W$ would be invalid. In other words, even if the limit were to exist at finite $\cV_{\cB_3}$, the supergravity approximation of the 4d $\cN=1$ effective theory would not be under parametric control in $\alpha'$. In this case, condition 2 in Definition~\ref{def:2fusionheunconnectedribbon} would not be satisfied.\footnote{The resulting limit can be viewed in analogy to the F-theory lift of the O-type A limits discussed in~\cite{Paper1}. There we also saw that there may exist an infinite distance limit once $g_s$ corrections are taken into account, but the perturbative description in terms of a Type IIB orientifold breaks down. In particular, the geometry in the F-theory lift is very different from what one would have expected classically.}
\end{enumerate}
Given the two possibilities above, we conclude that, indeed, the standard Sen-limit is K\"ahler-obstructed in the sense of Definition~\ref{def:2fusionheunconnectedribbon}. 

To understand the origin of the K\"ahler obstruction, we can consider the perturbative Type IIB description of the standard Sen-limit. For Type IIB orientifolds, the classical Einstein-frame divisor volumes are related to the string-frame divisor volumes via
\begin{equation}\label{VEDa0}
    \cV^{(0)}_{E}(D_a) = \frac{1}{g_s} \text{vol}(D_a) M_s^4\,,
\end{equation}
where $D_a$ is a divisor $D_a \in H_4(\cB_3)$ and $M_s$ is the Type IIB string scale. To keep the Einstein-frame divisor volumes constant in the $g_s\to 0$ limit, the string-frame divisor volumes have to be co-scaled to zero as well, 
\begin{equation}
    \text{vol}(D_a) M_s^4 \sim g_s \to 0\,. 
\end{equation}
Thus, even though all non-perturbative $g_s$ effects are suppressed in the standard Sen-limit at fixed Einstein-frame volumes, non-BPS worldsheet instanton corrections become unsuppressed since the limit corresponds to the small string-frame volume limit for $\cB_3$. 

The $\alpha'$ corrections in particular also correct the classical expression for the Einstein-frame divisor volumes. To infer the scaling of these corrections, we use that, from the perturbative Type IIB perspective, the candidate EFT string realizing the standard Sen-limit corresponds to a D7-brane wrapping the  Calabi--Yau threefold $V$ that arises as the double cover of $\cB_3$ in the Sen-limit, see also~\cite{Lanza:2021udy,Martucci:2022krl}. Due to the Chern--Simons coupling on the worldvolume of a D7-brane, the string obtained from the D7-brane wrapping $V$ has induced D3-brane string charge. Concretely, the charge corresponding to a D3-brane wrapping a generator $\widetilde{C}^a$ of the Mori cone of $V$ is given by
\begin{equation}
    Q_{D3|_{\widetilde{C}^a}}= -\frac{1}{24} \int_V c_2(V) \wedge \widetilde{D}_a \,,
\end{equation}
where $\widetilde{D}_a$ is the generator of the K\"ahler cone dual to $\widetilde{C}^a$. Since on a Calabi--Yau threefold $V$ the integral of $c_2(V)$ over elements in the K\"ahler cone is positive, the induced D3-brane charge on the D7-brane is negative, unless $V=T^6$ and the second Chern class vanishes identically. In case $V$ is an actual Calabi--Yau threefold, it is then clear that the candidate EFT string given by the D7-brane wrapping $V$ is not an EFT string, but we have to add D3-brane charge to obtain an actual EFT string.\footnote{This is analogous to the adjustment of the lattice of EFT 
string charges for K\"ahler moduli limits in heterotic compactifications as studied in \cite{Martucci:2022krl}, taking into account curvature corrections.}
Moreover, the negative D3-brane string charge on the candidate EFT string signals that the corrected Einstein-frame divisor volumes go to zero in the vicinity of this string. From the F-theory perspective, this means that there is a correction to the Einstein-frame divisor volumes of the form 
\begin{equation}\label{eq:correctionsIIB}
    \cV_E(D_a) = \cV^{(0)}_E(D_a) - \left(\frac{1}{24} \int_V c_2(V)\wedge D_a \right)\; {\rm Im}\,z+\dots\,.
\end{equation}
The classical term is given by \eqref{VEDa0}, which is kept constant for the EFT string realizing the standard Sen-limit. The corrections to the divisor volumes are hence of the form~\eqref{eq:quantumvolume} with $\alpha=1$ and $a$ given by the induced D3-brane charge. Using the Type IIB perspective, we hence confirm that the standard Sen-limit is K\"ahler-obstructed. In fact, this obstruction to the weak coupling limit is not a genuine $\cN=1$ effect as it can already be observed in the hypermultiplet moduli space of Type IIB compactifications on Calabi--Yau threefolds. As discussed in~\cite{Baume:2019sry}, limits of the same type as considered here are indeed not realized in the quantum moduli space.

Notice that the standard Sen-limit is special since we can describe it reliably using the perturbative Type IIB description. As stressed in~\cite{Paper1}, this is not the necessarily the case for limits involving also complex structure degenerations of the orientifold background $V/\Omega$. In these cases we have to use the F-theory description. 

\subsection{General \texorpdfstring{I$_n$}{In}-type limits}\label{ssec:localsen}
Apart from the standard Sen-limit, we can also consider general I$_n$-type limits in which the generic fiber develops a singularity of Kodaira type I$_n$ over some component(s) of the base. We now argue that also these limits are K\"ahler-obstructed. To see this, recall from Section~\ref{ssec:Kahleroverview} that the EFT strings associated with \emph{bona fide} infinite distance limits in $\cM_{\rm c.s.}(W)$ correspond either to critical strings or to supergravity strings in a higher-dimensional theory to which the theory decompactifies. However, as we argue in the following, this is not possible for EFT strings realizing I$_n$-type limits such that by Claim~\ref{claim:EFTstringobstruction} these limits are K\"ahler-obstructed. 

Let us assume first that the candidate EFT string corresponds to a supergravity string in a higher-dimensional theory. 
Since Minkowski vacua preserving four supercharges only exist in four dimensions, the higher-dimensional theory must  locally preserve at least eight supercharges.\footnote{Globally, there can be supersymmetry breaking defects. The $\frac{1}{2}$BPS EFT string preserves half of the local supercharges.}  
 This imposes additional constraints on the worldsheet theory of the string. In particular, the massless spectrum on the string must fill out multiplets of (at least) 2d $\cN=(2,2)$ or $\cN=(0,4)$ supersymmetry. To show that the worldsheet theory on the candidate EFT string cannot correspond to a supergravity string, we focus on the massless vector multiplets on the string worldsheet.
As discussed in Section~\ref{sssection:EFT-string-Sen}, the candidate EFT string associated with an I$_n$-type limit contains $\cN=(0,2)$ vector multiplets associated with a gauge group $G$ on the worldsheet.\footnote{The precise gauge group $G$ depends on the details of the geometry. To determine $G$, one has to identify the independent exceptional curve classes arising from the resolution of fibral I$_n$-singularities over the components of $\cB_{3,0}$. M2-branes wrapping these independent rational curves yield the W-bosons of the worldsheet gauge group $G$ in the F-theory limit, exactly as exemplified in the standard Sen-limit in Section~\ref{sssection:EFT-string-Sen}. Smooth fibers or fibers of type I$_{1}$ and II do not influence the worldsheet gauge group. It is then clear that in I$_n$-type limits a non-trivial gauge group $G$ must exist on the string worldsheet. The following argument is independent of which group $G$ is realised.}

Consider a supergravity string in $d>4$ dimensions.
 Such strings are, by definition, gravitationally coupled, and can form bound states of sufficiently high charge under the 2-forms to become black strings. 
 In the sequel we will constrain the worldsheet theory of the black string and, as a result, also the constituent candidate EFT string.
 In particular, we will argue that the black string should not carry a gauge group $G$ on its worldsheet, which then implies that also the original candidate EFT string cannot carry a gauge group $G$. This rules out the candidate EFT strings in I$_n$-type limits.

Coming back to a black string in $d>4$ dimensions, its near horizon geometry is AdS$_3\times S^{d-3}$. This means that the worldsheet theory on the black string flows to an interacting 2d CFT in the IR that is dual to the AdS$_3$ factor. The entropy of the black string is then related to the central charge of this CFT via the Cardy formula~\cite{Strominger:1996sh}. The UV theory on the string worldsheet can contain chiral multiplets, Fermi multiplets and vector multiplets associated with a worldsheet gauge group $G$. The overall worldsheet spectrum is constrained by bulk anomaly inflow onto the string, see e.g.~\cite{Kim:2019vuc,Katz:2020ewz}, which in particular fixes the difference $c_L-c_R$ of the central charges. In the IR, the interacting CFT that reproduces the black string entropy is a sigma model whose target space is given by the Higgs branch 
\begin{equation}
       \cM_{\rm Higgs} = \left(\text{chiral matter fields}\right)/G\,. 
\end{equation}
In addition, the IR CFT is characterised by a vector bundle $\mathbb{E}$ over $\cM_{\rm Higgs}$ associated with the Fermi multiplets. The central charges of the IR CFT are then given by 
\begin{equation}
    c_L =  \text{dim}_\mathbb{R}(\cM_{\rm Higgs}) + {\rm rk}(\mathbb{E}) \,,\qquad c_R = \frac32 \text{dim}_\mathbb{R} (\cM_{\rm Higgs})\,,
\end{equation}
where by anomaly matching the difference $c_L-c_R$ has to match the value derived in the UV via anomaly inflow. Since the fluctuations in 2d QFTs are unbounded and hence cannot be localised around a given point in the scalar field space~\cite{Coleman:1973ci}, we have to take into account the full Higgs branch to determine the IR CFT. This has two important consequences:
\begin{enumerate} 
\item Additional massless modes arising at the origin of the Higgs branch do not contribute to the central charge. In particular, this means that vector multiplets are merely constraints from the IR CFT perspective. 
\item For a given chiral multiplet on the string, we have to consider all possible values for its scalar fields such that in particular we can consider points in $\cM_{\rm Higgs}$ at which the gauge group on the string is maximally Higgsed.
\end{enumerate}

We argued in Section~\ref{sssection:EFT-string-Sen} that the EFT string realizing an I$_n$-type limit can carry zero-modes charged under a gauge group $G$ on the string worldsheet. Given our assumption that the candidate EFT string is a supergravity string, to compute the central charge of the IR CFT realized on the worldsheet of the black string obtained as a bound state of this candidate EFT string, we have to consider the full Higgs branch. 
 If there are chiral multiplets charged only under $G$, giving them a VEV (partially) Higgses the gauge algebra on the string. 
 The resulting degeneration then carries a smaller gauge algebra. Either this gives rise to a regular-fiber limit or to an I$_n$-type limit with reduced gauge algebra and without chiral matter charged only under this gauge algebra.
 We can therefore assume from now on that no chiral matter charged only under the worldsheet gauge group of the candidate EFT string is present.
Furthermore, if there is chiral matter charged also under the spacetime gauge group, giving a VEV to the scalars in the charged chiral multiplets on the string also Higgses part of the spacetime gauge symmetry. This can be interpreted as a recombination of the string with the spacetime-filling 7-branes supporting the spacetime gauge theory. In other words, the string in the original background does not exist altogether. Thus, our assumption that the candidate EFT string is a supergravity string of a higher-dimensional theory is wrong.

The above argument assumes that the worldsheet theory contains chiral multiplets that are charged under the worldsheet gauge symmetry $G$. It is also possible that the candidate EFT string only hosts $G$-charged Fermi multiplets or no charged matter multiplets at all. We can now again assume that the candidate EFT string is a higher-dimensional supergravity string and consider the black string formed by bound states of this candidate EFT string. In both cases, the gauge symmetry on the black string worldsheet would remain intact in the IR and the IR CFT would be a direct product 
\begin{equation}\label{eq:IR-CFT12}
    {\rm CFT}_1 \otimes {\rm CFT}_2\,,
\end{equation}
where CFT$_{1}$ describes the interacting CFT made up from the $G$-neutral multiplets whose central charge has to reproduce the entropy of the black hole via Cardy's formula. CFT$_2$ instead is the decoupled sector associated with $G$. The central charges of the latter are given by
\begin{equation}
    c_L({\rm CFT}_2) = n_{{\rm Fermi},G} \,,\qquad  c_R({\rm CFT}_2) =0\,,
\end{equation}
where $n_{{\rm Fermi},G}$ is the number of $G$-charged Fermi multiplets. If $n_{{\rm Fermi},G}$ is non-zero, the interacting CFT$_1$ cannot reproduce the entropy of the black string since
\begin{equation}
    c_L({\rm CFT}_1)-c_R({\rm CFT}_1)\neq c_L-c_R\,,
\end{equation}
i.e. the difference of central charges of the remaining interacting CFT$_1$ does not match the UV value derived from anomaly inflow onto the black string which includes the contribution from CFT$_2$. Thus, even if the charged matter is neutral under the spacetime gauge symmetry, the candidate EFT string cannot be a higher-dimensional supergravity string. 

The only option for the candidate EFT string realizing an I$_n$-type limit to be a supergravity string is thus that the gauge theory on the string has no charged matter altogether. However, also in this case, the IR CFT on a black string obtained as a bound state of the candidate EFT string factorises as in~\eqref{eq:IR-CFT12}, where now CFT$_2$ describes the decoupled vector multiplets with no contribution to the central charges. Thus, from the black string perspective, the existence of the gauge theory on the string worldsheet is redundant since the IR theory on the black string cannot detect whether or not such a gauge theory is realized. From the perspective of the candidate EFT string, this manifests itself geometrically in the degeneration: It implies that the degeneration that naively leads to an I$_n$-type limit can be transformed into a regular-fiber limit. To see this, notice that charged matter 
in the gauge theory on the string can only be absent if the base $\cB_3$ of the elliptic fibration degenerates into multiple components in the I$_n$-type limit;
 furthermore, the elliptic fiber $\cE$ must degenerate in codimension-zero over a component of the degenerate base $\cB_3$ that is not intersected by the O7-plane. 
Otherwise, there is always charged matter coming from the orientifold divisor that lies on $\cB_3$. Such configurations were encountered in~\cite{Paper1}, where it was argued that in this case the component over which $\cE$ degenerates can be blown down, whereby moving mutually local perturbative A-type 7-branes on top of each other. This then yields a semi-stable degeneration that is of regular-fiber type. 

To conclude, for I$_n$-type limits there are two possibilities:
\begin{enumerate}
    \item The vector multiplets of the worldsheet theory on the candidate EFT string have charged matter that is also charged under spacetime gauge symmetries. The string can then not give rise to a supergravity string, since either
    \begin{itemize}
        \item[(a)] there are charged chiral multiplets so that on generic points on its Higgs branch, the string would recombine with the spacetime-filling 7-branes, or,
        \item[(b)] only charged Fermi multiplets exist which decouple in the IR, such that the interacting CFT on the string worldsheet of a black string obtained as a bound state of the candidate EFT string fails to reproduce the entropy of this black string.
    \end{itemize}
     \item The vector multiplets on the string worldsheet either have no charged matter or matter only charged under the worldsheet gauge theory. In either case, the degeneration realizing the I$_n$-type limit can be turned into a regular-fiber limit by a birational transformation. 
\end{enumerate}

The above arguments demonstrate that the candidate EFT string associated with an I$_n$-type limit in $\cM_{\rm c.s.}(W)$ cannot be a supergravity string of a higher-dimensional theory due to the presence of the non-Abelian vector multiplets with charged matter on its worldsheet.\footnote{This can also be seen directly in concrete realizations of supergravity strings. For example, in M-theory realizations of 5d $\cN=1$ theories, supergravity strings with $\cN=(0,4)$ worldsheet supersymmetry are given by MSW strings~\cite{Maldacena:1997de} obtained from M5-branes wrapping semi-ample divisors $D$ of the Calabi--Yau threefold $X$ on which M-theory is compactified, see also~\cite{Katz:2020ewz}. Vector multiplets now arise from reducing the self-dual two-form on the M5-brane worldvolume theory over elements of $H^1(D)$. If $D$ is an ample divisor, then $b_1(D)=0$, such that the M5-brane string on $D$ does not contain any vector multiplets. If, instead, $D$ is semi-ample but not ample, it is either the fiber of a surface fibration of $X$ or a vertical divisor of an elliptic fibration of $X$~\cite{Oguiso}. In the former case, $b_1(D)\neq 0$ is possible if and only if it is an Abelian surface, in which case the string is a fundamental Type II string which also does not have vector multiplets with charged matter on its worldsheet. In the latter case, the supergravity string can be described in F-theory on $X$ as a D$3$-brane wrapping a movable curve. It was found in~\cite{Lawrie:2016axq} that also these strings do not carry vector multiplets on their worldsheet.} Moreover, for the same reason the candidate EFT strings cannot be critical strings since critical strings are known to not have any vector multiplets with charged matter on their worldsheet. Applying Claim~\ref{claim:EFTstringobstruction} to this case then establishes that these limits are K\"ahler-obstructed in the sense of Definition~\ref{def:2fusionheunconnectedribbon}.

\section{K\"ahler obstructions for regular-fiber limits}\label{sec:regularfiber}
We now turn to the second class of complex structure degenerations of elliptically fibered Calabi--Yau fourfolds, for which all double threefolds are themselves fibered by $\cE$. Unlike for the I$_n$-type limits discussed previously, we cannot give a general argument why these limits are K\"ahler-obstructed, but instead discuss the various types of degenerations separately. 

\subsection{Type II limits}\label{ssec:typeII}
We first analyse K\"ahler obstructions to regular-fiber type II limits. Using the logic underlying Claim~\ref{claim:EFTstringobstruction}, we argue that the worldsheet theory of the candidate EFT string associated with a general regular-fiber type II limit is incompatible with the expectation of the Emergent String Conjecture. We then show that in certain cases, namely those corresponding to the semi-stable degeneration limits underlying F-theory/heterotic duality, the quantum corrections are of the form as in Claim~\ref{claim:explicitobstructions}. This provides a second, independent argument why these limits are K\"ahler-obstructed in the sense of Definition~\ref{def:2fusionheunconnectedribbon}. 
\subsubsection{General discussion using EFT strings}
Consider a regular-fiber type II limit for an elliptically fibered Calabi--Yau fourfold $\cE\hookrightarrow W\to \cB_3$. For simplicity, we assume that the degeneration is of the form
\begin{equation}\label{eq:typeII}
    W \to W_0 =W_1 \cup_V W_2\,,
\end{equation}
where the intersection locus $V$ is a Calabi--Yau threefold that allows for an elliptic fibration with generic fiber $\cE$, i.e., $\pi_V: \cE\hookrightarrow  V\to \cB_2$.\footnote{For a type II degeneration with more than two components, all double threefolds $W_i\cap W_j$ have to be Calabi--Yau and moreover mutually isomorphic as complex manifolds, i.e., all $W_i\cap W_j$ are copies of $V$. Performing the counting in this case therefore simply leads to a ``double'' counting of WS modes, see also Conjecture~2 of~\cite{Hassfeld:2025uoy}.} Following the procedure outlined in Section~\ref{ssec:countingFM}, we can count the degrees of freedom on the candidate EFT string associated with this degeneration. The universal geometric modes provide two real, massless scalars $|\mathbf{z}_0|$ and $\arg(\mathbf{z}_0)$ on the string worldsheet in addition to a single real internal geometric mode $\Phi_1$ associated with the geometry of the type II degeneration. Since $H^1(V)=\emptyset$, there are no modes arising from $\cC_3$. The only additional scalar modes arise from $\cC_6$ reduced over $\pi_V$-vertical cycles. The resulting modes have definite chirality on the string in F-theory. The chirality is determined by the signature 
\begin{equation} \label{eq:1bmodes-typeII}
    \text{sgn}(H_{\rm vert}^4(V)) = (1,b)\,,
\end{equation}
for some $b\geq0$, where we used the Hodge index theorem for the base $\cB_2$ of $V$. Accordingly, there is a single right-moving scalar coming from $\cC_6$. Altogether, there are hence four right-moving real scalar degrees of freedom that are part of two chiral $\cN=(0,2)$ multiplets. Ths gives a right-moving central charge of $c_R=6$, indicating that the candidate EFT string for a regular-fiber type II limit is subcritical and not gravitational in nature. From Claim~\ref{claim:EFTstringobstruction} it then follows that regular-fiber type II limits are K\"ahler-obstructed.

It is instructive to compare this situation to the mode counting on an EFT string in Type IIB string theory compactified on a Calabi--Yau threefold that realizes a type II degeneration of the threefold. In F-theory language, this setup corresponds to the trivial fibration $V \times {\cal E}$ and its ${\cal N}=(0,4)$ supersymmetric EFT string discussed at the end of Section \ref{sec:regfibcounting}. In this case, the mode counting of \cite{Hassfeld:2025uoy} identified eight right-moving scalars, which are part of four ${\cal N}=(0,4)$ chiral multiplets and result in $c_R = 12$. This is indeed the correct value for a critical string, in agreement with the interpretation of the type II EFT string as an emergent heterotic string. The difference in counting comes from the fact that in the 4d ${\cal N}=2$ context, apart from the geometric deformation modes, the Type IIB 2-forms $(B_2, C_2)$ give rise to a full scalar field each. Furthermore, the 4-form $C_4$ contributes scalar modes with signature $(3,3+n)$ associated with the lattice of integral harmonic 2-forms on the double surface of the degeneration.\footnote{Here $n=16$ if the double surface is a K3 surface and $n=0$ if it is an abelian surface.} If we think of the 4-fold type II degeneration \eqref{eq:typeII} as the F-theory uplift of a type IIB orientifold, then the absence of modes from $(B_2, C_2)$ is an immediate consequence of the orientifold projection, under which $(B_2,C_2)$ are odd.
   Equivalently, the absence of these modes reflects the non-trivial twist of the elliptic fibration
   compared to the 4d 
${\cal N}=2$ setup in Section \ref{sec:regfibcounting}. 
 Similarly, the orientifold projection reduces the lattice of harmonic two-forms on the double surface to a lattice of signature $(1,b)$ in F-theory, see \eqref{eq:1bmodes-typeII}.

\subsubsection{Quantum corrections to the heterotic semi-stable degeneration limit} \label{ssec:hetstablelimit}
Let us now consider the special case in which the F-theory compactification on $W$ has a heterotic dual. 
 In this context, we will show that in the type II limit underlying Heterotic/ F-theory duality, complex structure moduli dependent quantum corrections to divisor volumes become uncontrollably large, see \eqref{eq:VS2_corr}, and require a co-scaling of the K\"ahler moduli for the theory to stay within the perturbative regime. The corrections are of the form advertised in Claim \ref{claim:explicitobstructions}
 and give a complementary explanation for the K\"ahler obstruction of the type II regular-fiber limits.

The existence of a heterotic dual requires that the base $\cB_3$ of $W$ be rationally fibered~\cite{Morrison:1996na,Morrison:1996pp}
\begin{equation} \label{eq:rhofibration}
\rho:\mathbb{P}^1\hookrightarrow \cB_3 \to \cB_2 \,.
\end{equation}
If the heterotic dual of the F-theory compactification is an $E_8\times E_8$ heterotic string, there must exist a type II limit in the complex structure moduli space of $W$ as in~\eqref{eq:typeII}, where now $V$ is an elliptic fibration over the base $\cB_2$ of $\cB_3$. For simplicity, we assume $V$ to be a smooth Weierstrass model. The Calabi--Yau threefold $V$ can then be identified with the compactification manifold $V_H$ of the heterotic $E_8\times E_8$ string. To understand the fate of the type II large complex structure limit at finite volume of $\cB_3$, we must analyse the analogue of this limit in the heterotic language. 

Let us denote by $z$ the complex structure modulus of $W$ associated with the type II degeneration described above. Heterotic/F-theory duality identifies 
\begin{equation}
    z = \int_\cE (B_2^H + \ii J_H)\equiv t_\cE\,,
\end{equation}
where $B_2^H$ is the heterotic two-form and $J_H$ is the K\"ahler form on $V$ measuring volumes w.r.t. the heterotic string scale $M_{\rm het}$. The heterotic volume moduli of the base $\cB_2$ of $V$ are identified with $\rho$-vertical divisors of $\cB_3$. For a curve $C_{i}\in H_2(\cB_2)$ we have
\begin{equation}
    t^i_B \equiv \int_{C_i} (B_2^H + \ii J_H) = \int\limits_{\rho^\ast(C_i)} \left(C_4 + \frac\ii2 J_{\cB_3}^2\right) \,,
\end{equation}
where $J_{\cB_3}$ is the K\"ahler form on $\cB_3$ measuring volumes w.r.t. the Type IIB string scale. While the 4d heterotic dilaton $S_{\rm het}$ is identified with the volume of the base $\cB_2$ of $\cB_3$ measured in Type IIB string units, the holomorphic gauge kinetic functions of the two $E_8$ factors are classically given by 
\begin{equation}\label{eq:gaugekin-het}
    f_{E_8^{(1)}} = k_1 \left[\int_{\cS_1} \left(C_4 + \frac\ii2 J_{\cB_3}^2\right)\right] \equiv k_1 S_1 \,,\qquad f_{E_8^{(2)}} = k_2\left[\int_{\cS_2} \left(C_4 + \frac\ii2 J_{\cB_3}^2\right)\right]\equiv k_2 S_2 \,,
\end{equation}
where $\cS_1$ and $\cS_2$ are the two sections of the rational fibration $\rho:\cB_3\to\cB_2$ satisfying 
\begin{equation}
    \cS_1 = \cS_2 + \rho^*c_1(\cT)\,,\qquad \cS_1\cdot_{\cB_3} \cS_2=0\,,
\end{equation}
and $k_{1,2}$ are the levels of the heterotic gauge group. In the above expression $\cT$ is the twist bundle of the rational fibration $\rho$. 

In the heterotic dual theory, the gauge couplings of the $E_8\times E_8$ gauge theory receive  threshold corrections~\cite{Dixon:1990pc} of the schematic form (see~\cite{Klaewer:2020lfg} for a recent discussion including the duality to F-theory) 
\begin{equation}
    \frac{16 \pi^2}{g_{{\rm YM},i}^2 } = k_i\, \Im S_{\rm het} + \Delta^{(i)}(M, \bar M)\,.
\end{equation}
Here, $M$ denotes the scalars in the 4d $\cN=1$ chiral multiplets in the heterotic theory and $S_{\rm het}$ is the heterotic 4d dilaton. The contributions to $\Delta$ can be split into those inherited from threshold corrections to the holomorphic gauge kinetic functions and those arising from non-holomorphic corrections to the K\"ahler potential, 
\begin{equation}
    \Delta^{(i)} = 2\pi \Im f_{\rm one-loop}^{(i)}(M) + \frac{c^{(i)}}{8\pi}K^H(M,\bar M) \equiv \Delta_0^{(i)} +\Delta_1^{(i)}\,,
\end{equation}
where $c^{(i)}$ is a one-loop coefficient of the $i$-th gauge group, see e.g.~\cite{Louis:1996ya}. The one-loop corrections encoded in $\Delta_0^{(i)}$ have been computed in~\cite{Dixon:1990pc},
\begin{equation}
    \Delta_0^{(i)} = \frac{1}{16\pi^2} \int_{\Gamma} \frac{{\rm d}^2\tau}{\tau_2}\left(\mathfrak{B}^{(i)}(\tau,\bar{\tau}) - b^{(i)}\right)\,.  
\end{equation}
Here, $b^{(i)}$ is the $\beta$-function coefficient of the $i$-th gauge group and $\tau$ is the complex structure parameter of the worldsheet $T^2$ integrated over its $SL(2,\mathbb{Z})$ fundamental domain $\Gamma$. Finally, the integrand $\mathfrak{B}^{(i)}$ is given by 
\begin{equation}
    \mathfrak{B}^{(i)}(\tau, \bar\tau) = |\eta(\tau)|^{-4} \sum_{\text{even}\; \mathbf{s}} (-1)^{s_1+s_2} \cdot \mathfrak{B}_{\rm ext}(\mathbf{s},\tau) \cdot \Tr_{s_1} \left(Q_{(i)}^2 (-1)^{s_2} q^H q^{\bar H}\right)\,,
\end{equation}
where $s_a=0,1$ denote NS, R boundary conditions of the string, $Q_{(i)}$ the charge of the state under the heterotic gauge group in the $i$-th factor and $H, \bar{H}$ are the left- and right-moving Hamiltonians of the internal CFT. A particularly simple case corresponds to the heterotic standard embedding for which the heterotic bundle in the first $E_8$ is identified with the tangent bundle of $V_H$ while the bundle in the second $E_8$ is trivial. In this case, the worldsheet has enhanced $\cN=(2,2)$ supersymmetry and the worldsheet integral can be evaluated explicitly. In fact, up to group-theoretic factors, the integral agrees with the genus-one free energy of the topological string computed in~\cite{Bershadsky:1993ta,Bershadsky:1993cx}. We are in particular interested in the dependence of the threshold corrections on the heterotic moduli close to infinite distance boundaries. From the behaviour of the genus-one free energy of the topological string for large values of the K\"ahler moduli, we infer 
\begin{equation}
    \Delta^{(i)}_0 \stackrel{t\to \infty}{\longrightarrow} \frac{b^{(i)}}{192 \pi^3} \int_{V_H} J_H \wedge c_2(V_H) \,.
\end{equation}
For standard embedding, the hidden gauge group $E_8^{(2)}$ remains unbroken such that the $\beta$-function coefficient is negative,~$b^{(2)}=-90$. The gauge coupling of the hidden $E_8^{(2)}$-factor gets corrected as 
\begin{equation}
    \frac{16\pi^2}{g_{{\rm YM},2}^2} = k_2 \Im S_{\rm het} - \frac{|b^{(2)}|}{192\pi^3} \int_{V_H} J_H \wedge c_2(V_H) + \Delta_1^{(2)}\,. 
\end{equation}
Notice that the second term is linear in the K\"ahler moduli of $V_H$. Since $\Delta_1^{(2)}$ is proportional to the tree-level K\"ahler potential of the heterotic string, the K\"ahler moduli only appear logarithmically in $\Delta_1^{(2)}$ such that the leading correction is encoded in the second term in the above expression. The corrections to the physical gauge coupling induce a strong coupling behaviour for large values of the heterotic K\"ahler moduli. The contribution of the K\"ahler modulus $t_\cE$ can be computed by first noticing that $\text{Im}\,t_{\cal E}$ is the coefficient of the shifted zero section of $p:V_H\to \cB_2$ to $J_H$. The shifted zero section is given by 
\begin{equation}
    Z_+ = Z_0 +p^*(c_1(\cB_2))\,,
\end{equation}
where $Z_0$ is the zero section of $p$. Using that $V_H$ is a smooth Weierstrass model, by adjunction we further have 
\begin{equation}
    c_2(V_H) = p^* c_2(\cB_2) +12 p^*(c_1(\cB_2))\wedge Z_+ - (p^*c_1(\cB_2))^2 \,,
\end{equation}
such that the $t_\cE$ dependent contribution to $\Delta_0^{(2)}$ is given by 
\begin{equation}
    - \Im(t_\cE)\cdot \frac{|b^{(2)}|}{192\pi^3} \int_{V_H} Z_+ \wedge c_2(V_H) = - \Im(t_\cE)\cdot \frac{|b^{(2)}|}{192\pi^3} \int_{\cB_2} \left(c_2(\cB_2)+ 11 c_1^2(\cB_2)\right)\,.
\end{equation}
The integral over the characteristic classes of $\cB_2$ is positive, e.g., for Hirzebruch or del-Pezzo surfaces such that the $t_\cE$-dependent threshold corrections increase the gauge coupling in the hidden $E_8^{(2)}$. In particular, it is not possible to take the $t_\cE\to \ii \infty$ limit while maintaining a perturbative heterotic description of the theory. 

Via~\eqref{eq:gaugekin-het} we see that the heterotic gauge coupling can be identified with the action of a D3-brane instanton wrapping $\cS_2$. As discussed in~\cite{Klaewer:2020lfg}, a subset of the threshold corrections to the heterotic gauge coupling are encoded in the classical F-theory geometry. Concretely, the geometry encodes the contribution to the threshold corrections linear in the K\"ahler moduli of the base $\cB_2$ of $V_H$ such that for large $t^i_B$ we identify
\begin{equation}
    \text{Im}(S_1)-\Im(S_2) = \int_{\cB_3} J_{\cB_3}^2\wedge \rho^*(c_1(\cT)) = \sum_{i=1}^{h^{1,1}(\cB_2)}\Im \, t^i_B \left(\left|\partial_{\Im t^i_B} \Delta_0^{(1)}\right| +\left|\partial_{\Im t^i_B}\Delta_0^{(2)}\right|\right) \,,
\end{equation}
where we used $b^{(2)}=-|b^{(2)}|$.
Instead, the dependence of the threshold corrections on $t_\cE$ is not captured by the classical geometry of the F-theory base but has to arise as a quantum correction to the D3-brane instanton action. For standard embedding and with $b^{(2)}=-90$ we then have
\begin{equation} \label{SD3S-action}
    \Im S_{D3|\cS_2} = \Im S_2 - \Im\,z\cdot \frac{15}{32\pi^3} \int_{\cB_2} \left(c_2(\cB_2)+ 11 c_1^2(\cB_2)\right)\,.
\end{equation}
In the large $\Im z$ limit realizing the type II degeneration for the F-theory fourfold $W$, the instanton action of the D3-brane wrapping $\cS_2$ hence becomes unsuppressed. Correspondingly, the $E_8^{(2)}$ gauge theory realized on $\cS_2$ becomes strongly coupled. This strong coupling behaviour of the $E_8$ gauge theory is the F-theory analogue of the strong coupling singularity of the heterotic string with asymmetric instanton embedding first discussed in~\cite{Witten:1996mz}. Whereas classically this strong coupling singularity acts as a finite-distance boundary of the moduli space, it gets resolved at the quantum level by non-perturbative effects from brane instantons as discussed in~\cite{Cvetic:2024wsj}, implying that the moduli space continues into a strongly coupled phase. In genuine $\cN=1$ compactifications, the nature of this phase and in particular whether it features infinite distance limits along the original direction has not yet been investigated. However, in closely related heterotic $\cN=2$ setups, it was shown in~\cite{Cvetic:2025nfx} that the strong-coupling phase that is present due to the non-perturbative effects in four dimensions indeed does not feature a non-compact direction replacing the classical infinite distance direction.\footnote{Concretely, in the setup studied in~\cite{Cvetic:2025nfx}, the classical infinite distance limit corresponds to a decompactification limit of the heterotic string to Ho\v{r}ava--Witten M-theory. At the quantum level, the HW interval is replaced by a domain wall interpolating between the visible 9-brane and a supersymmetric AdS$_5$ vacuum. As gravity is confined to a regime close to the visible brane, the resulting theory is not a theory with gravity propagating in a five-dimensional bulk theory. For this reason, the original large interval limit does not correspond to a decompactification limit and does not feature a light tower of KK-states or an emergent string.} 

Conversely, we can interpret the corrections in~\eqref{SD3S-action} as corrections to the quantum volume of $\cS_2$ as 
\begin{equation}\label{eq:VS2_corr}
    \cV_{\cS_2} = \underbrace{\text{Im}\,(S_2)}_{=\cV_{\cS_2}^{(0)}}  - a \;\text{Im}\,z \,,
\end{equation}
where $a$ can be read off from \eqref{SD3S-action}. The corrections to $\cV_{\cS_2}$ are hence of the form given in~\eqref{eq:quantumvolume} with $\alpha=1$ such that using Claim~\ref{claim:explicitobstructions} we conclude that the type II limit corresponding to the heterotic semi-stable degeneration limit is indeed K\"ahler-obstructed.

\subsection{Type III limits}\label{ssec:typeIII}
Another class of regular-fiber limits corresponds to type III limits. Here, the Calabi--Yau fourfold $W$ degenerates as 
\begin{equation}
    W\to W_0 = \bigcup_{i=1}^n W_i\,,
\end{equation}
where $n\geq 3$ and $W^{(k+1)}=\emptyset$ for $k>2$. The triple surfaces $W_{i_0 i_1 i_2}$ are K3 surfaces which are elliptically fibered with generic fiber $\cE$, i.e., the fiber of $\pi: W\to \cB_3$ is smooth over generic points in the base of $W_{i_0i_1i_2}$. In the following, we do not provide a general argument why all possible type III limits of F-theory fourfolds have to be obstructed, but focus on two relevant cases. In the first case, we assume the type III degeneration to be `minimal' in the sense that the central fiber $W_0$ is the union of three components $W_{1,2,3}$. In this case, we can compute the central charge of the worldsheet theory of the candidate EFT string. Based on this, we will see that the string is not a 6d supergravity string, contrary to what one would expect for a type III limit in the complex structure moduli space based on the behaviour of the classical effective action in this limit. This shows that the corresponding limits are K\"ahler-obstructed via Claim~\ref{claim:EFTstringobstruction}. The second example are type III degenerations for fourfolds that are Calabi--Yau threefold fibered and for which the Calabi--Yau threefold fiber undergoes a type III degeneration.\footnote{At the end of Section~\ref{sssec:typeIIIexplicit} we comment on generalisations of the obstructions to arbitrary degenerations of the Calabi--Yau threefold fiber.} In the adiabatic limit, the corrections to certain divisor volumes can be computed within the 6d parent theory and are of the form~\eqref{eq:quantumvolume} such that by Claim~\ref{claim:explicitobstructions} also these limits are K\"ahler-obstructed. This argument in fact generalises to all type II, III or IV limits, of regular fiber type (unless we are in type IV) or I$_n$ type, which arise as corresponding limits of a Calabi--Yau threefold fiber. 

\subsubsection{EFT strings and minimal type III degenerations}
Consider a type III degeneration of a Calabi--Yau fourfold $W$ that is of minimal type as introduced above. For simplicity, we assume that $W$ can be described as a smooth Weierstrass model over some base $\cB_3$ with generic fiber $\cE$. In this type III limit in the complex structure moduli space of $W$, the fourfold splits into three components $W_1,W_2,W_3$ which intersect over the double threefolds 
\begin{equation}
    V_{12} = W_1 \cap W_2 \,,\quad V_{13} =W_1\cap W_3\,\quad V_{23} = W_2\cap W_3 \,. 
\end{equation}
In addition, there is a single triple surface 
\begin{equation}
    W_{123} = W_1 \cap W_2 \cap W_3\,. 
\end{equation}
All double threefolds and triple surfaces are themselves elliptically fibered with generic fiber $\cE$. We denote the respective projections associated with these fibrations by
\begin{equation}
    \pi_{ij}: \cE \hookrightarrow V_{ij} \to \cB_{2,ij}\,,\quad \pi_{123}:\cE \hookrightarrow W_{123} \to \mathbb{P}^1\,. 
\end{equation}
According to our count of degrees of freedom in Section~\ref{ssec:countingFM}, the candidate EFT string associated with this classical type III degeneration has four massless real scalars on its worldsheet. These are the universal $|\mathbf{z}_0|$ and $\arg \mathbf{z}_0$-modes and the two internal deformations $\Phi_{1,2}$. Since $H^1(V_{ij})=H^1(W_{123})=\emptyset$, there are no scalar modes coming from the M-theory 3-form $\cC_3$. Instead, there are modes arising from $\cC_6$ when reduced over elements of $H^4(V_{ij})$ and $H^4(W_{123})$. As discussed in Section~\ref{ssec:countingFM}, only modes arising from $\pi$-vertical four-forms survive the F-theory lift to modes on the string worldsheet, i.e., from forms that are the pull-back of two-forms on the base of $V_{ij}$ or $W_{123}$. The chirality of the modes on the string worldsheet is then determined by the properties of the two-form. 

Concretely, since the base of $W_{123}$ is simply a $\mathbb{P}^1$, the M-theory mode arising from $\cC_6$ reduced over $H^4(W_{123})$ yields a real scalar field on the string in the F-theory uplift. Furthermore, the signature of the base of $\cB_{2,ij}$ of $V_{ij}$ is ${\rm sgn}(H^2(\cB_{2,ij}))= (1,d_{ij})$ for some $0\leq d_{ij}$. For this reason, $\cC_6$ reduced over each of the double threefolds yields exactly one right-moving scalar degree of freedom, in addition to a number of left-moving ones, whose number depends on $d_{ij}$.
In total, the M-theory $\cC_6$-form hence contributes four right-moving scalars in addition to the four geometric right-moving scalars (see Appendix~\ref{sapp:III-ex} for the analogous counting in Type IIB). These scalars form part of the 2d $\cN=(0,2)$ chiral multiplets such that the total right-moving central charge for the candidate EFT string realizing a minimal type III limit is $c_R=12$. By contrast, for the left-moving central charge we obtain the bound 
\begin{equation}\label{boundcL}
    5\leq c_L\,.
\end{equation} We now argue that these central charges are inconsistent with the asymptotics of the classical effective action, thereby indicating that there have to be significant quantum corrections arising in the type III limit. 

To this end, we focus first on the effective action obtained from the classical 4d $\cN=1$ K\"ahler potential. For simplicity, let us assume that the type III limit corresponds to a one-parameter limit obtained by sending $z\to \ii \infty$ while keeping all other moduli of the theory fixed. In this case, the K\"ahler potential behaves to leading order as 
\begin{equation}
    K= -2\log(\Im z) + \dots\,.   
\end{equation}
Choosing a scaling of the form $\Im z\sim \lambda$, the distance between the loci of points corresponding to $\lambda=\lambda_0$ and $\lambda=\l_1 $ in the classical moduli space grows logarithmically as
\begin{equation}
    \Delta = \int\limits_{\lambda_0}^{\lambda_1} \sqrt{\frac12 \frac{\mathrm{d}^2 K }{\mathrm{d} \lambda^2}} \mathrm{d} \lambda = \log \lambda_1/\lambda_0+\cdots\,. 
\end{equation}
In particular, the coefficient of the leading logarithm is one. Under the assumption that the classical theory correctly describes the physics in the type III limit, the quantum gravity cutoff as a function of $\Delta$ has to scale as
\begin{equation}\label{scalingtypeIII}
    \frac{\Lambda_{\rm QG}}{M_{\rm Pl}} \sim e^{-\Delta}\,.
\end{equation}
Here we used that the quantum gravity cutoff has to be linear in $\Im z$ if the limit is unobstructed \cite{Martucci:2024trp}, as discussed around \eqref{eq:FGBscaling}. The scaling of the quantum gravity cutoff as a function of the distance is characteristic for a decompactification from four to six dimensions~\cite{vandeHeisteeg:2023ubh}. If the limit exists, the candidate EFT string realizing this limit then has to be a supergravity string of a six-dimensional supergravity theory. 

However, as we now argue, the candidate EFT string cannot be a supergravity string in a 6d theory of supergravity. To see this, it is sufficient to consider BPS supergravity strings in minimal 6d $\cN=(1,0)$ supergravity. The goal is to show that such strings cannot have $c_R=12$. For a BPS string in a 6d $\cN=(1,0)$ theory carrying charges $Q^\alpha$ under the 2-forms of the 6d bulk theory, the central charges of the string worldsheet theory are determined via anomaly inflow and are given by \cite{Kim:2019vuc}
\begin{equation}
    c_L = 3Q^2 -9Q\cdot a+6\,,\qquad c_R=3Q^2 -3Q\cdot a+6\,.
\end{equation}
Here, $a^\alpha$ encodes the contribution to the gravitational anomaly polynomial and the inner product is w.r.t. the intersection form with signature $(1,n_T)$, where $n_T$ is the number of 6d tensor multiplets. Since all quantities in the above expression are integers, the difference of left- and right-moving central charges satisfies the quantisation condition
\begin{equation}\label{eq:cLminusCR}
    c_L - c_R \in 6\mathbb{Z}\,. 
\end{equation}
For $c_R=12$ this constrains the left-moving central charge to be $c_L = 12+6k$ for $-1\leq k\in \mathbb{Z}$. In terms of $k$ we then obtain
\begin{equation}
    Q\cdot a = -k\,,\quad Q^2 =2-k \,. 
\end{equation}
Notice that for $k=2$ the string would correspond to a critical heterotic string. However, for critical EFT strings the relation between the moduli space distance and tension setting the quantum gravity cutoff does not scale as in~\eqref{scalingtypeIII} such that the candidate EFT string realizing the type III limit cannot be a critical string. Instead, for $k=-1,0,1$ we obtain subcritical strings which therefore cannot be supergravity strings of the 6d $\cN=(1,0)$ theory.\footnote{For $k=0$ and $k=1$ the central charges are 
$(c_L,c_R)=(12,12)$ and $(c_L,c_R)=(18,12)$, which for non-chiral $\cN=(2,2)$ theories would signal a critical and, respectively, a supercritical string. For the chiral $\cN=(0,2)$ theories we consider, these central charges correspond to a subcritical string. Even if we allowed for enhanced $\cN=(2,2)$ supersymmetry, these cases can be excluded: The case $(c_L,c_R)=(12,12)$ is not possible because a critical EFT string would imply an emergent string limit, in contradiction with the behaviour of the EFT, while $(c_L,c_R)=(18,12)$ is incompatible with lef-right symmetric supersymmetry. } Finally, for $k>2$, the string has $Q^2+Q\cdot a <-2$ and therefore does not satisfy the unitarity bounds for 6d supergravity strings~\cite{Kim:2019vuc}.\footnote{Another way to see this is that in the F-theory realization of 6d $\cN=(1,0)$ theories, $Q^2$ corresponds to the self-intersection of the curve wrapped by the D3-brane that gives rise to the supergravity string. For $k>2$, the corresponding curve would have negative self-intersection such that it is shrinkable at finite distance implying that the associated string is not a supergravity string.} We thus conclude that for the minimal type III degenerations, the candidate EFT strings do not correspond to 6d supergravity strings such that by Claim~\ref{claim:EFTstringobstruction} these limits are~K\"ahler-obstructed.

A simple class of fourfolds that feature a minimal type III limit are obtained as O-type B orientifolds of minimal type III limits of Calabi--Yau threefolds. An example for this based on the hypersurface Calabi--Yau threefold $\mathbb{P}^4_{2,2,2,3,3}[18]/(\mathbb{Z}_6\times\mathbb{Z}_2^2)$ has been discussed in detail in~\cite{Paper1}.

\subsubsection{K\"ahler obstructions from explicit quantum corrections}\label{sssec:typeIIIexplicit}
\begin{figure}[t]
    \centering
    \includegraphics[width=0.9\linewidth]{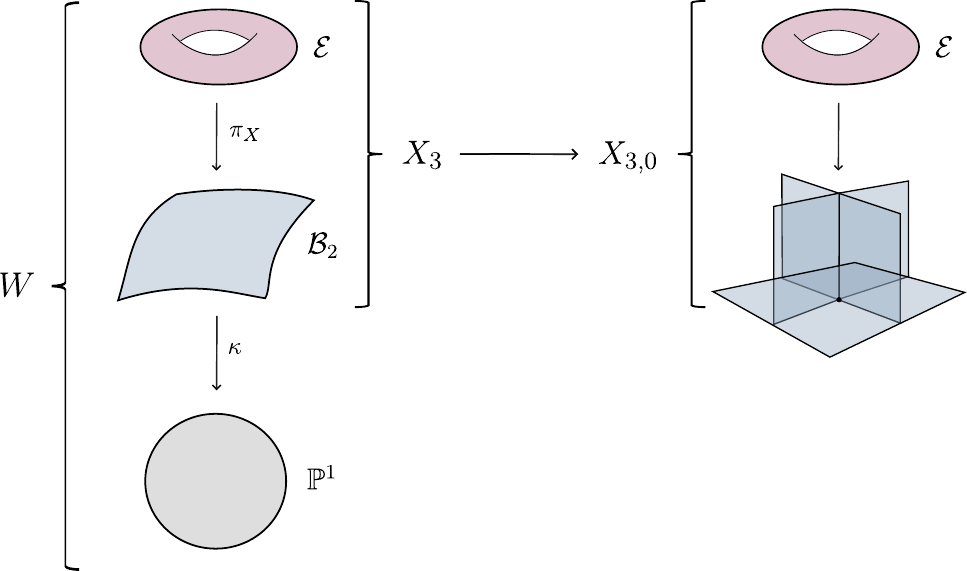}
    \caption{The left-hand side depicts an elliptic Calabi--Yau fourfold $W$ that in addition allows for a compatible Calabi--Yau threefold fibration. The right-hand side illustrates a regular-fiber type III limit of $W$ that is induced by a degeneration of the generic $X_3$-fiber over $\mathbb{P}^1$, such that the elliptic fiber $\mathcal{E}$ remains smooth in the limit. Here we have depicted the minimal case in which $X_{3,0}$ splits into three components $X_1,X_2,X_3$. }
    \label{fig:regular-III}
\end{figure}

We now consider a second class of type III limits in which the Calabi--Yau fourfold $W$ allows for a fibration
\begin{equation} \label{eq:adiabatic}
    \kappa: X_3 \hookrightarrow W \to \mathbb{P}^1\,,
\end{equation}
where the generic fiber $X_3$ is itself a Calabi--Yau threefold. As always in F-theory, $W$ has to allow for a genus-one fibration $\pi:\cE\hookrightarrow W\to \cB_3$. In the following, we assume the fibrations $\pi$ and $\kappa$ to be compatible. In particular, this implies that the generic fiber of $\kappa$ is itself elliptically fibered $\pi_X: \cE\hookrightarrow  X_3\to \cB_2$ such that the generic fiber of $\pi_X$ is the same as the generic fiber of $\pi$. The setup is depicted on the left-hand side of Figure~\ref{fig:regular-III}.

For such a Calabi--Yau fourfold, we consider a type III limit in the complex structure moduli space of $X_3$ that is of regular-fiber type. This means that $X_3$ undergoes a degeneration 
\begin{equation}
    X_3 \to X_{3,0} = \bigcup_{i=1}^n X_i\,,
\end{equation}
where $X_i\cap X_j\cap X_k\cap X_l=\emptyset$ for all $i,j,k,l\in \{1,\dots,n\}$ and at least one triple curve $X_i\cap X_j \cap X_k$ is non-empty. All non-empty triple curves are copies of the smooth elliptic fiber $\cE$ of $\pi_X:X_3\to \cB_2$. This limit in the complex structure moduli space of $X_3$ induces a type III limit for $W$ in which the generic fiber of $\kappa$ degenerates. This is the analogue of a Sen-limit for Calabi--Yau threefold fibrations.\footnote{Let us stress that by assumption the fiber of $\pi$ remains finite in this type III limit even though the generic fiber of $\kappa$ is not smooth.} The minimal case in which $X_{3,0}$ splits into three components is depicted on the right-hand side of Figure~\ref{fig:regular-III}. 

\begin{figure}
\centering
\begin{tikzpicture}[
  >=stealth,
  every node/.style={align=center},
  row sep=1.5cm,
  column sep=2cm
]

\matrix (m) [matrix of nodes]{
F-theory on $W$ 
& F-theory on $X_3$ 
& IIA on $X_3$ 
& IIB on $X_3$ \\

$\mathcal{M}_{\rm c.s.}(W)$
&
$\begin{array}[t]{c}
\mathcal{M}_{\rm c.s.}(X_3)\\
\cap\\
\mathcal{M}^{6{\rm d},\,\mathcal{N}=(0,1)}_{\rm HM}
\end{array}$
&
$\begin{array}[t]{c}
\mathcal{M}_{\rm c.s.}(X_3)\\
\cap\\
\mathcal{M}^{\rm IIA}_{\rm HM}
\end{array}$
&
$\begin{array}[t]{c}
\mathcal{M}_{\rm c.s.}(X_3)\\
\parallel\\
\mathcal{M}^{\rm IIB}_{\rm VM}
\end{array}$ \\
};

\foreach \x in {1,2,3,4}{
  \draw[<->] (m-1-\x) -- (m-2-\x);
}

\draw[->] (m-1-1) -- (m-1-2)
  node[midway, above] {$\mathcal{V}_{\mathbb{P}^1}\to\infty$};

\draw[->] (m-1-2) -- (m-1-3)
  node[midway, above] {$T^2$ comp.}
  node[midway, below] {};

\draw[<->] (m-1-3) -- (m-1-4)
  node[midway, above] {$c$\,-map};

\draw[<-{Hooks[left,length=1mm]}] ([xshift=27pt,yshift=-11.1pt]m-2-1.north) -- ([xshift=-32pt,yshift=-12.6pt]m-2-2.north);

\draw[-] ([xshift=30pt,yshift=-10.4pt]m-2-2.north) -- ([xshift=-34pt,yshift=-10.4pt]m-2-3.north);
\draw[-] ([xshift=30pt,yshift=-12.6pt]m-2-2.north) -- ([xshift=-34pt,yshift=-12.6pt]m-2-3.north);

\draw[-] ([xshift=30pt,yshift=-40.4pt]m-2-2.north) -- ([xshift=-34pt,yshift=-40.4pt]m-2-3.north);
\draw[-] ([xshift=30pt,yshift=-42.6pt]m-2-2.north) -- ([xshift=-34pt,yshift=-42.6pt]m-2-3.north);

\draw[-] ([xshift=30pt,yshift=-10.4pt]m-2-3.north) -- ([xshift=-34pt,yshift=-10.4pt]m-2-4.north);
\draw[-] ([xshift=30pt,yshift=-12.6pt]m-2-3.north) -- ([xshift=-34pt,yshift=-12.6pt]m-2-4.north);

\draw[<-{Hooks[left,length=1mm]}] ([xshift=26pt,yshift=-41.5pt]m-2-3.north) -- ([xshift=-26pt,yshift=-41.5pt]m-2-4.north);

\end{tikzpicture}
\caption{Dualities used in excluding regular-fiber type III limits in Calabi--Yau three-fibered Calabi--Yau fourfolds. After taking the adiabatic limit of large base $\mathbb{P}^1$, the theory is best described as F-theory on the generic fiber Calabi--Yau threefold. Our interest lies in the (geometric) hypermultiplet moduli space of this 6d theory, which is the same as the (geometric) hypermultiplet moduli space as probed by Type IIA on the generic fiber Calabi--Yau threefold. Finally, the $c$-map relates this to the vector multiplet moduli space of Type IIB compactified on the same threefold.}
\label{fig:typeIII-dualities}
\end{figure}
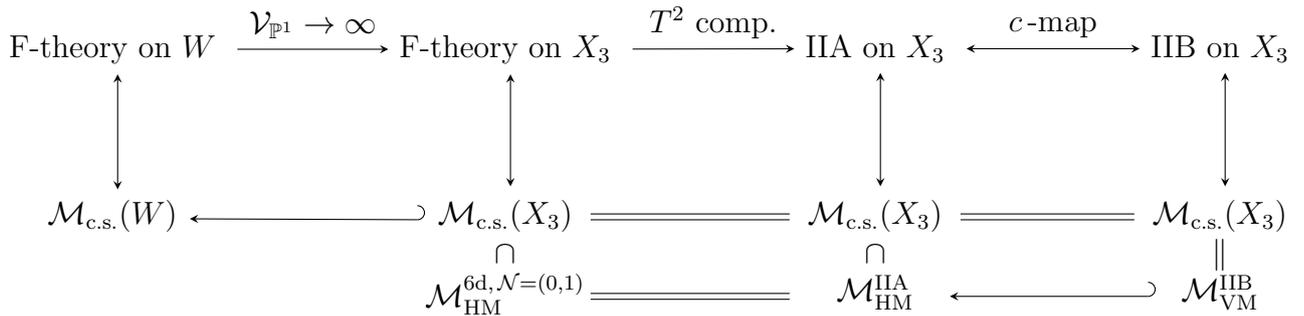

The main result of this section is that this limit by itself is K\"ahler-obstructed in the spirit of Claim \ref{claim:explicitobstructions}. In particular, we will show that in this limit there are quantum corrections of the form \eqref{eq:quantumvolume} to the classical volume of the divisor ${\cal B}_2$ in Figure~\ref{fig:regular-III}.
 To achieve this, we will make use of a chain of dualities that eventually relates the current setup to Type II Calabi--Yau compactifications, see Figure~\ref{fig:typeIII-dualities}. Similar chains of dualities have been used, e.g., in~\cite{Jockers:2009ti,Garcia-Etxebarria:2012bio} to determine quantum effects in F-theory compactifications.

As a first step, unlike in previous cases, we superimpose the considered type III degeneration of $W$ with a limit in the K\"ahler moduli space of $W$ corresponding to the limit of large volume, $\cV_{\mathbb{P}^1}\to \infty$, for the $\mathbb{P}^1$ that is the base of the $\kappa$-fibration while keeping the volume of the generic fiber $X_3$ fixed. This is the limit for $\kappa$ in which the variation of the fiber $X_3$ of $\kappa$ is adiabatic. In the limit $\cV_{\mathbb{P}^1}\to \infty$, the theory is best described by a six-dimensional $\cN=(1,0)$ theory obtained as an F-theory compactification on $X_3$. This limit is depicted in the first two columns of Figure~\ref{fig:typeIII-dualities}. We now argue that already in this 6d theory, the type III limit for $X_3$ is obstructed if the volume of the base $\cB_2$ of the fibration $\pi_X$ is kept constant. 

To show this, we recall that the complex structure moduli of $X_3$ form part of the hypermultiplet sector of the 6d $\cN=(1,0)$ effective theory obtained from F-theory on $X_3$. Its hypermultiplet moduli space  has quaternionic dimension $h^{2,1}(X_3)+1$. The saxionic parts of the scalars in the $h^{2,1}(X_3)$ hypermultiplets can be identified with the complex structure moduli $z^i$, $i=1,\dots, h^{2,1}(X_3)$ of $X_3$.
In addition, there is the universal hypermultiplet, whose saxionic component is the volume $\cV_{\cB_2}$ of $\cB_2$.
In the following, we ignore all the axions and focus on the saxionic components of the hypermultiplets as these are the most relevant ones for the study of classical infinite distance limits. 

Recall that the limit we are interested in corresponds to a limit in the complex structure moduli space of $X_3$ at fixed volume of $\cB_2$. For simplicity, we take a one-parameter limit obtained as 
\begin{equation}
    z \to \ii \infty\,.
\end{equation}
We thus take a one-dimensional limit in the geometric part of the hypermultiplet moduli space while keeping the universal hypermultiplet constant. This type of limits has been investigated in different contexts in~\cite{Marchesano:2019ifh,Baume:2019sry,Alvarez-Garcia:2021pxo}. More specifically, the analysis of~\cite{Marchesano:2019ifh,Baume:2019sry} studies classical infinite distance limits in the K\"ahler moduli space of Type IIB Calabi--Yau threefold compactifications  at constant 4d dilaton, which are related to the ones considered here via mirror symmetry. 
For our purposes, it is crucial that the hypermultiplet moduli space of F-theory on $X_3$ is identical to the hypermultiplet moduli space of F-theory on $X_3\times  T^2$. The reason is that the geometric moduli of the $T^2$ factor form part of the vector multiplet sector of the resulting 4d $\cN=2$ effective action and therefore decouple from the hypermultiplet sector. Thus, the hypermultiplet sector is insensitive to the additional decompactification, see again Figure~\ref{fig:typeIII-dualities}. For this reason, we can first discuss the obstruction to type III limits of $X_3$ in the Type IIA compactification, which we then translate into an obstruction in the 6d theory. 

To summarise the discussion so far: Obstructions to a type III degeneration of the generic fiber $X_3$ of the F-theory fourfold $W$ (at fixed $\cV_{\cB_2}$ and large $\cV_{\mathbb{P}^1}$) can be inferred from obstructions to limits in the hypermultiplet moduli space of {\it Type IIA} compactified on $X_3$ (at fixed universal hypermultiplet). 

The relevant corrections have been computed using twistor techniques in~\cite{Alexandrov:2008gh}. We will not review the details of this computation here, but refer to the original references and the review~\cite{Alexandrov:2011va} for details. In the Type IIA formulation, the relevant corrections in the type III limit arise from D2-brane instantons wrapping certain three-cycles. 
The instanton corrections to the hypermultiplet moduli space are encoded in the so-called contact potential $e^\Phi$, which can be viewed as the quantum corrected version of the 4d Type IIA dilaton. If we denote by $e^{\Phi_0}$ the classical 4d dilaton, then the corrected 4d dilaton is  given by~\cite{Alexandrov:2008gh}
\begin{equation}\label{eq:contactpotential}
    e^{\Phi} = e^{\Phi_0} + \frac{\chi(X_3)}{192\pi} + \frac{1}{8\pi^2} \sum_{\gamma\in H_3(X_3)} n_\gamma \sum_{m>0} \frac{|W_\gamma|}{m} K_1(2\pi m|W_\gamma|)\,.
\end{equation}
Here, $\chi(X_3)$ is the Euler characteristic of $X_3$ and the sum runs over integer three-cycles $\gamma$ of $X_3$, for which $n_\gamma$ is associated the BPS invariant. Moreover, $W_\gamma$ is the action of a D2-brane instanton on $\gamma$ and $K_1$ denotes the Bessel function. For simplicity, we set all RR-axions to zero in the above expression. 

For instantons to become relevant for the physics of the type III limit, their classical action has to go to zero in the limit, $|W_\gamma|\to 0$. To understand the scaling of $W_\gamma$, we use  the $c$-map to map Type IIA compactified on $X_3$ to Type IIB compactified on $X_3$; this is depicted in the two rightmost columns of the diagram in Figure~\ref{fig:typeIII-dualities}. Since the D$2$-brane action $W_\gamma$ is identified with the mass of a D$3$-brane wrapping $\gamma$ under the $c$-map, we can infer the scaling of the former from the scaling of the latter.\footnote{Indeed, Type IIA on $X_3\times S^1$ is T-dual to Type IIB on $X_3\times S^1$. Under T-duality, BPS particles obtained from Type IIB D3-branes wrapping 3-cycles $\gamma\in H_3(X_3)$ map to D2-brane instantons on the same 3-cycles in the Type IIA frame.} At constant $e^{\Phi_0}$, the scaling of the D2-brane instanton action in complex structure limits is then given by the behaviour of the central charge of the Type IIB BPS-particles in asymptotic limits of the vector multiplet moduli space. Let us denote by $\cI_{\rm III}$ the subset of three-cycles $\gamma$ for which the instanton action vanishes at the fastest rate in the type III limit.\footnote{As discussed in~\cite{Baume:2019sry}, the relevant instantons for possible obstructions are always those whose action vanishes at the fastest parametric rate. On the $c$-dual Type IIB side, these instantons correspond to those BPS particles obtained from wrapped D$3$-branes which become massless at the fastest parameteric scale. As shown in~\cite{Monnee:2025ynn}, these particles are purely electric, which in turn means that the leading instantons on the Type IIA side are mutually local.\label{fn:mutlocal}} Crucially, these instantons are always mutually local such that their contribution to the corrected contact potential is indeed as given in~\eqref{eq:contactpotential}. Keeping only the contribution of these instantons, we approximate 
\begin{equation}
    e^\Phi\simeq e^{\Phi_{\rm III}} = e^{\Phi_0} +\frac{\chi(X_3)}{192\pi} + \frac{1}{8\pi^2} \sum_{\gamma\in \cI_{\rm III}} n_\gamma \sum_{m>0}\frac{|W_\gamma|}{m} K_1(2\pi m|W_\gamma|)\,.  
\end{equation}
Through the connection of D2-brane instantons in Type IIA on $X_3$ to D3-brane BPS particles of Type IIB on $X_3$, we can use the results of~\cite{Monnee:2025ynn} to determine the elements of $\cI_{\rm III}$ and the associated BPS indices. 
 The key point is
  that general type III limits for Calabi--Yau compactifications of Type IIB string theory are decompactification limits to a 6d $\cN=(1,0)$ theory of gravity. The tower of massless states in this limit then corresponds to the KK modes of the 6d theory. For Type IIB compactified on $X_3$, the Hodge numbers $h^{2,1}(X_3)$ and $h^{1,1}(X_3)$ count the number of massless multiplets that are uncharged under a possible gauge group $G$ of the 6d theory. More precisely, we have 
\begin{equation}
    h^{2,1}(X_3)=n_T + {\rm rk}\,G +2  \,,\qquad h^{1,1}(X_3) = n_H^{(0)}-1\,,
\end{equation}
where $n_T$ is the number of tensors in the 6d theory and $n_H^{(0)}$ the number of neutral hypers. The extra 2 in $h^{2,1}$ accounts for the two KK photons arising in the compactification to 4d. We can then split the lattice $\cI_{\rm III}$ into charge sectors 
\begin{equation}
    \cI_{\rm III} = \bigoplus_{\mathbf{q}} \cI_{\rm III}^{(\mathbf{q})}\,.
\end{equation}
Here, $\mathbf{q}$ is the charge of a state under the Cartan subgroup of $G$. The central charge of the D3-brane BPS particles becoming massless in type III limits of Type IIB on $X_3$ is characterised by the charge $\mathbf{q}$ in addition to the two KK momenta $(k,l)$ along the two directions that decompactify in this limit.\footnote{4d states that are charged in addition under those ${\rm U}(1)$'s that descend from tensor multiplets in the 6d theory do not become massless.} In the $z \to \ii \infty$ limit, the mass of these states scales as
\begin{equation}
    m_{k,l,\mathbf{q}} = \frac{1}{\Im z}\left|k+\tau  l + \mathbf{q} \cdot \vec{\zeta} \,\right|\,.
\end{equation}
Here, $\tau$ is the complex structure of the torus $\cE$ that is the triple curve arising in the type III degeneration of $X_3$ and the complex vector $\vec{\zeta}\in \mathbb{C}^{{\rm rk}\,G}$ encodes the Coulomb branch parameters of the 6d gauge group after compactification to 4d. For the limit to be a type III limit, the Coulomb branch parameters have to satisfy $|\zeta_i|\lesssim 1$ for $i=1,\dots,{\rm rk}\,G$.  Since $(k,l)$ are just the KK-momenta, the BPS invariant for a state with charge $(k,l,\mathbf{q})$ is independent of $(k,l)$ such that we can simply denote it by $n^{(0)}_{\mathbf{q}}$. In particular, for the $\mathbf{q}=0$ sector we have~\cite{Monnee:2025ynn}
\begin{equation}
    n^{(0)}_0 = -2(n_T + {\rm rk}\,G +3 - n_H^{(0)}) = \chi(X_3)\,.
\end{equation}
The ``$+3$'' accounts for the graviphoton and the two KK photons obtained after compactification to 4d. With this input, we can now rewrite the expression for $e^{\Phi_{\rm III}}$ as\footnote{Notice that the $\tau$ used in~\cite{Alexandrov:2008gh} and the $\tau$ used here are unrelated. More concretely, $\tau_{loc.\,cit.}=(\Im z)^{-1}$. Moreover, notice that the integer $k$ can be viewed as the analogue of the D$(-1)$ charge in a mirror description in case the type III limit allows for such a mirror description.} 
\begin{equation*}
    e^{\Phi_{\rm III}} = e^{\Phi_0} +\frac{\chi(X_3)}{192\pi} + \frac{1}{8\pi^2} \sum_{\mathbf{q}} n^{(0)}_{\mathbf{q}} \sum_{k,l} \sum_{m>0}\frac{|k+\tau l+\mathbf{q}\cdot \vec\zeta |}{\text{Im}\,z} \frac{K_1\left(2\pi m\, (\text{Im}\,z)^{-1}|k+\tau l+\mathbf{q}\cdot \vec\zeta |\right)}{m} \,. 
\end{equation*}
As in the Type IIB case discussed in~\cite{Alexandrov:2008gh}, the above expression is the Poisson resummation over $n$ of
\begin{equation}\begin{aligned}
   & e^{\Phi_{\rm III}} = e^{\Phi_0} +\frac{\chi(X_3)}{192\pi} \\&+ \frac{(\Im\,z)^{-2}}{32\pi^2} \sum_{\mathbf{q}}n^{(0)}_{\mathbf{q}}\sum_{l=0}^\infty\sum_{(m,n)\neq (0,0)} \frac{1+2\pi |m z^{-1} +n| (l\tau_2 +\mathbf{q}\cdot \vec\zeta_2)}{|m z^{-1} +n|^3} e^{-2\pi |m| (\Im\,z)^{-1}(l\tau_2 +\mathbf{q}\cdot \vec\zeta_2) }\,,
\end{aligned}\end{equation}
where $\tau_2=\Im \tau$ and $\vec\zeta_2=\Im\vec\zeta$. Let us again set all RR-axions to zero. In the limit $z\to \ii \infty$, only terms with $n=0$ contribute significantly such that we can approximate 
\begin{equation}\begin{aligned}
    e^{\Phi_{\rm III}}-e^{\Phi_0} \simeq\; \frac{(\Im\,z)}{16\pi^2} \sum_{\mathbf{q}} n^{(0)}_{\mathbf{q}} \sum_{l=0}^\infty \sum_{m>0} \frac{1+2\pi m| z^{-1}| (l\tau_2 +\mathbf{q}\cdot \vec\zeta_2)}{m^3} e^{-2\pi m (\Im\,z)^{-1} (l\tau_2 +\mathbf{q}\cdot \vec\zeta_2)}\,, 
\end{aligned}\end{equation}
where we also dropped the constant term proportional to $\chi(X_3)$.
We can then split the above sum into contributions from $l=0$ and from $l\neq 0$ and find, to leading order in the limit $z\to \ii \infty$, 
\begin{equation}\begin{aligned}
    e^{\Phi_{\rm III}}- e^{\Phi_0}\simeq\;& 
    \frac{(\Im\,z)}{16 \pi^2}  \left(  \zeta(3) \chi(X_3) + \sum_{\mathbf{q}\neq \vec{0}} n^{(0)}_{\mathbf{q}} \, {\rm Li}_3\left(e^{-2\pi (\Im\,z)^{-1}\mathbf{q} \cdot \vec{\zeta}_2}\right) \right)\\& + \frac{(\Im\,z)}{16\pi^2}\sum_{\mathbf{q}} n^{(0)}_{\mathbf{q}}  \sum_{l,m>0}\frac{1+2\pi m |z^{-1}|(l\tau_2 +\mathbf{q}\cdot \vec\zeta_2)}{m^3} e^{-2\pi m (\Im\,z)^{-1} (l\tau_2 +\mathbf{q}\cdot \vec\zeta_2)}  \,. 
\end{aligned}\end{equation}
The last sum is divergent in the limit $\Im z \to \infty$, and we can estimate the rate of divergence using 
\begin{equation}
    \sum_{l,m>0} \frac{e^{-xm (l\tau_2+q) }}{m^3} = \frac{\pi^4}{90 x \tau_2} + \cO(x^0)\,. 
\end{equation}
To leading order we then find
\begin{equation}
      e^{\Phi_{\rm III}} -e^{\Phi_0}=\frac{\pi}{1440} \frac{(\Im\,z)^2}{\tau_2}\left( \chi(X_3) +\sum_{\mathbf{q}\neq \vec{0}} n^{(0)}_{\mathbf{q}}\right)+\cO(\Im\,z )\,. 
\end{equation}
We can further evaluate the last bracket using that the massless $W$-bosons contribute $-2$ to the BPS index and massless charged hypermultiplets $+2$.\footnote{Supersymmetry demands that the gauge representation of a charged hypermultiplet is quaternionic. This means that for each representation $q$ of $G$, also its complex conjugate $\Bar{q}\equiv-q$ is part of the hypermultiplet.} We can then evaluate
\begin{equation}\begin{aligned}
    \chi(X_3) + \sum_{\mathbf{q}\neq \vec0} n_{\mathbf{q}} &= -2(n_T + {\rm rk}\,G +3 - n_H^{(0)}) -2 (\text{dim}\,G-{\rm rk}\,G) + 2 n_H^{{\rm charged}} \\&= -2(3+ n_T +n_V -n_H)\,.
\end{aligned}\end{equation}
Here, $n_V$ and $n_H$ respectively count the total number of vector and hypermultiplets in the 6d $\cN=(1,0)$ theory that arises in the type III limit of \emph{Type IIB} compactified on $X_3$. For this 6d $\cN=(1,0)$ theory, gravitational anomaly cancellation imposes 
\begin{equation}
    273-29n_T = n_H-n_V\,.
\end{equation}
For \emph{Type IIA} compactified on $X_3$, this means that the corrected contact potential can be written as
\begin{equation}
      e^{\Phi_{\rm III}}-e^{\Phi_0} =\frac{\pi}{1440} \frac{(\Im\,z)^2}{\tau_2}\left(540-60n_T \right)+\cO(\Im\,z)\,. 
\end{equation}
We thus see that for $n_T\neq 9$, the correction to the contact potential in the type III limit diverges as $(\Im\,z)^2$; the sign is determined by the number of tensor multiplets arising in the 6d theory to which \emph{Type IIB} on $X_3$ decompactifies in this limit. 

Using the duality between Type IIA on $X_3$ and F-theory on $X_3\times T^2$, the above correction to the Type IIA contact potential can be translated into a correction to the 6d dilaton $e^{\phi_6}$. In 6d F-theory compactifications, the 6d dilaton can be identified with the volume of the base $\cB_2$ in 10d Einstein-frame units and the corrections to $e^{\phi_6}$ can be interpreted as quantum corrections to the classical volume 
\begin{equation}\label{VB_2corr}
\cV_{\cB_2} \simeq \cV_{\cB_2}^{(0)} + \frac{\pi}{1440}\frac{(\Im z)^2}{\tau_2}\left(540-60n_T \right) +\cO(\Im\,z) \,.
\end{equation}
For the $X_3$-fibered Calabi--Yau fourfold $W$, the 6d dilaton becomes a K\"ahler modulus of $W$. The correction~\eqref{VB_2corr} then translates to a correction to the volume of the divisor $\cB_2$ of $\cB_3$ which is of the form~\eqref{eq:quantumvolume}, implying a K\"ahler obstruction for the limit $z\to \ii \infty$. Indeed, regardless of the sign, in the generic case the coefficient of the $(\text{Im}\,z)^2$ correction is non-zero, such that perturbative control is lost in the $\Im\,z\to \infty$ limit. This indicates that we leave the regime in the quantum moduli space of F-theory on $W$ in which we can trust the supergravity approximation. The nature of the breakdown of the supergravity approximation is different depending on the value of $n_T$:
\begin{itemize}
    \item If $n_T <  9$, the correction to $\cV_{\cB_2}$ is positive. This means that while perturbative control over the supergravity approximation is lost, the quantum corrections simply induce an additional infinite distance limit (the large volume limit for $\cB_2$) in the K\"ahler moduli space of $W$. 
    
    \item If $n_T>9$, the correction to $\cV_{\cB_2}$ is negative. This means that we lose the supergravity approximation and enter the small volume regime for $\cB_2$. Again, perturbative control over the supergravity is lost. From our analysis it is not clear whether this regime still has a non-compact infinite distance direction at the quantum level. 

    \item If $n_T = 9$, the leading correction cancels. This does not exclude subleading corrections at $\cO(\Im\,z)$ which still diverge in the limit. Depending on their sign, the same interpretations as before hold.
\end{itemize}

We thus conclude that in the simple class of Calabi--Yau fourfolds that are CY3-fibered, a regular-fiber type III limit cannot be taken while maintaining control over the classical supergravity approximation of the 4d $\cN=1$ theory. Notice that our analysis was performed in the adiabatic limit for the $\kappa$-fibration which by itself is a limit in the K\"ahler moduli space. Our point here is that the type III limit induces an \emph{additional} limit in the K\"ahler moduli space of $W$ on top of the $\kappa$-adiabatic limit.

\paragraph{Comment on the origin of the corrections.} One might wonder about the physical origin of the corrections to the volume $\cV_{\cB_2}$ from a Type IIB orientifold perspective. In this picture, one can split the $g_s$ corrections into contributions that are inherited from the $\cN=2$ parent theory, corresponding to Type IIB compactified on a Calabi--Yau threefold, and genuine $\cN=1$ corrections that arise due to the orientifolding procedure. The $\cN=2$ corrections include in particular D1-D$(-1)$-brane instanton corrections to the hypermultiplet moduli space as computed in~\cite{Alexandrov:2008gh}. However, unlike the corrections appearing in~\eqref{VB_2corr}, these only depend on the string coupling and the curve volumes of the Calabi--Yau threefold and are independent of the complex structure parameters. From the Type IIB orientifold perspective, the corrections appearing in~\eqref{VB_2corr} should hence be interpreted as genuine $\cN=1$ effects induced by the orientifold action. On the other hand, we used the same corrections computed in~\cite{Alexandrov:2008gh} to determine the corrections in~\eqref{VB_2corr} which we claim to be inherited from an $\cN=2$ parent theory. From the Type IIB orientifold perspective this looks like a contradiction, in particular because the duality chain in Figure~\ref{fig:typeIII-dualities} maps the D2-brane instantons of Type IIA string theory that enter in~\eqref{eq:contactpotential} to $[p,q]$-string instantons wrapping certain curves inside $\cB_2$ in F-theory. However, as we will clarify in the following, this is not a contradiction, but rather a consequence of the fact that a 4d $\cN=1$ theory can be obtained from two distinct $\cN=2$ theories, only one of which corresponds to Type IIB compactified on a Calabi--Yau threefold. In other words, this illustrates yet again that by sticking to a Type IIB orientifold perspective of 4d $\cN=1$ theories one misses crucial information about the full 4d $\cN=1$ theory. 

To understand the two $\cN=2$ origins of the 4d $\cN=1$ theories obtained from F-theory/Type IIB Calabi--Yau orientifolds, let us begin with F-theory compactified on $\cE \times K3\times T^2$, or equivalently, Type IIB on K3$\times T^2$. Here, we denote by $\cE$ the F-theory torus such that its complex structure $\tau_\cE$ encodes the Type IIB axio-dilaton. The resulting 4d theory preserves $\cN=4$ supersymmetry, which is realized by four pairs of Weyl spinors $(Q_\alpha^I, \bar{Q}_{\dot \alpha}^I)$, $I=1,2,3,4$. There are now two ways of breaking this $\cN=4$ symmetry to $\cN=2$. 
 These correspond to the two possible ways of turning the direct products in $\cE \times K3\times T^2$ into non-trivial fibrations; each of the two operations removes a different subset of covariantly constant spinors of the original fourfold:
\begin{itemize}
    \item Let us start by replacing the first product by a non-trivial fibration. For the F-theory fourfold and the preserved supercharges this means
    \begin{equation}\begin{aligned}\label{breaking1}
       \cE \times K3\times T^2 \qquad &\longrightarrow \qquad  \underbrace{\cE \to \cB_2}_{=:X_3} \times T^2 \,,\\
       (Q_\alpha^{1,2,3,4}, \bar{Q}_{\dot\alpha}^{1,2,3,4}) \qquad &\longrightarrow \qquad (Q_\alpha^{1,2}, \bar{Q}_{\dot \alpha}^{1,2})\,. 
    \end{aligned}\end{equation}
    Here, $X_3$ is a Calabi--Yau threefold and w.l.o.g. we assumed that the first two pairs of Weyl spinor supercharges survive the $\cN=4\to \cN=2$ breaking. The Type IIB dilaton encoded in $\tau_\cE$ now varies non-trivially over the base $\cB_2$. The details of the non-trivial fibration are encoded in the complex structure of $X_3$. This indicates that the 6d dilaton and the complex structure moduli of $X_3$ interact and are hence in the same sector of the 4d $\cN=2$ moduli space. This is the $\cN=2$ parent theory that we considered to derive~\eqref{VB_2corr}. 
    \item If instead we turn the second product into a non-trivial fibration, the resulting F-theory geometry and preserved supercharges are
    \begin{equation}\begin{aligned}\label{breaking2}
       \cE \times K3\times T^2 \qquad &\longrightarrow \qquad  \cE \times \underbrace{(K3\to \mathbb{P}^1)}_{=:\tilde{X}_3} \,,\\
       (Q_\alpha^{1,2,3,4}, \bar{Q}_{\dot\alpha}^{1,2,3,4}) \qquad &\longrightarrow \qquad\quad (Q_\alpha^{1,3}, \bar{Q}_{\dot \alpha}^{1,3})\,. 
    \end{aligned}\end{equation}
    The resulting F-theory compactification can be interpreted as Type IIB compactified on a Calabi--Yau threefold $\tilde{X}_3$ which in general is different from $X_3$. Notice that the subset of the original $\cN=4$ supercharges that is preserved in this $\cN=2$ theory differs from the supercharges preserved in the breaking in~\eqref{breaking1}. The properties of the non-trivial K3-fibration of $\tilde{X}_3$ are encoded in the complex structure moduli of $\tilde{X}_3$. The complex structure of $\cE$ being constant is a signal that the dilaton and the complex structure of $\tilde{X}_3$ decouple and are hence in different sectors of the $\cN=2$ theory, as is well-known to be the case for Type IIB compactifications on Calabi--Yau threefolds. This is the $\cN=2$ parent theory from the perspective of Type IIB Calabi--Yau orientifolds. 
\end{itemize}
By replacing the remaining product in~\eqref{breaking1} and~\eqref{breaking2} by a non-trivial fibration, we can obtain the same 4d $\cN=1$ theory in which only $(Q_\alpha^1,\bar{Q}_{\dot \alpha}^1)$ are realized as preserved supercharges:
\begin{equation}\begin{aligned}
    \underbrace{\cE \to B_2}_{=:X_3} \times T^2 \quad &\longrightarrow \quad \cE \to B_2 \to \mathbb{P}^1 \quad \longleftarrow \quad \cE \times \underbrace{(K3\to \mathbb{P}^1)}_{=:\tilde{X}_3} \\
    (Q_\alpha^{1,2}, \bar{Q}_{\dot \alpha}^{1,2}) \;\quad &\longrightarrow\qquad  (Q_\a^1,\bar{Q}_{\dot \a}^1)\qquad \,\longleftarrow \qquad  (Q_\alpha^{1,3}, \bar{Q}_{\dot \alpha}^{1,3})\,. 
\end{aligned}\end{equation}
Following the arrow from right to left corresponds to orientifolding the Type IIB compactification on $\tilde{X}_3$. Viewing the $\cN=1$ theory as such a Calabi--Yau orientifold, it is clear why the corrections to $\cV_{B_2}$ in~\eqref{VB_2corr} are a genuine $\cN=1$ effect: the $\cN=2$ algebra preserved in the breaking~\eqref{breaking1} only shares an $\cN=1$ subalgebra with the $\cN=2$ algebra preserved in the breaking~\eqref{breaking2}. 

\paragraph{Generalisation to arbitrary limits of the $X_3$-fiber.} The logic followed in this section can be generalised to all complex structure limits in which the Calabi--Yau threefold fiber $X_3$ appearing in \eqref{eq:adiabatic} undergoes an infinite distance degeneration. This analysis therefore directly points to the origin of the quantum K\"ahler obstructions also for type II and III degenerations of the fiber $X_3$, both of regular-fiber and I$_n$-type, as well as type IV degenerations of $X_3$ which are necessarily of I$_n$-type. In the Type IIB duality frame  appearing on the right in Figure~\ref{fig:typeIII-dualities}, the leading quantum corrections come, in all these cases, from the three-cycles on $X_3$ whose volume vanishes at the fastest rate, as analysed in~\cite{Hassfeld:2025uoy,Monnee:2025ynn}.

For example, for Type IIB string theory compactified on $X_3$, a type II limit of $X_3$ corresponds to an emergent string limit \cite{Hassfeld:2025uoy}. If the double surface of the degenerating $X_3$ is a K3, the asymptotic duality frame is that of the heterotic string on $K3_{\rm het} \times T^2_{\rm het}$. 
 In this duality frame, the BPS particles that induce the corrections are the winding states along $T^2_{\rm het}$, whose multiplicity is encoded in certain meromorphic Jacobi forms.
 Even without working out the resulting corrections explicitly, we can conclude that, in the $c$-dual hypermultiplet moduli space of Type~IIA on $X_3$, these will lead to a quantum correction analogous to~\eqref{VB_2corr}. 
 Combined with the discussion in Section \ref{ssec:hetstablelimit}, we have therefore found a direct argument for the K\"ahler obstruction of regular-fiber type II degenerations whenever the F-theory base ${\cal B}_3$ is  rationally fibered over base ${\cal B}_2$ as in~\eqref{eq:rhofibration}, or surface fibered over $\mathbb P^1$ with fiber ${\cal B}_2$, as in the left picture in Figure~\ref{fig:regular-III}; in the latter case the analysis also applies to I$_n$-type limits.\footnote{For type II degenerations of the standard Sen-type, already the discussion in Section \ref{ssec:globalSen} identifies quantum corrections directly in the effective field theory.}

\subsection{Type IV limits}\label{ssec:typeIV}
We are left with type IV limits of the regular-fiber kind. In these limits, the Calabi--Yau fourfold undergoes a degeneration such that the components of $W_0$ still have vanishing quintuple intersection, but some quadruple intersections $W_{ijkl}$ are non-zero and copies of the smooth elliptic fiber $\cE$ of $W$. For these limits, our arguments detecting K\"ahler obstructions invoked in the previous cases do not apply. Based on the classical effective action in a type IV limit, the candidate EFT string realizing such a limit should correspond to a supergravity string in a five-dimensional theory since, classically, the type IV limit is a decompactification limit to five dimensions, see~\cite{Monnee:2025ynn}. However, unlike for the other limits analysed in this work, there is no obvious reason why the worldsheet theory on the candidate EFT string realizing a regular-fiber type IV singularity of an elliptically fibered fourfold is incompatible with the string being a 5d supergravity string. In particular, similar arguments to those presented in Sections~\ref{ssec:localsen} and~\ref{ssec:typeIII} cannot be applied in this case for the following reasons:
\begin{itemize}
    \item Unlike the strings realizing I$_n$-type limits discussed in Section~\ref{ssec:localsen}, the candidate EFT string realizing the classical regular-fiber type IV limit does not have any (non-)Abelian vector multiplets on its worldsheet. A similar reasoning as in Section~\ref{ssec:localsen} can therefore not be applied to argue for an obstruction to this limit. 
    \item For 5d supergravity strings, the central charges are less constrained than their 6d counterparts. In particular, the analogue of~\eqref{eq:cLminusCR} in 5d only tells us that the left- and right-moving central charge differ by an integer, which is satisfied for the EFT strings in type IV limits. The central charges of type IV candidate EFT strings hence do not hint towards an apparent inconsistency. 
\end{itemize}
Notice that this does not mean that these strings \emph{have} to be 5d supergravity strings and, accordingly, the regular-fiber type IV limits are not K\"ahler-obstructed. It just means that Claim~\ref{claim:EFTstringobstruction} cannot be straight-forwardly applied to show a K\"ahler obstruction.\footnote{In Section~\ref{ssec:mirror}, we comment on the possibility of applying Claim~\ref{claim:EFTstringobstruction} to EFT strings realizing the mirror duals of regular-fiber type IV limits in orientifolds of Type IIA.}

Instead of considering the worldsheet theory on the candidate EFT strings, we could also try to use Claim~\ref{claim:explicitobstructions} to show that regular-fiber type IV limits are K\"ahler-obstructed by computing corrections to the F-theory K\"ahler moduli as we did in the case of type II and III regular-fiber limits. In the examples of type II and III limits, for which we had access to these corrections, we used a specific fibration structure of $\cB_3$ to obtain a dual description in which the relevant corrections are computable. In particular, the duality singled out certain K\"ahler moduli for which the dependence of the corrections on the complex structure modulus that is sent to infinity can be computed in the dual frame: 
\begin{itemize}
    \item For type II limits that correspond to the standard semi-stable degeneration limit underlying F-theory/heterotic duality, we used this duality to infer the corrections to the volume of the base of the rationally fibered F-theory base. Geometrically, the divisor for which the duality determines the corrections to the volume modulus is transverse to the degeneration associated with the type II limit. 
    \item For type III limits in CY3-fibered Calabi--Yau fourfolds for which the generic fiber undergoes a type III degeneration, we used the duality to a 6d theory to infer corrections to the volume modulus of the base of the CY3-fiber. In this type III limit, the base of the threefold itself degenerates such that the degeneration occurs parallel to the divisor for which the duality determines the corrections to the volume modulus.  
\end{itemize}
For regular-fiber type IV limits, the situation is different since the entire base $\cB_3$ of $W$ degenerates. This does not single out a particular divisor because each divisor has directions parallel and transverse to the degeneration. For this reason, there is no obvious duality that can be used to compute the dependence of the corrections to certain divisor volumes on the complex structure modulus that induces the type IV limit. Instead, one would have to compute corrections to divisor volumes directly in F-theory, which is beyond the scope of this paper. 

For the reasons outlined above, the techniques used in this paper do not lead to a clear verdict on the fate of regular-fiber type IV limits in the quantum moduli space of F-theory compactifications. Let us stress that these are the only type of limits where our arguments do not establish a K\"ahler obstruction for at least a subclass of limits. However, notice that by continuity regular-fiber type IV limits that are obtained as enhancements of the kind of type II and III limits considered in this paper also have to be K\"ahler-obstructed. We hope to investigate quantum corrections arising in general regular-fiber type IV limits in future work.

\section{Comments on obstructions in Type IIA orientifolds}\label{ssec:mirror}
While our focus,
both in this work and in~\cite{Paper1}, has been on corrections to the effective action of Type IIB orientifolds/F-theory, we now comment on the mirror dual picture.
 By mirror symmetry, a perturbative Type IIB orientifold with O7/O3-planes on $V$ is dual to an orientifold of Type IIA compactified on the mirror Calabi-Yau threefold $\hat{V}$ with O6-planes. 

From the dual Type IIA perspective, it becomes even more evident why 4d ${\cal N}=1$ infinite distance limits descending from the vector multiplet moduli space of the 4d $\cN=2$ parent theory are obstructed by unsuppressed quantum corrections. To see this, we recall that on the Type IIA side, the vector multiplet moduli space is spanned by the K\"ahler moduli of $\hat{V}$. Importantly, the four-dimensional dilaton,
\begin{equation}
    e^{-2\phi_{4,\rm IIA}} = \frac{ \cV_{\hat{V}}}{g_s^2}\,,
\end{equation}
is part of a hypermultiplet. Here, $\cV_{\hat{V}}$ is the volume of $\hat{V}$ measured in Type IIA string units, which is a function of the K\"ahler moduli 
\begin{equation}
    t^i = \int_{C^i} (B_2 + \ii  J_{\hat{V}}) \,.
\end{equation}
In this expression, $C^i$ are generators of the Mori cone of $\hat{V}$, $B_2$ is the Type IIA NS-NS 2-form and $J_{\hat{V}}$ the K\"ahler form on $\hat{V}$. Infinite distance limits in the vector multiplet sector, at fixed hypermultiplets, are then obtained by sending some of the K\"ahler moduli to infinity, $t^i\to \ii \infty$, while co-scaling the 10d dilaton $g_s\to \infty$ to keep the four-dimensional dilaton constant. In the 4d $\cN=2$ theory obtained from Type IIA on $\hat{V}$, this strong-coupling regime can be taken reliably due to the factorisation of the moduli space between the hyper- and vector multiplet sector; in particular there are no $g_s$ corrections to the two-derivative effective action for the vector multiplet sector. The large $g_s$ limit then simply corresponds to the M-theory limit of Type IIA string theory, in which we obtain M-theory compactified on $\hat{V}$. 

Consider now an orientifold of this theory given by 
\begin{equation}
    \Omega_{\rm IIA} = (-1)^{F_L} \Omega_p \tilde{\sigma}\,,
\end{equation}
where $\tilde{\sigma}$ is an anti-holomorphic isometric involution of $\hat{V}$. The 4d $\cN=1$ tree-level effective action resulting from this orientifold projection has been derived in~\cite{Grimm:2004ua}. Since the K\"ahler form $J_{\hat{V}}$ is odd under $\tilde{\sigma}$ and $B_2$ is odd under $(-1)^{F_L}\Omega_p$, the K\"ahler moduli that survive the orientifold action are counted by $h^{1,1}_-(\hat{V})$. In addition, there are $h^{2,1}(\hat{V})+1$ chiral multiplets from expanding the combination 
\begin{equation}
    \Omega_c = C_3 + 2\ii\, \text{Re}(C\hat \Omega_3)\,,
\end{equation}
where $C_3$ is the Type IIA 3-form, $\hat \Omega_3$ the holomorphic $(3,0)$-form on $\hat{V}$ and the normalisation $C$ is given by 
\begin{equation}
    C = e^{-\phi_{4,\rm IIA}} \left(\int_{\hat{V}}\hat{\Omega}_3\wedge\hat{\bar{\Omega}}_3\right)^{-1}\,. 
\end{equation}
The holomorphic coordinates associated with complex structure variations of $\hat{V}$ are then obtained from the expansion
\begin{equation}
    \Omega_c = N^K_{(0)} \alpha_K \,,
\end{equation}
where $\alpha_K$ is a basis of $H^3_+(\hat{V}) $. The moduli $N^K_{(0)}$ control the classical holomorphic gauge kinetic functions of the 6-branes in the Type IIA orientifold and also the classical actions of BPS D2-brane instantons wrapping special Lagrangian 3-cycles. For this reason, the moduli $N^K_{(0)}$ are the analogue of the complexified classical divisor volumes $T_a$ in the Type IIB/F-theory dual. The analogue of the Type IIB/F-theory complex structure deformations are instead the K\"ahler moduli $t^i$, $i=1,\dots , h^{1,1}_-(\hat{V})$. The pure complex structure infinite distance limit in the Type IIB orientifold/ F-theory compactifications discussed in this paper is mirror dual to $t^i\to \ii \infty$ while keeping all $N^K_{(0)}$ constant.\footnote{Strictly speaking, only those limits in the complex structure moduli space on the Type IIB side that are connected to the Large Complex Structure point are mirror dual to large volume limits in the Type IIA moduli space. We restrict to these for the rest of the discussion of corrections in Type IIA orientifolds.} In particular, this requires $e^{\phi_{4,\rm IIA}}$ to remain constant in the large volume limit.
As in the Type IIA compactification on $\hat{V}$, this can only be achieved if the 10d string coupling is co-scaled to infinity, $g_s\to \infty$. 

However, after orientifolding, there is no non-renormalisation theorem ensuring the absence of quantum corrections in the $g_s\to\infty$, i.e., M-theory, limit. In fact, if $g_s$ corrections to the Type IIA effective action are present, these become unsuppressed in this limit. The results of this paper for the mirror dual Type IIB/F-theory setup imply that such $g_s$ corrections are indeed non-zero also on the Type IIA side. Concretely, our results summarised by Claim~\ref{claim:explicitobstructions} indicate that there are corrections to the complex structure moduli $N^K$ of the form 
\begin{equation}\label{NKcorrections}
    \Im{N^K} = \Im{N^K_{(0)}} + a_{K,i}(\Im t^i)^{\alpha_{K,i}} +\dots\,, \qquad 0\neq a\in \mathbb{R}\,,\;\alpha>0\,,
\end{equation}
which become unsuppressed in the $t^i\to \ii \infty$ limit and dominate over the classical term for
\begin{equation}
    \Im{N^K_{(0)}} \prec (\Im t^i)^{\alpha_{K,i}}\,. 
\end{equation}
For toroidal orientifolds, the corrections to the gauge kinetic function on D6-branes, and hence to $N^K$, have been computed in~\cite{Lust:2003ky,Akerblom:2007np,Blumenhagen:2007ip} which indeed are of the form~\eqref{NKcorrections}.  For the Type IIA orientifold this means that the decompactification limit to M-theory is obstructed. This is similar to the obstruction to taking the Ho\v{r}ava--Witten limit for compactifications of the heterotic string with asymmetric instanton embedding~\cite{Witten:1996mz}, see also~\cite{Cvetic:2024wsj,Cvetic:2025nfx}. 

Also in the Type IIA orientifold setups one could argue for the quantum obstruction to the limits considered here using a strategy similar to the one underlying Claim~\ref{claim:EFTstringobstruction}. In this case, the candidate EFT strings are NS5-branes wrapping movable divisors in $\hat{V}/\Omega_{\rm IIA}$ as has been discussed already in~\cite{Lanza:2021udy} and more recently in~\cite{Grieco:2025bjy}. Before orientifolding, the strings obtained from NS5-branes on movable divisors of $\hat{V}$ descend from M5-branes in M-theory wrapping the same divisor and are hence indeed supergravity strings of a higher-dimensional theory (or critical strings). In particular, the worldsheet theory on the Type IIA EFT strings is the same as the worldsheet theory on the supergravity strings in M-theory. However, the worldsheet theory on the NS5-brane strings can be sensitive to the orientifold projection. For this reason, the worldsheet theory of candidate EFT strings of the 4d $\cN=1$ theory may differ from the worldsheet theory on the M5-brane supergravity strings in M-theory, and our results on the mirror dual Type IIB side suggest so. As a consequence, by Claim~\ref{claim:EFTstringobstruction}, the corresponding limit in the K\"ahler moduli space of the Type IIA orientifold should be obstructed and must, at the very least, be accompanied by a co-scaling of the Type IIA complex structure moduli $N^K_{(0)}$. It would be very interesting to work out the exact action of the Type IIA orientifold on the NS5-branes and in particular the interplay between the worldsheet theory and the O6-planes. In particular, this provides a promising avenue to argue for the obstruction of regular-fiber type IV limits via an application of Claim~\ref{claim:EFTstringobstruction} in the mirror dual theory to which we hope to return in the future.

\section{Discussion and implications for model building} 
\label{sec:pheno}

In this work and in the companion paper~\cite{Paper1}, we have investigated quantum obstructions to infinite distance limits in the classical moduli space of 4d $\cN=1$ compactifications of Type IIB string theory/F-theory. We have focused on infinite distance limits in the complex structure moduli space of Calabi--Yau threefolds in Type IIB orientifolds or of Calabi--Yau fourfolds in F-theory. Our results demonstrate that these limits are generically $g_s$- and/or K\"ahler-obstructed in the sense of Definitions~\ref{def:2fusionheunconnected} and~\ref{def:2fusionheunconnectedribbon}. 

The reason why these limits are obstructed, whereas infinite distance limits in, for instance, the F-theory K\"ahler moduli space are unobstructed, is a consequence of supersymmetry: Even in theories with minimal supersymmetry there exist non-renormalisation theorems ensuring that certain holomorphic couplings are not renormalised beyond a specific order in perturbation theory. The existence or obstruction of infinite distance limits in the moduli space is then tied to the behaviour of these holomorphic couplings. In 4d $\cN=1$ theories an example of such a coupling is the holomorphic gauge kinetic function, which is not corrected beyond one-loop but receives non-perturbative corrections from BPS instantons. 
 For an infinite distance limit in the classical moduli space to be unobstructed at the quantum level the corrections due to BPS instantons have to vanish asymptotically. 

In theories with extended supersymmetry, the factorisation of the moduli space can ensure that the non-perturbative corrections to the gauge kinetic function vanish identically.  However, in 4d $\cN=1$ theories, there is no such factorisation of the moduli space. In our concrete case of F-theory/Type IIB orientifolds, BPS instantons arise from D$(-1)$ and D3-brane instantons wrapping holomorphic divisors in the base $\cB_3$ of the F-theory fourfold. For this reason, the absence of non-perturbative corrections requires weak-coupling, $g_s\to 0$, and large Einstein-frame divisor volumes. Thus, in particular, classical infinite distance limits in the F-theory K\"ahler moduli space are not obstructed as long as all divisor volumes are parametrically large. 

By contrast, our results  imply that infinite distance limits in the classical complex structure moduli space of F-theory/Type IIB orientifolds (and their Type IIA mirror duals) are different because in these limits the BPS instanton actions are not necessarily suppressed due to threshold corrections to the classical instanton actions. Concretely, our results can be summarised as follows: 
\begin{itemize}
    \item[(a)] Consider Type IIB string theory compactified on a Calabi--Yau threefold $V$ and let $\Omega$ be an orientifold projection leading to O3- and O7-planes. For an infinite distance complex structure degeneration $\phi\to\infty$ of $V$, the local geometry of the singular Calabi--Yau was established in~\cite{Hassfeld:2025uoy,Monnee:2025ynn}. Depending on the interplay of $\Omega$ with this local geometry, the orientifold is said to be of {\it O-type A/B} with respect to the degeneration $\phi\to\infty$. Based on the F-theory uplift of these orientifolds, infinite distance O-type A limits (in which by definition the O7-plane lies on one of the double surfaces of the degeneration)  are found to be $g_s$-obstructed in the sense of Definition~\ref{def:2fusionheunconnected} ~\cite{Paper1}.
    
    \item[(b)] The infinite distance limits of the 4d $\cN=1$ moduli space that survive in the $g_s$-corrected moduli space are best described directly in the F-theory uplift of the Type IIB orientifold. While this moduli space is $g_s$-exact, there are further non-perturbative corrections -- both in $\alpha'$ and mixed $\alpha',g_s$ -- obstructing asymptotic regions of the classical F-theory complex structure moduli space in the sense of Definition~\ref{def:2fusionheunconnectedribbon}. These obstructions can either be detected by considering the worldsheet theory on candidate EFT strings realizing the classical infinite distance limits or by studying the corrections to BPS instanton actions. The results of Sections~\ref{sec:Sentype}-\ref{sec:regularfiber} imply that both strategies are consistent with each other such that, in particular, in general infinite distance limits in the complex structure moduli space of F-theory, there are divisor volumes that are corrected as in~\eqref{eq:quantumvolume}.
\end{itemize}
In the remaining part of this section, we discuss some implications of our analysis for string theory model building, see~\cite{Cicoli:2023opf,McAllister:2025qwq} and references therein on the current status. Concretely, in Section~\ref{ssec:phenoOfolds}, we begin with the implications of the analysis in~\cite{Paper1} for model building scenarios based on the large complex structure regime of perturbative Type IIB orientifolds. In Section~\ref{ssec:asymaccexp}, we analyse the implications of K\"ahler obstructions for proposals to realize asymptotic accelerated expansion in string theory. Finally, in Section~\ref{ssec:Kahlerstab}, we turn to the problem of K\"ahler moduli stabilisation in light of the results of this work.

\subsection{Type IIB orientifolds at Large Complex Structure}\label{ssec:phenoOfolds}
Despite much effort, identifying a string vacuum that correctly captures all properties of our universe remains an open problem. At a technical level, the challenge arises because full computational control can only be achieved for highly unrealistic string theory compactifications. String theory based model building hence has to identify a middle ground in which computational control can be achieved while important features of our universe can still be realised. In this context, a well-studied corner of string theory is the large complex structure (LCS) regime in Type IIB orientifolds $V/\Omega$ at weak string coupling $\mathrm{Im}\,\tau\gg1$, see~\cite{Cicoli:2023opf,McAllister:2025qwq} for reviews.

At tree-level in the string coupling, the $\cN=1$ closed string moduli K\"ahler potential takes the form 
\begin{equation}\label{eq:KN=1-cl}
    K_{\rm cl}=-\log\left(i\int_Y\Omega_3\wedge\Bar{\Omega}_3\right)-\log\left(-i(\tau-\Bar{\tau})\right)-2\log\left(\cV_E(V)\right)\,.
\end{equation}
Here $\Omega_3$ is the holomorphic 3-form on $V$, $\tau$ is the axio-dilaton with $\Im\tau=g_s^{-1}\gg1$ and $\cV_E(V)$ is the Einstein-frame volume of $V$. Correspondingly, the closed string moduli space factors at tree-level as
\begin{equation}
    \cM_{\cN=1}^{\rm cl}=\cM_{\rm c.s.}(V/\Omega)\times\cM_\tau\times\cM_{\rm K}(V/\Omega)\,.
\end{equation}
This factorisation is a remnant of the extended $\cN=2$ supersymmetry of the parent Type IIB compactification on $V$. Using this $\cN=2$ parent theory and mirror symmetry at LCS to a large volume compactification of Type IIA on the mirror Calabi--Yau of $V$, the first and third term in~\eqref{eq:KN=1-cl} can be computed exactly, including all perturbative as well as non-perturbative corrections in $\alpha'$ that survive the orientifold projection. As discussed for example in~\cite{McAllister:2024lnt}, the search for (semi-)realistic (A)dS string vacua is commonly performed at the level of the effective action derived from~\eqref{eq:KN=1-cl}.

One of the main outcomes of the companion analysis~\cite{Paper1} is that the effective action derived from the K\"ahler potential~\eqref{eq:KN=1-cl} at string tree-level misses potentially dangerous corrections in $g_s$. This applies most prominently to O-type A limits which are $g_s$-obstructed, see Definition~\ref{def:2fusionheunconnected}. To connect this analysis to Type IIB model building, recall that from the perspective of the underlying Calabi--Yau threefold $V$, the LCS limit is a type IV degeneration and as such of O-type A with respect to any orientifold projection $\Omega$. As the F-theory lift of an O-type A orientifold shows, the classical factorisation of the moduli space is badly broken once $g_s$ corrections are taken into account, meaning that the K\"ahler potential~\eqref{eq:KN=1-cl} does not correctly capture the physics of the LCS limit. As already stressed in Section~4.4 of~\cite{Paper1}, it is important to note that while our analysis is phrased in terms of strictly asymptotic regimes in moduli space, the $g_s$-obstructions to O-type A limits affect also the interior of moduli space. Concretely, the effective action derived from~\eqref{eq:KN=1-cl} can at best be trusted in the {\it parametric} regime
\begin{equation}\label{eq:gs-bound}
    \frac{|\tau|}{\Im(z)}\gg1\quad{\rm as\,\,}1\ll \Im(z)<\infty\,,
\end{equation}
where $z$ is the complex structure modulus of $V$ realizing the LCS limit as $\Im(z)\to\infty$. Notice that, as discussed in~\cite{Paper1}, there are cases for which the strict limit $\Im(z)=\infty$ is never within the regime of validity of the effective action derived from~\eqref{eq:KN=1-cl}.
 These are precisely the O-type A limits, for which even at $g_s = 0$ the orientifold picture breaks down.

The expression on the left hand side of this parametric bound is a control parameter for the perturbative Type IIB approximation to the effective action. The role of this parameter is not directly obvious from the effective 4d $\cN=1$ action since there is no BPS instanton for which the action is given by~\eqref{eq:gs-bound}. Instead, from the perturbative Type IIB perspective, the ratio in~\eqref{eq:gs-bound} is a genuine quantum gravitational control parameter whose role only becomes visible if the details of the underlying Calabi--Yau geometry are considered. 

As such, this parameter is commonly overlooked when performing a control analysis in the LCS regime of Type IIB orientifolds. Notice furthermore that for fixed string coupling $g_s$, this bound works against the common lore that computational control is gained in the LCS regime. Indeed, while the computation of the periods of $\Omega_3$ is simple in the $\Im(z)\gg1$ regime, it is also in this regime that $g_s$ effects become important (and potentially even unsuppressed), which makes it inaccessible to the perturbative Type IIB description. 

To perform a quantum-gravitational control analysis in a given perturbative Type IIB orientifold, the couplings of the 4d $\cN=1$ theory have to be computed directly in the F-theory uplift of the theory and subsequently compared to the perturbative Type IIB prediction. Typically, the dimension of $\cM_{\rm c.s.}$ of a smooth Weierstrass model is very large such that a general computation of the exact periods of the fourfold uplift of the orientifold is challenging. To delineate the exact regime of validity of the string tree-level 4d $\cN=1$ effective action one thus has to go case-by-case, which clearly goes beyond the scope of this paper. In any event, the important implication of the analysis in~\cite{Paper1} for model building is that a purely classical (in $g_s$) analysis of the LCS regime in Type IIB orientifolds at finite string coupling is not justified.

\subsection{Flux potentials in F-theory}\label{ssec:asymaccexp}
The problem of $g_s$ corrections is avoided if we directly work with F-theory compactifications. In this work, we have considered F-theory compactifications on an elliptically fibered Calabi--Yau fourfold $W$ in the absence of fluxes. The possible fluxes are best described via their dual M-theory description in terms of four-form fluxes $G_4$ satisfying~\cite{Witten:1996md}
\begin{equation}
    G_4 + \frac{c_2(W)}{2}\in H^4(W,\mathbb{Z})\,. 
\end{equation}
Of particular interest for four-dimensional model building are fluxes satisfying ${J\wedge G_4=0}$, where $J$ is the K\"ahler form on $W$. These fluxes induce a superpotential~\cite{Gukov:1999ya}
\begin{equation}
    W_F = \int_W G_4 \wedge \Omega_4 \,.
\end{equation}
After uplift to F-theory, the superpotential gives rise to a scalar potential 
\begin{equation}
    V_F = M_{\rm Pl}^2\, e^{K_F} \left(g^{i\bar \jmath} \cD_i W_F \cD_{\bar \jmath}\bar W_F - 3|W_F|^2 \right)\,,
\end{equation}
where $\cD_i$ is the K\"ahler covariant derivative and the index runs over all chiral fields in the theory. The superpotential and the resulting scalar potential are derived within the supergravity approximation to the 4d effective theory. In particular, in the supergravity approximation the K\"ahler potential is given by the classical K\"ahler potential in~\eqref{eq:KFclass}. One then finds
\begin{equation}\label{eq:potentialSUGRA}
    V_F =  \frac{M_{\rm Pl}^2}{\cV_{\cB_3}^2\|\Omega_4\|^2} \left(g^{i\bar \jmath} \cD_i W_F \cD_{\bar \jmath}\bar W_F - 3|W_F|^2 \right)\,,
\end{equation}
where
\begin{equation}
    \|\Omega_4\|^2 \equiv \int_W \Omega_4 \wedge\bar{\Omega}_4\,. 
\end{equation}
As discussed in detail in~\cite{Grimm:2019ixq}, the complex structure dependence of $V_F$ depends on the chosen fluxes. Suppose we consider a one-parameter limit in the classical complex structure moduli space $\cM_{\rm c.s}(W)$ given by $z\to \ii \infty$. Depending on the type of the limit, the complex structure dependence of $V_F$ is bounded as 
\begin{equation}\label{VFscaling}
    V_F \precsim \frac{M_{\rm Pl}^2}{\cV_{\cB_3}^2}\, V_0 \,(\Im\,z)^{d}\,,
\end{equation}
where we used the nilpotent orbit theorem~\cite{schmid} to estimate $W_F\precsim(\ii\Im z)^d$, $V_0$ is the flux-dependent coefficient of the leading contribution to the scalar potential and $d=1,\dots,4$ for a singularity of primary singularity type II, \dots, V. If we were able to take the limit $z\to \ii \infty$ within the validity of the supergravity approximation to the low-energy EFT, the above scaling would be problematic. The reason is that consistency of the effective theory requires that the potential is bounded as~\cite{Hebecker:2018vxz,Scalisi:2018eaz,vandeHeisteeg:2023uxj} 
\begin{equation}\label{eq:VFbound}
    V_F \lesssim \Lambda_{\rm QG}^2\,,
\end{equation}
which is clearly violated by~\eqref{VFscaling} in the limit $\Im z\to \infty$ at constant $\cV_{\cB_3}$ and non-zero $V_0$. Suppose that we could trust the effective action derived from supergravity in this regime, i.e., that there were no corrections to the effective action arising in the large $\Im z$ limit at constant $\cV_{\cB_3}$. The fact that we can turn on fluxes leading to non-zero $V_0$ in~\eqref{VFscaling} then means that supergravity is not self-consistent as~\eqref{eq:VFbound} is violated and the effective theory should anyway not be trusted.

The results in this paper resolve this contradiction  since we saw that the limit $\Im z\to \infty$ at finite $\cV_{B_3}$ cannot be taken within the regime of validity for the supergravity effective action. Instead, to remain within the supergravity regime we have to co-scale the Einstein-frame volume of at least one divisor $D$ in $\cB_3$ as 
\begin{equation}\label{eq:VDZalpha}
    \cV_D \succsim (\Im\,z)^\alpha\,, 
\end{equation}
for some $\a>0$.\footnote{Notice that in~\cite{Grimm:2025cpq} it has already been stressed that the volume $\cV_{\cB_3}$ of the F-theory base can depend on the complex structure moduli of the fourfold $W$. However, the kind of complex structure dependence of the base volume is different from what we discuss in this paper. We consider additive corrections to the volume of divisors, whereas~\cite{Grimm:2025cpq} considers multiplicative corrections such as the factor of the dilaton appearing in the definition of the Einstein-frame volume that enters the K\"ahler potential.} In the regime where the supergravity approximation is valid, the quantum gravity cutoff is given by the 10d Planck scale 
\begin{equation}
    M_{10}^2 = \cV_{\cB_3}^{-1} M_{\rm Pl}^2\,,
\end{equation}
such that for $\text{Im}\,z\to \infty$ 
\begin{equation}
    \frac{V_F}{\Lambda_{\rm QG}^2} \precsim \frac{1}{\cV_{\cB_3}}\, V_0 \,(\Im\,z)^{d} \,.
\end{equation}
For a fixed choice of fluxes, the condition~\eqref{eq:VFbound} is hence satisfied if 
\begin{equation}\label{eq:VB3Imz}
    \cV_{\cB_3} \succsim (\Im z)^{d}\,.
\end{equation}
To decide whether such a co-scaling is indeed required by the corrections to the divisor volume we have to explicitly determine $\alpha$ in~\eqref{eq:VDZalpha}. As we mentioned repeatedly in this paper, this is in general a very difficult task. However, using dualities we computed the corrections to the volume of specific divisors $D_0$ for examples of type II and III limits in $\cM_{\rm c.s.}(W)$. From~\eqref{eq:correctionsIIB},\eqref{eq:VS2_corr}, and \eqref{VB_2corr} we find that in these cases the volume of the relevant divisors $D_0$ has to be co-scaled as in~\eqref{eq:VDZalpha} with $\alpha=d$. Using 
\begin{equation}
    \cV_{\cB_3} \succsim \cV_{D_0}\succsim \Im(z)^d \,,
\end{equation}
we find that in these examples the condition~\eqref{eq:VB3Imz} is satisfied. Hence, the condition for staying within the supergravity approximation imposed by the correction automatically guarantees that the flux-induced scalar potential parametrically satisfies~\eqref{eq:VFbound}. In other words, at least in the examples where the explicit corrections can be computed via duality, these corrections ensure that the supergravity approximation is self-consistent.\footnote{A similar situation arises for the non-perturbative scalar potential of 4d $\cN=1$ toroidal compactifications of the heterotic string. Using modular invariance, it can be shown~\cite{Cvetic:1991qm,Gonzalo:2018guu,Leedom:2022zdm} that the scalar potential diverges in certain classical infinite distance regimes. However, as shown in~\cite{Cvetic:2024wsj}, this divergence of the scalar potential is accompanied by non-perturbative corrections to the effective action becoming unsuppressed. This indicates that the original perturbative description of the theory becomes invalid similar to what happens for the F-theory setups discussed here.}

Our results, however, are not just relevant for flux choices that classically lead to a divergent scalar potential in the asymptotic regimes of $\cM_{\rm c.s.}(W)$. Instead, for phenomenological applications scalar potentials that classically decay in asymptotic regimes of $\cM_{\rm c.s.}(W)$ are more interesting as they could be cosmological models that feature asymptotic accelerated expansion. This possibility was analysed in detail in the F-theory context in~\cite{Calderon-Infante:2022nxb}, where in particular a candidate for a flux potential based on the supergravity expression~\eqref{eq:potentialSUGRA} was identified that can lead to asymptotic accelerated expansion along its gradient flow. However, as stressed in~\cite{Calderon-Infante:2022nxb}, whether or not this gives rise to an actual string theory realisation of cosmic acceleration depends on whether full moduli stabilisation can be achieved, including the K\"ahler moduli. The results in our paper indicate that even if full K\"ahler moduli stabilisation can be achieved, this does not yet guarantee that these models lead to accelerated expansion. Indeed, if all classical divisor volumes are set to some fixed value, we have seen that asymptotic limits in $\cM_{\rm c.s.}(W)$ (such as those underlying the proposal for accelerated expansion in~\cite{Calderon-Infante:2022nxb}) cannot be described within the supergravity approximation. However, the potential~\eqref{eq:potentialSUGRA} used in the analysis of~\cite{Calderon-Infante:2022nxb} has been derived within supergravity. Hence, even if there exist asymptotic limits in the quantum moduli space of the kind considered in~\cite{Calderon-Infante:2022nxb} realized at constant classical volume, these lie outside the regime of validity of the supergravity approximation underlying the analysis of asymptotic accelerated expansion.

\subsection{K\"ahler moduli stabilisation}\label{ssec:Kahlerstab}
Instead of considering cosmological solutions that dynamically realize asymptotic limits in the complex structure moduli space of F-theory, one can also attempt to find actual vacua in the interior of moduli space. For F-theory compactifications, the complex structure moduli can be classically stabilised by fluxes. However, K\"ahler moduli stabilisation is famously hard in Type IIB/F-theory compactifications. The reason is that there are no fluxes that can stabilise the K\"ahler moduli at tree-level such that one has to rely on perturbative and non-perturbative corrections to stabilise the K\"ahler moduli. For these to have a sizable effect, they cannot be entirely negligible in the effective action. Therefore, scenarios such as KKLT~\cite{Kachru:2003aw} and LVS~\cite{Balasubramanian:2005zx,Conlon:2005ki}, which rely on quantum corrections to achieve K\"ahler moduli stabilisation, cannot be realized in the strict asymptotic regime of the moduli space. The challenge of K\"ahler moduli stabilisation via quantum corrections to the effective action is thus to identify points in the interior of the moduli space where the quantum corrections required for moduli stabilisation are strong enough to create a non-trivial vacuum, while at the same time quantum corrections that would spoil the validity of the effective action can still be safely ignored. 

In the context of AdS vacua in the LVS and KKLT scenarios, the separation of quantum corrections can potentially be achieved since the non-perturbative corrections arise from rigid subsectors of the full theory of gravity. The reason is that, for a D3-brane instanton to contribute to the non-perturbative superpotential it has to wrap a rigid (or rigidified) divisor in the base of the Calabi--Yau fourfold. Rigid sectors can be decoupled from the theory of gravity. Geometrically this amounts to shrinking the corresponding divisor without shrinking the entire manifold.\footnote{The actual statement of decoupling is that the gauge coupling of a 7-brane gauge theory can be made parametrically stronger than the gravitational coupling. In 4d this means that the gauge kinetic function for this 7-brane gauge theory can vanish. Without fluxes, this is the case for shrinkable divisors. In the presence of fluxes, one has to take into account flux-induced corrections to the gauge kinetic function. If a divisor is rigidified by fluxes, as discussed for D3-brane instantons for example in~\cite{Bianchi:2011qh}, the gauge kinetic function can then also vanish even if the classical volume of the divisor cannot vanish.} In other words, the proposed scenarios attempt to find gravitational vacua within the regime of validity of the supergravity approximation using as central input non-perturbative effects in a field theory sector. 

As the volumes of all movable, and hence gravitational, divisors have to be stabilised by the non-perturbative field theory effects as well, also these divisor volumes cannot be arbitrarily large. In other words, the field theory sector cannot completely decouple from gravity. To check whether the point in moduli space at which the proposed vacuum is realized lies within the regime of validity of the supergravity approximation, one has to check whether possible corrections to the effective action are under control. Recall that in Type IIB/F-theory compactifications the classical effective action is derived from dimensional reduction of the 10d supergravity effective action. Already at the level of the supergravity, one can infer the strength of non-perturbative corrections to a given coupling in the effective action. These arise from instantons, i.e., solutions to saddle points of the classical Euclidean action. Consider the coefficient, $F_n$, of some operator in the effective action. At the two-derivative level we could for example consider the K\"ahler covariant superpotential
\begin{equation}
    G= K + \log|W|^2\,,
\end{equation}
but $F_n$ could equally well correspond to some higher-derivative term in the effective action. Schematically, we can then write 
\begin{equation}\label{eq:Gcorr}
    F_n = F_n^{\rm cl} + \delta F_n = F_n^{\rm cl} + \sum_{a}  \cA_a^{(n)} e^{- S_{\rm cl}^{a}}\,,
\end{equation}
where $F_n^{\rm cl}$ is the classical term obtained from dimensional reduction of the corresponding term in the 10d effective action. The sum in the last term runs over all possible instantons with classical instanton action $S_{\rm cl}^a$ and $\cA_a^{(n)}$ encodes the quantum fluctuations around the $a$-th Euclidean saddle. For Type IIB/F-theory compactifications, an important class of Euclidean saddles corresponds to a Euclidean D3-brane wrapping a divisor ${D_a \in H_4(\cB_3)}$ with classical action 
\begin{equation}
    S_{D3|_{D_{a}}} =2\pi  \left|\cV_{D_a}^{(0)} + \ii \int_{D_a} C_4\right|\,.
\end{equation}
Thus, the classical divisor volumes $\cV_{D_a}^{(0)}$ are good candidates for control parameters to check whether the corrections in~\eqref{eq:Gcorr} are small and the effective action is well-described by the dimensional reduction of the 10d effective action. In particular, if all classical divisor volumes are large, the volume of $\cB_3$ is large as well and we have to reproduce the classical 10d supergravity effective action. However, this is a purely classical control analysis, since the actual quantum nature of the corrections in \eqref{eq:Gcorr} is encoded in the one-loop determinant $\cA_a^{(n)}$. From a field theory perspective, the $\cA_a^{(n)}$ can simply be constants. Indeed, \cite{Demirtas:2021nlu} identifies rigid, i.e., field theory divisors with constant $\cA_a^{(n)}$.

In quantum gravity, we instead expect the $\cA_a^{(n)}$ to be moduli-dependent. In fact, the leading contribution to $\cA_a^{(n)}$ is expected to be moduli-dependent reflecting that quantum gravitational theories are \emph{not} field theories. The results presented in this paper imply that this is indeed the case. To see this, we interpret the logarithm of a non-zero $\cA_a^{(n_0)}$ as a correction to the instanton action (setting the axions to zero for simplicity)
\begin{equation}
    S_{{\rm D3}|_{D_a}} = 2\pi \cV_{D_a}^{(0)} -  \log |\cA_a^{(n_0)}|^2\,. 
\end{equation}
In other words, $\log \cA_a^{(n_0)}$ is a contribution to the quantum volume of the divisor. The corrections to the divisor volumes can schematically be split as (see e.g.~\cite{McAllister:2024lnt})
\begin{equation}
    \cV_{D_a} = \cV_{D_a}^{(0)} + \delta \cV_{D_a}^{\cN=2} + \delta \cV_{D_a}^{\cN=1}\,,
\end{equation}
where the second term encodes the $\cN=2$ corrections to the divisor volume already present if the elliptic fibration of the Calabi--Yau fourfold is trivial, i.e., in Type IIB compactifications on Calabi--Yau threefolds. Instead, the third term encodes genuine $\cN=1$ corrections to the divisor volumes which are absent in Type IIB Calabi--Yau compactifications. This term is generated by genuine quantum gravitational effects of the 4d $\cN=1$ theory. Computing these corrections directly in Type IIB orientifolds is very difficult. In this work, we analysed examples in F-theory where the moduli-dependence of $\delta\cV_{D_a}^{\cN=1}$ can be computed using duality and showed that they are very sensitive to the complex structure sector,
\begin{equation}\label{eq:deltaN1}
    \delta\cV_{D_a}^{\cN=1} \sim (\Im z)^\alpha\,\quad \alpha>0\,,
\end{equation}
for $\Im\,z\to \infty$. Moreover, the K\"ahler obstruction argued for via the worldsheet theory of the candidate EFT strings implies that in general asymptotic limits in $\cM_{\rm c.s.}(W)$ a correction of the form~\eqref{eq:deltaN1} arises for some divisors of $\cB_3$. Via the relation between the divisor volume and the D3-brane instanton action, the correction $\log \cA_a^{(n_0)}$ to the instanton action gets related to $\delta \cV_{D_a}^{\cN=1}$. 

Notice that even though the instanton may correct the effective action at a higher-derivative level, the shift in the D3-brane instanton actions implies that also the two-derivative effective action is corrected. The reason is that the holomorphic D3-brane instanton actions correspond to the complex scalar fields in the 4d $\cN=1$ multiplets. Therefore, a correction to the volume as in~\eqref{eq:deltaN1} will manifest itself in a perturbative correction to the K\"ahler potential. Indeed, since the K\"ahler potential in 4d $\cN=1$ is not related to a holomorphic object, it can receive corrections at all orders in perturbation theory. To ensure validity of the effective action, this perturbation theory must be under control. The results in this paper imply that in the full theory of quantum gravity, the control parameter for validity of the supergravity action is hence not simply given by the classical divisor volume $\cV_{D_a}^{(0)}$. If quantum gravity was describable as an ordinary quantum field theory, the classical divisor volume could be a good control parameter. However, since quantum gravity is very different from a quantum field theory, actual control over the supergravity effective action requires 
\begin{equation}
     \frac{\cV_{D_a}^{(0)}}{(\Im \,z)^\alpha}\gg 1\,,
\end{equation}
which ensures that the perturbative corrections to the two-derivative action are small compared to the tree-level term and that instanton corrections to any coupling in the full effective action are suppressed. This illustrates, once again, that a simple supergravity control analysis is insufficient since quantum gravity behaves very differently from the naive supergravity expectation. 

The tension between supergravity and quantum gravitational control analysis implied by our results challenges customary attempts for K\"ahler moduli stabilisation, which typically rely on a supergravity control analysis. In particular, the quantum corrections discussed in this paper may exacerbate the difficulties in finding controlled string vacua following the KKLT~\cite{Kachru:2003aw} or LVS~\cite{Balasubramanian:2005zx,Conlon:2005ki} approaches. This would be consistent with (yet completely independent of) the holographic arguments questioning the existence of controlled KKLT-like AdS vacua put forward in~\cite{Lust:2022lfc}. Unfortunately, so far, our results only indicate that the supergravity control analysis is insufficient and that an actual quantum gravity control analysis will considerably narrow down the regime of validity of the supergravity approximation. However, apart from some examples where we can compute $\delta\cV_{D_a}^{\cN=1}$ using dualities, we do not yet have a reliable tool to actually implement the quantum gravitational control analysis in a practical way. This would require a theory that encodes all $\alpha'$ corrections in the 4d $\cN=1$ theory similar to what F-theory achieves for the pure $g_s$ corrections to perturbative Type IIB orientifolds.  

\subsubsection*{Acknowledgements}
We thank Bj\"orn Hassfeld, Seung-Joo Lee, Severin L\"ust and Luca Martucci for useful discussions. JM thanks the string theory group at IFT, Madrid, for hospitality and interesting discussions. This work is supported in part by Deutsche Forschungsgemeinschaft under Germany’s Excellence Strategy EXC 2121 Quantum Universe 390833306, by Deutsche Forschungsgemeinschaft through a German-Israeli Project Cooperation (DIP) grant “Holography and the Swampland” and by Deutsche Forschungsgemeinschaft through the Collaborative Research Center 1624 “Higher Structures, Moduli Spaces and Integrability.” MW acknowledges support by  Deutsche Forschungsgemeinschaft through the Emmy Noether program 557478919. 

\appendix

\section{EFT strings in Type IIB Calabi--Yau compactifications}\label{app:WSmodesIIB}
In this appendix, we develop the mode counting on complex structure EFT strings of Type IIB string theory compactified on Calabi--Yau threefolds. For the special case of Tyurin type II degenerations of Calabi--Yau threefolds, this was pioneered in~\cite{Hassfeld:2025uoy}. Here, we extend this procedure to general Calabi--Yau threefold degenerations including those of type III and IV. We start by outlining the general strategy which we then exemplify for infinite distance limits in the complex structure moduli space of the mirror quintic in section~\ref{appssec:quintic} and the mirror of $\mathbb{P}^4_{1,1,1,6,9}[18]$ in section~\ref{sapp:P11169}. We further comment on the effect of orientifolding on the spectrum of EFT strings in section~\ref{appssec:Ofolds}. \\ 

Consider a semi-stable degeneration of a Calabi--Yau threefold $V$ as studied in~\cite{Hassfeld:2025uoy,Monnee:2025ynn,Monnee:2025msf,Paper1}. The family $V_z \hookrightarrow \cV \to \mathbf{D}$ can be viewed as a 4d EFT string configuration with the coordinate on the disk $\mathbf{D}$ identified with the coordinate $u$ transverse to the string. The degrees of freedom that propagate along the EFT string correspond to modes associated with the intersections of the components $V_i$ of $V_0$. Recall from~\cite{Monnee:2025ynn,Paper1} that infinite distance limits in the complex structure moduli space of Calabi--Yau threefolds can be classified by the dimension of the dual graph of the degeneration or, equivalently, the largest integer $d$ such that $V_{i_0\dots i_d}$ is non-empty. Concretely, for type II degenerations this means that only double surfaces $S_{ij}=X_i\cap X_j$ appear with all triple intersections of the components $V_i$ empty. Instead, type III degenerations are characterised by non-vanishing triple curves $C_{ijk}= X_i\cap X_j \cap X_k$ for some $i\neq j\neq k$ but vanishing quadruple intersections. Type IV singularities also have non-vanishing quadruple points $P_{ijkl}= X_i\cap X_j\cap X_k \cap X_l$.

Paralleling our analysis in Section~\ref{ssec:countingFM}, the massless degrees of freedom on the EFT strings associated with any degeneration of type II, III or IV have three different origins. 

\paragraph{Universal geometric mode.} Irrespective of the type of the degeneration there is a universal geometric zero mode on the string associated with the position of the string in the 4d spacetime. As in~\cite{Hassfeld:2025uoy} we denote this mode by $\mathbf{z}_0=|\mathbf{z}_0| e^{\ii \arg{\mathbf{z}_0}}$ providing two real scalars on the worldsheet, which we choose to be the non-compact scalar $|\mathbf{z}_0|$ and the compact scalar $\arg(\mathbf{z}_0)$. 

\paragraph{Internal geometric modes.} In addition to the mode describing the motion of the string in the four-dimensional extended spacetime, there are additional modes associated with the location of the intersection inside the normal crossing variety $V_0$. The number of geometric modes is determined by the type of degeneration as it corresponds to the dimension of the dual graph of the degeneration. The dual graph of type II degenerations being one-dimensional, the EFT string realizing a type II limit has a single real scalar degree of freedom $\Phi_1$ coming from the internal geometric sector. Instead, EFT strings associated with type III degenerations have two real scalar modes, $\Phi_1, \Phi_2$, on their worldsheet and type IV degenerations three real scalar modes $\Phi_1,\Phi_2,\Phi_3$ arising from the internal geometry of the degeneration. 

\paragraph{Modes from $p$-forms.} In addition to the geometric modes, there are massless degrees arising from modes of the $p$-forms of 10d supergravity that localise to the intersections of the components of $V_0$. We therefore consider the harmonic forms in $H^\bullet(S_{ij})$ and $H^\bullet(C_{ijk})$. As in~\cite{Hassfeld:2025uoy}, we obtain a degree of freedom propagating along the string by reducing a $p$-from potential $C_p$ over a harmonic $(p-2)$-form $\omega_{p-2}$ as 
\begin{equation}
    C_p = \mathsf{B}_{(2)}\wedge \omega_{p-2}+\dots \,,
\end{equation}
The ${\rm d}z\wedge {\rm d}\bar{z}$ component of the two-form $\mathsf{B}_{(2)}$ then gives rise to a scalar mode $b$ that can be interpreted as a mode on the string. In Type IIB we can thus consider $C_4$, $C_6$ and $B_6$ and reduce them as 
\begin{equation}
    C_6 = \widetilde{\mathsf{C}}^\alpha\wedge  \omega^{(4)}_\alpha \,,\quad B_6 = \widetilde{\mathsf{B}}^\alpha\wedge \omega^{(4)}_\alpha\,  \qquad \omega_\a^{(4)} \in \bigoplus_{i<j} H^4(S_{ij}) 
\end{equation}
and 
\begin{equation}
    C_4 = \mathsf{C}^a \wedge \omega_a^{(2)} + \mathsf{C}^i \wedge \omega_i^{(2)}\,,\quad \text{with} \quad \omega_a^{(2)} \in \bigoplus_{i<j} H^2(S_{ij})\,,\;\omega_i^{(2)} \in \bigoplus_{i<j<k} H^2(C_{ijk})\,.   
\end{equation}
For type II degenerations, there are no triple curves such that all modes come from the two-forms on the double surfaces $S_{ij}$. Using that these are all expected to be K3 surfaces,~\cite{Hassfeld:2025uoy} showed that the EFT string associated with a type II degeneration is a critical heterotic string. Since for types III and IV singularities the surfaces $S_{ij}$ intersect non-trivially, the expansions above generally overcounts the number of two-forms in this case, as the forms in $\bigoplus_{i<j} H^\bullet(S_{ij})$ and $\bigoplus_{i<j<k}H^\bullet(C_{ijk})$ are not all independent but satisfy linear relations. The exact counting of massless degrees of freedom on the string worldsheet must take into account these linear relations, which depend on the details of the degenerate geometry.

The string worldsheet has 2d $\cN=(0,4)$ supersymmetry such that the degrees of freedom must organize into full multiplets of the right-moving supersymmetry algebra. So far, we have only considered the scalar degrees of freedom. By reducing the fermionic partners of the 10d $p$-form potentials, we generate fermions that complete the right-moving degrees scalars into (twisted) hypermultiplets.  In the following, we discuss the counting in concrete examples of type III and IV degenerations. Let us notice that, in case the degeneration in the complex structure limit has a mirror dual description in terms of a large volume limit in a geometric compactification of Type IIA string theory, the counting of the zero modes and hence the central charge of the string worldsheet theory can be compared to the mirror dual of the Type IIB EFT string corresponding to Type IIA NS5-branes wrapping certain divisors in the mirror Calabi--Yau threefold $\hat{V}$. For type II limits, the mirror map to Type IIA NS5-branes has been discussed in detail in~\cite{Hassfeld:2025uoy}. For more general limits, we notice that the central charges on the mirror dual string obtained by wrapping an NS5-brane on a nef and effective divisor $\hat{D}$ in $\hat{V}$ are given by~\cite{Maldacena:1997de}
\begin{align}\label{centralchargesIIA}
        c_R = \hat{D}^3 + \frac12 c_2(\hat{V})\cdot \hat{D}\,,\qquad     c_L = \hat{D}^3 + c_2(\hat{V})\cdot \hat{D}\,,
\end{align}
where $c_2(\hat{V})$ is the second Chern class of $\hat{V}$. These properties of $\hat{D}$ guarantee that the resulting string is (super-)ciritical. 

In the following, we apply the procedure outlined above to two EFT strings arising in concrete Calabi--Yau compactifications of Type IIB string theory and furthermore discuss an EFT string point of view on the results of the companion paper~\cite{Paper1}.

\subsection{Type IV example: Mirror quintic}\label{appssec:quintic}
The mirror quintic is given by the hypersurface
\begin{equation}\label{quintic}
     P = x_1^{5}+x_2^{5}+x_3^{5}+x_4^5+x_5^5 -5\psi\, x_1 x_2 x_3 x_4 x_5=0\,
\end{equation}
inside $\mathbb{P}^4/\mathbb{Z}_5^3$, and the large complex structure limit $\psi\to \infty$ corresponds to a type IV singularity. In this limit, the threefold degenerates into a union of five threefolds, 
\begin{equation}\label{eq:mquintic-degeneration}
    V_0 = \bigcup_{i=1}^5 \{x_i=0\}\,.
\end{equation}
We are interested in the worldsheet theory of the EFT string realizing the limit $\psi\to \infty$. Before we come to the mode counting using the strategy outlined above, we discuss the mirror of this EFT string. To this end, we consider Type IIA compactified on the quintic threefold in $\mathbb{P}^4$, i.e., we do not orbifold the ambient space by $\mathbb{Z}_5^3$. The mirror dual of the EFT string realizing the large complex structure limit of the mirror quintic corresponds to a Type IIA NS5-brane wrapping the hyperplane divisor $\hat{D}$ on the quintic. Using the topological data of the quintic, 
\begin{equation}
    \hat{D}^3 =5\,,\qquad c_2\cdot \hat{D} =50\,,
\end{equation}
we find from~\eqref{centralchargesIIA} that the left- and right-moving central charges of the string obtained from the NS5-brane wrapped on $\hat{D}$ are 
    $c_R= 30$ and $c_L =55$.
The string worldsheet theory thus contains five (twisted) hypermultiplets contributing $c_R({\rm hyper})=6$ each. Therefore, the NS5-brane string associated with the large volume limit of the quintic is a supercritical string. 

Returning to the EFT string in Type IIB string theory, we now describe how the counting outlined above can be applied to the degeneration of the mirror quintic in~\eqref{eq:mquintic-degeneration}. In the following, we focus on the right-moving sector. The EFT string realizing the large complex structure limit for the mirror quintic contains five real scalars corresponding to the universal and internal geometric modes $|\mathbf{z}_0|, \arg(\mathbf{z}_0), \Phi_{1,2,3}$. To count the independent two-forms (and thus worldsheet scalars on the string), we have to determine the number of independent double surfaces and triple curves and their cohomology classes. 

The degenerate geometry consists of $\binom{5}{2}=10$ double surfaces $S_{ij}$, the same number of triple curves $C_{ijk}$ as well as five quadruple points. The double surfaces have the topology of $\mathbb{P}^2$, whereas the triple curves are $\mathbb{P}^1$s. The double surfaces satisfy linear relations as divisors in the components $X_i=\{x_i=0\}$. To determine these, we notice that for $\psi\to \infty$, the zero section of the polynomial $P$ defined in~\eqref{quintic} restricts trivially to $X_i=\{x_i=0\}$, i.e., 
\begin{equation}
    \text{div}(P|_{X_i}) = 0\in \text{Div}(X_i)\,. 
\end{equation}
On the other hand, we have 
\begin{equation}
    \text{div}(P|_{X_i}) = X_i \cap\sum_{i\neq j} X_j\,,
\end{equation}
implying that for each $i=1,\dots,5$ there is a linear relation satisfied by the $S_{ij}=X_i\cap X_j$ in $\text{Div}(X_i)$. The number of independent double surfaces is thus reduced to $10-5=5$.\footnote{Over $\mathbb{Q}$, not all linear relations are independent since they satisfy themselves one linear relation, showing that over $\mathbb{Q}$ there are only four relations for the double surfaces. Here, however, we work over $\mathbb{Z}$ so that we have five independent relations.} Similarly, the number of independent triple curves is also reduced to five. 

From the five independent double surfaces we obtain five pairs of two-forms upon reducing $C_6$ and $B_6$ along the top-form on the surfaces whose ${\rm d}z\wedge{\rm d}\bar{z}$ component gives real scalar fields $\beta^\a$ and $\gamma^\a$, $\a = 1,\dots, 5$ along the string. 
To determine the number of worldsheet modes coming from $C_4$, we first consider the triple curves. Since there are five independent triple curves, these give rise to five two-forms $C^i$, $i=1,\dots, 5$, upon reducing $C_4$ along them. On the worldsheet, these give rise to five real scalars $c^i$. Inside $S_{ij}\simeq \mathbb{P}^2$, the triple curves $C_{ijk}$ (if non-zero) are representatives of the hyperplane class. Thus, reducing $C_4$ over the elements in $H^2(S_{ij})$ does not yield any additional degrees of freedom on the string. In total, there are therefore 15 additional modes in the right-moving sector coming from the Type IIB $p$-forms.  Together with the geometric modes, these form the scalar degrees of freedom of five (twisted) hypermultiplets. Along with their fermionic partners, these indeed give $c_R=30$, consistent with the prediction from mirror symmetry. 

\subsection{Type III example: \texorpdfstring{$\mathbb{P}^4_{1,1,1,6,9}[18]/(\mathbb{Z}_{18}\times \mathbb{Z}_6)$}{P11169[18]/(Z18xZ6)}}\label{sapp:III-ex}
To study an EFT string realizing a type III degeneration, we consider the mirror $V$ of the (resolved) Calabi--Yau threefold $\hat{V}=\mathbb{P}^4_{1,1,1,6,9}[18]$. We provide details about this manifold in Appendix~\ref{sapp:P11169}. As established there, the degeneration
\begin{equation}
    V_0\equiv V_{\phi\to\infty}=\bigcup_{i=1}^3\{x_i^6=0\}/\Tilde{G}_{\rm GP}
\end{equation}
is a semi-stable type III degeneration. For the mode counting on the EFT string realizing this limit at its core, the geometry of the intersections of the components $X_i$ is important. We first note that by adjunction the triple curve $C_{123}=X_1\cap X_2\cap X_3$ is an elliptic curve,
\begin{equation}
    c_1(C_{123})=(18-3\times 6)H\vert_{C_{123}}=0\,.
\end{equation}
To specify the geometry of the double surfaces $S_{ij}$, $i<j$, we find (again by adjunction) $c_1(S_{12})=6H\vert_{S_{12}}$, where $H$ is a generic hyperplane of $V_0$. With $H=\{x_3=0\}$ we conclude $K_{S_{12}}=-6C_{123}$. As $g(C_{123})=1$, adjunction for $C_{123}\subset S_{12}$ tells us that $0=K_{S_{12}}\cdot C_{123}+C_{123}^2=-5C_{123}^2$, i.e., $C_{123}$ is an elliptic fiber inside $S_{12}$. From the Hodge-Deligne diamond associated with the degeneration we get $h^{2,0}(S_{12})=0$. Furthermore, $h^{1,0}(S_{12})=g(\Sigma)$, where $\Sigma$ is the base curve of the elliptic fibration of $S_{12}$. The projection $S_{12}\rightarrow \Sigma$ is given by
\begin{equation}
    [0:0:x_3:x_4:x_5]\mapsto [x_4:x_5]\,,
\end{equation}
showing that $\Sigma\simeq\mathbb{P}^1_{6,9}\simeq\mathbb{P}^1$. Hence, $h^{1,0}(S_{12})=0$, so that the Kodaira dimension of $S_{12}$ is $\kappa(S_{12})=-\infty$ and $S_{12}$ is a rational elliptic surface, i.e., $S_{12}\simeq{\rm dP}_9$.\footnote{Alternatively, Theorem 6.12 of~\cite{schuett2010ellipticsurfaces} can be used to determine the full Hodge diamond of $S_{12}$. It indeed coincides with that of ${\rm dP}_9$.}\\

Using our strategy outlined at the beginning of this section, we now describe the worldsheet theory on the EFT string realizing the $\phi\to \infty$ limit. As before, it is instructive to consider first the mirror dual of this string corresponding to a Type IIA NS5-brane wrapping a divisor in the mirror threefold. The mirror $\hat{V}$ of $V$ is an elliptic fibration over $\mathbb{P}^2$. The string dual to the EFT string realizing the type III limit corresponds to the NS5-brane wrapping the vertical divisor $\hat{D}_h$ over the hyperplane class $h$ of $\mathbb{P}^2$. Using $\hat{D}_h^3=0$ and $c_2\cdot \hat{D}_h=36$, the central charges are of the string are given by 
\begin{equation}\label{eq:centralchargesIII}
    c_R = 18\,,\qquad c_L=36\,. 
\end{equation}
Notice that the string can be viewed as a bound state of three E-strings, each contributing $c_R=6$ and $c_L=12$. The contributions to $c_L$ come from the left-moving $E_8$-algebra on each of the three E-strings. 

To reproduce the central charge directly for the Type IIB EFT string, we first collect the geometric modes. The modes transverse to the degeneration $\mathbf{z}_0$ and $\Phi_1,\Phi_2$ provide in total four real scalars on the string worldsheet. To count the modes arising from localised $p$-forms, we first notice that only two out of the three double surfaces $S_{12}$, $S_{13}$ and $S_{23}$ are independent as divisors in $V_0$. We thus obtain four real scalars $\beta^\alpha$, $\gamma^\alpha$, $\alpha=1,2$, from reducing $B_6$ and $C_6$ over the top forms of these two independent surfaces. The counting of the modes arising from $C_4$ is more involved. We first notice that the single triple curve $C_{123}$ yields a single real scalar $b^1$ arising from the component of $C_4$ along the unique element in $H^2(C_{123})$. Each of the three double surfaces is a ${\rm dP}_9$ such that 
\begin{equation}
    \bigoplus_{i<j} H^2(S_{ij},\mathbb{Z}) = U^{\oplus 3} \oplus (-E_8)^{\oplus 3}\,,
\end{equation}
which has signature
\begin{equation}
    {\rm sgn}\,\left[\bigoplus_{i<j} H^2(S_{ij})\right] = (3,27) \,. 
\end{equation}
There is a second linear relation as the elliptic triple curve $C_{123}$ is contained in all three double surfaces. The contribution of this elliptic curve to the massless degrees of freedom is therefore already accounted for by $b^1$. In $\bigoplus_{i<j} H^2(S_{ij},\mathbb{Z})$, we thus have to consider the complement of this two-form which is given by 
\begin{equation}\label{eq:H2null}
   \left[\bigoplus_{i<j} H^2(S_{ij},\mathbb{Z})/\sim\right] = (0)^{\oplus 3 }\oplus (-E_8)^{\oplus 3}\,. 
\end{equation}
What remains of the hyperbolic $U$-planes are three null-directions. Reducing $C_4$ along these null-directions gives modes $b^i$, $i=2,3,4$, without a definite chirality and hence three real scalars on the worldsheet of the EFT string. In summary, we thus have 12 right-moving and 36 left-moving scalar degrees of freedom which, taking into account the right-moving supersymmetry, reproduces the central charges in~\eqref{eq:centralchargesIII}. 

\subsection{EFT strings and orientifolds}\label{appssec:Ofolds} 

In the previous subsection we saw that -- similar to the type II limits analysed in~\cite{Hassfeld:2025uoy} -- also for limits of type III and IV the number of double surfaces arising at the degeneration are crucial in order to determine the zero mode spectrum on the EFT string associated with the degeneration. More specifically, an EFT string realizing a type III or IV degeneration can be viewed as a bound state of multiple strings, one for each double surface $V_{i_0i_1}$ of the degeneration.

Taking an orientifold of the Calabi--Yau threefold $V$ introduces two effects that have an impact on the worldsheet spectrum of the original EFT string. First, since the orientifold projects out certain Type IIB supergravity fields, also the corresponding zero modes are projected out. For O-type B limits as introduced in~\cite{Paper1} this is the only relevant effect and since these limits lift trivially to F-theory, we refer to the main text for an analysis of the worldsheet spectrum of the associated EFT string.

Apart from this truncation due to the orientifold action on the Type IIB supergravity fields, in O-type A limits there is a second effect acting on the EFT string worldsheet as in these limits the number of double surfaces arising at the degeneration is reduced.  Correspondingly, the EFT string realizing the orientifolded limit in the $\cN=1$ moduli space is very different from the original EFT string in the $\cN=2$ moduli space. For type II O-type A limits, there is no double surface and hence the EFT string disappears altogether. Thus, also the EFT string perspective confirms that for type II limits, there is no infinite distance limit left in the F-theory uplift of an O-type A orientifold as the resulting type I singularity of the fourfold is at finite distance in moduli space and therefore not associated with an EFT string. For type III and IV limits in O-type A orientifolds, the EFT string does not have to disappear altogether but only some components of the bound state survive the orientifold action. Thus, there is still a string associated with this limit although its worldvolume theory is very different from the original EFT string in the 4d $\cN=2$ parent theory. In line with the general discussion of the F-theory uplift of O-type A orientifolds in Section~4 of~\cite{Paper1}, the EFT strings associated with O-type A limits are best described directly in F-theory, as done in the main text of this article.
\section{Details on \texorpdfstring{$\mathbb{P}^4_{1,1,1,6,9}[18]/(\mathbb{Z}_{18}\times\mathbb{Z}_6)$}{P11169[18]/(Z18xZ6)}}\label{sapp:P11169}
In this section, we consider the 2-parameter mirror $V$ of the resolved Calabi--Yau threefold $V'=\mathbb{P}^4_{1,1,1,6,9}[18]$. The only singularities on the Calabi--Yau hypersurface are those inherited from the ambient space $X=\mathbb{P}^4_{1,1,1,6,9}$,  which occur along the curve $C=\{x_1=x_2=x_3=0\}$ and at the two points $p_2=[0:0:0:1:0]$ and $p_3=[0:0:0:0:1]$. As mentioned in~\cite{Candelas:1994hw}, the curve intersects the hypersurface in the point $p_1=[0:0:0:1:-1]$, while $p_{2,3}$ do not lie on the hypersurface. To arrive at a smooth Calabi--Yau we therefore have to perform a single blow-up of the point $p_1$. To do this, we first blow-up the ambient space along the curve $C$,
\begin{equation}
    \pi:\Tilde{X}={\rm Bl}_C(X)\longrightarrow X,
\end{equation}
such that $\Tilde{X}\subset\mathbb{P}^4_{1,1,1,6,9}\times\mathbb{P}^2_{[y_1:y_2:y_3]}$ is defined by the equations $x_iy_j=x_jy_i$ for $i,j\in\{1,2,3\}$. The exceptional divisor $E$ is a $\mathbb{P}^2$-bundle over $C$ with fiber coordinates $[y_1:y_2:y_3]$. 

Inside the original weighted projective space $X$, we consider the zero locus
\begin{equation}
    \{P=x_1^{18}+x_2^{18}+x_3^{18}+x_4^3+x_5^2-18\psi x_1x_2x_3x_4x_5-3\phi x_1^6x_2^6x_3^6+\,270\,\,\text{terms}=0\}\subset X\,,
\end{equation}
defining a singular Calabi--Yau threefold. The smooth threefold $\Hat{V}=\mathbb{P}^4_{1,1,1,6,9}[12]$ is given by the proper transform of this zero locus in the resolved ambient space $\Tilde{X}$. Expressed in local coordinates in a neighbourhood intersecting $E$ such that $E=\{t=0\}$ and $x_i=ty_i$ for $i\in\{1,2,3\}$, this proper transform (which due to the genericity of $P$ coincides with the total transform) is given by
\begin{equation}\label{eq:P11169-P-blow-up}
    \Hat{V}=\{\Tilde{P}=t^{18}(y_1^{18}+y_2^{18}+y_3^{18})+x_4^3+x_5^2-18\psi t^3y_1y_2y_3x_4x_5-3\phi t^{18}y_1^6y_2^6y_3^6+\,270\,\,\text{terms}=0\}\,.
\end{equation}
As expected, the intersection of $\Hat{V}$ with $E$ is given by the $\mathbb{P}^2$-fiber of $E$ above the point $p_1$, which therefore constitutes a second divisor class on the hypersurface defined by~\eqref{eq:P11169-P-blow-up}. This hypersurface defines the smooth Calabi--Yau threefold $\hat{V}=\mathbb{P}^4_{1,1,1,6,9
}[18]$ with Hodge numbers $h^{1,1}(\Hat{V})=2$, $h^{2,1}(\Hat{V})=272$.

The mirror $V$ of $\hat{V}$ is given by $\{\Tilde{P}=0\}/\Tilde{G}_{\rm GP}$, where $\Tilde{G}_{\rm GP}$ describes the action of the Greene--Plesser group $G_{\rm GP}=\mathbb{Z}_6\times\mathbb{Z}_{18}$ in the blow-up. Generators of $G_{\rm GP}$ are given by~\cite{Candelas:1994hw}
\begin{equation}
    g^{(1)}=(0,1,3,2,0),\quad g^{(2)}=(1,-1,0,0,0)\,.
\end{equation}
Generators of $\Tilde{G}_{\rm GP}$ that leave $C$ invariant act on our chosen local coordinates $[t:y_1:y_2:y_3:x_4:x_5]$ as
\begin{equation}
    \Tilde{g}^{(1)}=(0,0,1,3,2,0),\quad\Tilde{g}^{(2)}=(1,0,-2,-1,0,0)\,.
\end{equation}
Similar to~\cite{Candelas:1994hw} one can show that the terms written explicitly in~\eqref{eq:P11169-P-blow-up} are the only degree 18 terms invariant under $\Tilde{G}_{\rm GP}$. Thus,
\begin{equation}
    V=\{t^{18}(y_1^{18}+y_2^{18}+y_3^{18})+x_4^3+x_5^2-18\psi t^3y_1y_2y_3x_4x_5-3\phi t^{18}y_1^6y_2^6y_3^6=0\}/\Tilde{G}_{\rm GP}\,,
\end{equation}
where the quotient introduces new cyclic quotient singularities. Since we are interested in complex structure degenerations, their resolution is not of importance to us.

\paragraph{Type III degeneration.} The complex structure moduli space of the smooth Calabi--Yau $V$ is two-dimensional and parametrised by $(\psi,\phi)$. In Appendix~\ref{app:WSmodesIIB}, we consider the type III degeneration $\phi\to\infty$ of $V$, for which we now show that it is semi-stable in the sense of Deligne--Mumford. To this end, we first notice that, at the level of the unresolved Calabi--Yau $V'=\{P=0\}$, the degeneration gives rise to a threefold $V'_0=\{x_1^6x_2^6x_3^6=0\}$ which does contain the two singular points $p_2$ and $p_3$. Moreover, these singular points lie on the triple curve $C$ of the degeneration. Thus, we first have to blow-up the singularities in the triple curve, then the residual singularity at the level of double surfaces and finally all singularities that are still left from the threefold perspective. In the following we focus on the blow-ups of $p_2\in C$. The analysis for $p_3$ works the same. 
\begin{itemize}
    \item The singularity $(C,p_2)$ is of type $\frac{1}{6}(3)$, which we resolve by introducing the exceptional divisor $E_{(2),1}=\mathbb{P}^2_{[u_1:u_2:u_3]}=\{t=0\}$ with $x_i=tu_i$ for $i=1,2,3$. The proper transform of $V'_0$ is then given by $\{u_1^6u_2^6u_3^6=0\}$.
    \item Notice that the singularity is not fully resolved yet. Indeed, the double surface $\{u_1^6=u_2^6=0\}$ intersects $E_{(2),1}$ in the point $[0:0:u_3=1]\in E_{(2),1}$. On the normal coordinates there is a residual $\mathbb{Z}_3$-action, yielding a $\frac{1}{3}(1,1,1)$ singularity in $E_1$.\footnote{The tangent directions, including $x_4$, are fixed, which means that only those $\lambda\in\mathbb{Z}_9$ with $\lambda^3=1$ have a residual action on the normal coordinates.} This can again be blown-up by introducing an exceptional divisor $E_{(2),12}=\mathbb{P}^2_{[v_0:v_1:v_2]}=\{s_{12}=0\}$ with $s_{12}v_0=t$ and $s_{12}v_i=u_i$, $i=1,2$. We find similar relations for the other two double surfaces. 
\end{itemize}
The total transform of $V'_0$ after these blow-ups is then given by
\begin{equation}
    \{t^{18}s_{12}^{12}s_{23}^{12}s_{13}^{12}z_1^6z_2^6z_3^6=0\}=\{z_1^6z_2^6z_3^6=0\}+18E_{(2),1}+12E_{(2),12}+12E_{(2),23}+12E_{(2),13},
\end{equation}
from which the proper transform is again obtained by subtracting the exceptional contributions. Translating this to the resolved mirror $V$ constructed before, we find
\begin{equation}\label{eq:total-tr-P11169}
    \lim_{\phi\to\infty}V=\left(\bigcup_{i=1}^3\{z_i^6=0\}+18E_{(2),1}+12(E_{(2),12}+E_{(2),23}+E_{(2),13})\right)/\Tilde{G}_{\rm GP},
\end{equation}
which coincides with the total transform of the central fiber of the degeneration $\phi\to\infty$ in the unresolved $V_{\rm sing}=\mathbb{P}^4_{1,1,1,6,9}[18]/G_{\rm GP}$. As we are interested in the proper transform, however, we have to subtract from~\eqref{eq:total-tr-P11169} the exceptional contributions. Thus, the central fiber of the type III degeneration $\phi\to\infty$ in the smooth mirror $V$ of $\hat{V}$ is given by
\begin{equation}
    V_{\phi\to\infty}=\bigcup_{i=1}^3\{z_i^6=0\}/\Tilde{G}_{\rm GP},
\end{equation}
where due to the action of $\Tilde{G}_{\rm GP}$ all components $V_i=\{z_i^6=0\}/\Tilde{G}_{\rm G}$ are reduced so that $V_{\phi\to\infty}$ only has simple normal crossing singularities and the degeneration  of $V$ in the type III limit, $\phi\to\infty$, is semi-stable.

\bibliography{papers_Max}
\bibliographystyle{JHEP}

\end{document}